\documentclass{aa}  

\usepackage{graphicx}
\usepackage{txfonts}
\usepackage[colorlinks=true,allcolors=blue]{hyperref}

\usepackage{ragged2e}

\newcommand{\oiii}{[O\,{\sc{iii}}]}
\newcommand{\sii}{[S\,{\sc{ii}}]}
\newcommand{\nii}{[N\,{\sc{ii}}]}

\newcommand{\civ}{C\,{\sc{iv}}}
\newcommand{\heii}{He\,{\sc{ii}}}
\usepackage{xspace}
\newcommand{\hb}{H$\beta$}
\newcommand{\ha}{H$\alpha$}
\newcommand{\lya}{Ly$\alpha$}
\newcommand{\emsm}{{\it emsm}}
\newcommand{\drizzle}{{\it drizzle}\xspace}
\newcommand{\lbqs}{{LBQS 0302$-$0019}}
\newcommand{\kms}{{km s$^{-1}$}}

\begin{document}

    \title{GA-NIFS: The ultra-dense, interacting environment of a dual AGN at $z\sim3.3$ revealed by JWST/NIRSpec IFS}

    \titlerunning{GA-NIFS: JWST/NIRSpec view of \lbqs}
   \author{M. Perna
          \inst{\ref{iCAB}}\thanks{e-mail: mperna@cab.inta-csic.es}
          \and
          S. Arribas\inst{\ref{iCAB}}
          \and
          M. Marshall\inst{\ref{iNRC}}
          \and 
          F. D'Eugenio\inst{\ref{iKav},\ref{iCav}}
          \and 
          H. \"Ubler\inst{\ref{iKav},\ref{iCav}}
          \and 
          A. Bunker\inst{\ref{iOxf}}
          \and 
          S. Charlot\inst{\ref{iSor}}
          \and 
          S. Carniani\inst{\ref{iNorm}}
          \and 
          P. Jakobsen\inst{\ref{iDAWN},\ref{iBohr}}
          \and
          R. Maiolino\inst{\ref{iKav},\ref{iCav}, \ref{iUCL}}
          \and
          B. Rodr\'iguez Del Pino\inst{\ref{iCAB}}
          \and
          C. J. Willott\inst{\ref{iNRC}}
          \and 
          T. B{\"o}ker\inst{\ref{iESOba}}
          \and
          C. Circosta\inst{\ref{iESA}, \ref{iUCL}}
          \and
          G. Cresci\inst{\ref{iOAA}}
          \and 
          M. Curti\inst{\ref{iESOge}}
          \and
          B. Husemann\inst{\ref{iEUMETSAT}}
          \and
          N. Kumari\inst{\ref{iAURA}}
          \and
          I. Lamperti\inst{\ref{iCAB}}
          \and
          P. G.~P\'erez-Gonz\'alez\inst{\ref{iCAB}}
          \and
          J. Scholtz\inst{\ref{iKav},\ref{iCav}}
          }

   \institute{Centro de Astrobiolog\'ia (CAB), CSIC--INTA, Departamento de Astrof\'\i sica, Cra. de Ajalvir Km.~4, 28850 -- Torrej\'on de Ardoz, Madrid, Spain\label{iCAB}
        \and 
        National Research Council of Canada, Herzberg Astronomy \& Astrophysics Research Centre, 5071 West Saanich Road, Victoria, BC V9E 2E7, Canada\label{iNRC}
        \and
        Kavli Institute for Cosmology, University of Cambridge, Madingley Road, Cambridge, CB3 0HA, UK\label{iKav}
        \and
        Cavendish Laboratory - Astrophysics Group, University of Cambridge, 19 JJ Thomson Avenue, Cambridge, CB3 0HE, UK\label{iCav}
        \and
        Department of Physics, University of Oxford, Denys Wilkinson Building, Keble Road, Oxford OX1 3RH, UK\label{iOxf}
        \and
        Sorbonne Universit\'e, CNRS, UMR 7095, Institut d’Astrophysique de Paris, 98 bis bd Arago, 75014 Paris, France\label{iSor} 
        \and
        Scuola Normale Superiore, Piazza dei Cavalieri 7, I-56126 Pisa, Italy\label{iNorm}
        \and 
        Cosmic Dawn Center (DAWN), Copenhagen, Denmark  Niels Bohr Institute, University of Copenhagen, Jagtvej 128, DK-2200, Copenhagen, Denmark\label{iDAWN}
        \and 
        Niels Bohr Institute, University of Copenhagen, Jagtvej 128, DK-2200, Copenhagen, Denmarkk\label{iBohr}
        \and
        Department of Physics and Astronomy, University College London, Gower Street, London WC1E 6BT, UK\label{iUCL}
        \and
        European Space Agency, c/o STScI, 3700 San Martin Drive, Baltimore, MD 21218, USA\label{iESOba}
        \and
        European Space Agency (ESA), European Space Astronomy Centre (ESAC), Camino Bajo del Castillo s/n, 28692 Villanueva de la Cañada, Madrid, Spain\label{iESA}
        \and 
        INAF - Osservatorio Astrofisco di Arcetri, largo E. Fermi 5, 50127 Firenze, Italy\label{iOAA}
        \and 
        European Southern Observatory, Karl-Schwarzschild-Straße 2, 85748, Garching, Germany\label{iESOge}
        \and
        EUMETSAT - Fertigung für Verteidigung und Raumfahrt, Darmstadt, Germany\label{iEUMETSAT}
        \and
        AURA for European Space Agency, Space Telescope Science Institute, 3700 San Martin Drive. Baltimore, MD, 21210\label{iAURA}
             }

   \date{Received September 15, 1996; accepted March 16, 1997}

 
  \abstract
   {\lbqs \ is a blue quasar (QSO) at $z\sim 3.3$ that hosts powerful outflows and resides in a complex environment consisting of an obscured active galactic nucleus (AGN) candidate and multiple companions, all within 30 kpc in projection.  
   }
   {We aim to characterise this complex system using JWST NIRSpec Integral Field Spectrograph (IFS) observations obtained as part of the NIRSpec IFS GTO
programme `Galaxy Assembly with NIRSpec IFS' (GA-NIFS); 
   these data cover the QSO rest-frame optical emission lines with a spatial resolution of $\sim 0.1\arcsec$ and a sampling of 0.05\arcsec ($\sim 380$ pc) over a contiguous sky area of $\sim 3\arcsec \times 3\arcsec$ ($23\times 23$ kpc$^2$).}
   {We developed a procedure to correct for the spurious oscillations (or `wiggles') in NIRSpec single-spaxel spectra caused by the spatial under-sampling of the point spread function. We performed a QSO–host decomposition with the {\sc{QDeblend3D}} tools. We used multi-component kinematic  decomposition of the optical emission line profiles to infer the physical properties of the emitting gas in the  QSO environment. 
   }
   {The QSO--host decomposition allows us to identify both a low- and a high-velocity component. The former possibly traces a warm rotating disk with a dynamical mass $M_{dyn}\sim 10^{11}$ M$_\odot$ and a rotation-to-random motion ratio $v_{\rm{rot}}/\sigma_0 \sim 2$. The other kinematic component traces a spatially unresolved ionised outflow with a velocity of $\sim 1000$ \kms \ and an outflow mass rate of $\sim 10^4$ M$_\odot$ yr$^{-1}$. 
   We clearly detect eight companion objects close to \lbqs. 
   For two of them, we detect a regular velocity field that likely traces rotating gas, and we infer individual dynamical masses of $\approx 10^{10}$ M$_\odot$. Another companion shows evidence of gravitational interaction with the QSO host. Optical line ratios confirm the presence of a second, obscured AGN $\sim 20$ kpc from the primary QSO; the dual AGN dominates the ionisation state of the gas in the entire NIRSpec field of view.}
   {This work has unveiled in unprecedented detail the complex environment of \lbqs, which includes its host galaxy, a close obscured AGN, and nine interacting companions (five of which were previously unknown), all within 30 kpc of the QSO. Our results support a scenario where  mergers can trigger dual AGN and can be important drivers of rapid early supermassive black hole growth.  
   }

   \keywords{quasars: supermassive black holes -- quasars: emission lines -- Galaxies: high-redshift -- Galaxies: interactions -- Galaxies: active -- ISM: jets and outflows}

   \maketitle
%

\begin{figure*}
\begin{center}
\includegraphics[scale=0.55,trim= 35 20 20 22,clip]{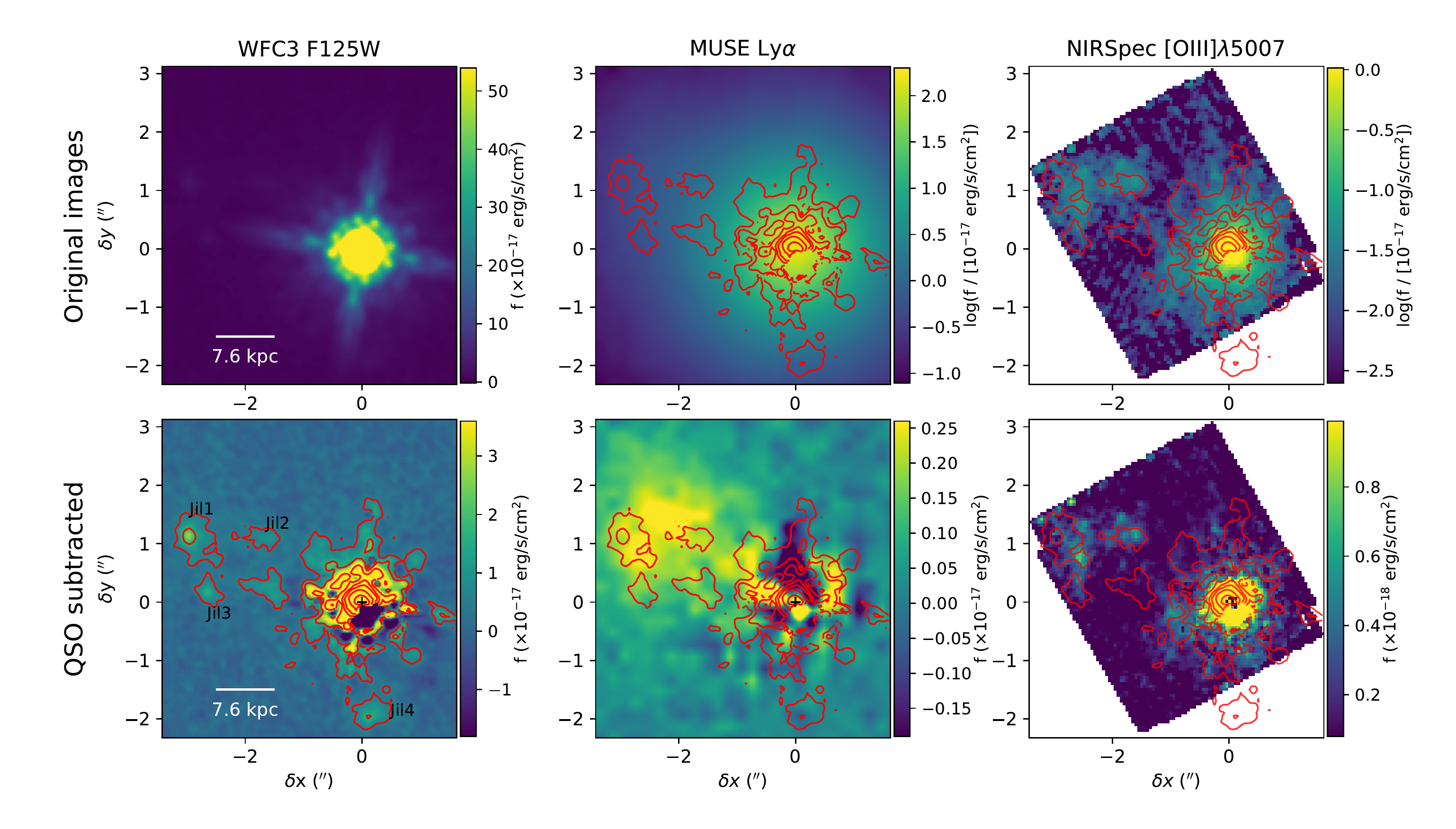}

\caption{Original (top) and PSF-subtracted (bottom) images of the QSO \lbqs \ and its close neighbouring galaxies as observed from space- and ground-based telescopes. In the left panels, we show the HST WFC3 near-infrared images from \citet{Husemann2021}, with contours in the bottom panel showing the Jil1-to-4 galaxies discovered by Husemann et al. The middle panels present the MUSE \lya \ emission before (top) and after (bottom) the QSO PSF subtraction, from \citet{Husemann2018a}, with contours from the PSF-subtracted HST image. The right panels show the \oiii $\lambda5007$ emission from JWST/NIRSpec observations; see Sect. \ref{sec:QSOsubtraction} for details on the QSO PSF subtraction. North is up, and east is to the left.}
\label{fig:LBQS_images_qsosub}
\end{center}
\end{figure*}

\section{Introduction}

The \textit{James Webb} Space Telescope (JWST) promises to reveal a new view of galaxy formation in the early Universe.
Thanks to its unprecedented sensitivity and spectroscopic capability in the near- and mid-infrared wavelengths, the rest-frame optical nebular emission lines (e.g.   H$\beta$, \oiii $\lambda\lambda$4959,5007, H$\alpha$, and \nii $\lambda\lambda$6548,6583) of star-forming galaxies and active galactic nuclei (AGN) can, for the very first time, be directly detected and resolved across early cosmic epochs, from cosmic noon ($z\sim2-3$; e.g. \citealt{Schreiber2020})  to the epoch of re-ionisation  ($z\gtrsim  7$;  e.g. \citealt{Robertson2022, Curtis2022}). Early Release Observations and Cycle 1 General Observer and Guaranteed Time Observations (GTO) programme results have clearly demonstrated the power of JWST’s spectroscopic observations (e.g. \citealt{Brinchmann2022, Bunker2023, Cameron2023, Cresci2023, Curti2023, Kocevski2023, Tacchella2022, Vayner2023}), promising many exciting discoveries over the coming years.

All cosmological models of hierarchical structure formation predict the existence of multiple supermassive black holes (SMBHs) inside many galaxies, consequences of previous merging events (\citealt{Hopkins2007, Colpi2014, Volonteri2021}). These events can be revealed by the detection of dual AGN separated by up to a few kiloparsecs.  The observational search for close dual quasars (QSOs) at $1 < z < 3$ (i.e. at the peak of QSO activity) is particularly important for constraining the merger process in cosmological models because the effects of mergers are believed to be the most significant in the high-luminosity, close-separation regime (e.g. \citealt{Hopkins2008,VanWassenhove2012}).
Unfortunately, only very few dual AGN have been confirmed observationally at such high $z$ (e.g. \citealt{Chen2022,Chen2022b,Lemon2022,Mannucci2022}); whether these systems are intrinsically rare or are simply undiscovered is not yet known. The study of the few dual AGN known so far at high $z$ is therefore of paramount importance for testing the predictions of the cosmological models in these early epochs of the Universe.
In this paper we use data from the JWST/NIRSpec Integral Field Spectrograph \citep[IFS;][]{Jakobsen2022, Boker2022} of the optically luminous QSO \lbqs, one of the rare QSOs at high $z$ with a close AGN (\citealt{Husemann2018a}).

The QSO \lbqs \ (RA\ $3^{\rm h}4^{\rm m}49.93^{\rm s}$, Dec.\ $-0^{\circ}8^{\prime}13.10\arcsec$, J2000) at $z \sim 3.3$ has been intensively targeted for studies of the intergalactic medium along our line of sight (LOS). It is one of the rare ultraviolet-transparent luminous QSOs that allows the He\,{\sc{ii}} \lya \ absorption of the intergalactic medium to be investigated in detail: \citet{Worseck2021} inferred for \lbqs \ a large proximity zone, 13.2 Mpc, caused by the enhanced ionising photon flux around the QSO (e.g. \citealt{Jakobsen1994}), which implies a long active phase of more than $11$ Myr for this QSO.

Analysing archival observations from the Multi Unit Spectroscopic Explorer (MUSE; \citealt{Bacon2010}) on the Very Large Telescope (VLT), \citet{Husemann2018a} report the detection of a \lya \ nebula surrounding \lbqs \ out to tens of kiloparsecs that is associated with various high ionisation lines. In particular, these authors report the serendipitous discovery of an obscured AGN -- dubbed Jil (Klingon for neighbour) -- about 20 kpc from the QSO, inferred from \lya, C\,{\sc{iv}}$\lambda$1549, He\,{\sc{ii}}$\lambda$1640, and C\,{\sc{iii}}]$\lambda$1909 ultraviolet emission-line diagnostics. The He\,{\sc{ii}} line luminosity, L(\heii) $\sim 1.7\times 10^{42}$ erg s$^{-1}$, was inconsistent with being induced by \lbqs \ given the compact, point-like spatial distribution of this line emission and its corresponding small cross-section. The He\,{\sc{ii}} luminosity can more easily be explained by the presence of an AGN of about 1/500--1/1000 the luminosity of \lbqs \ (corresponding to a bolometric luminosity of L$_{\rm{AGN}} \sim 10^{45}$ erg s$^{-1}$), if located within the compact region emitting He\,{\sc{ii}}. 

Follow-up ground-based Ks-band imaging and near-infrared spectroscopy are presented in \citet{Husemann2018b}, who successfully detected Jil's host galaxy emission, with an estimated stellar mass of $\sim 10^{11}$ M$_\odot$, and the optical \oiii$\lambda5007$ line (\oiii \ hereinafter), with $L($\oiii $) \sim 2.5\times 10^{42}$ erg/s. However, no other rest-frame optical lines were detected. 
Finally, \citet{Husemann2021} present \textit{Hubble} Space Telescope (HST) Wide-Field Camera 3 (WFC3) near-infrared imaging of the QSO, revealing the presence of close multiple companion objects: emission from Jil was resolved into two sources separated by $\sim 1\arcsec$ ($\sim 8$ kpc), Jil1 and Jil2, while two additional sources were dubbed Jil3 and Jil4. They also constrained stellar ages and masses for the two most prominent companions, Jil1 with $t_* = 252_{-109}^{+222}$ Myr and log(M$_*$/M$_\odot$) = $11.2_{-0.1}^{+0.3}$, and Jil2, associated with the compact \heii \  emission, with $t_* = 19_{-14}^{+74}$ Myr  and log(M$_*$/M$_\odot$) = $9.4_{-0.4}^{+0.9}$. These early near-infrared (HST) and optical (MUSE) observations are presented in Fig. \ref{fig:LBQS_images_qsosub} to display the complex environment of \lbqs.

\lbqs \ also hosts a powerful outflow: \citet{Shen2016}, after analysing near-infrared slit spectroscopy, reported the presence of an ionised outflow traced by \oiii, with a velocity of $\gtrsim 1000$ \kms. A velocity offset of \civ \  relative to the centroid of the \hb\ broad line region (BLR) and \oiii \  narrow line region (NLR) of 400-600 \kms\ is also reported by \citet{Coatman2017} and \citet{Zuo2020}; such a significant displacement of the \civ \ to the blue suggests the presence of strong nuclear outflows in the BLR of \lbqs \ (see also e.g. \citealt{Vietri2020}).

In this manuscript we present the JWST/NIRSpec IFS observations of \lbqs\  to study the rest-frame optical lines  and  characterise its intergalactic  and interstellar medium. NIRSpec data enable us to  shed light on the gravitational interaction between the Jil sources and the QSO host galaxy, as well as the  possible accretion onto the QSO host through the circumgalactic medium and the ejection of material through powerful outflows. 
The paper is outlined as follows. In Sect. \ref{sec:observations} we describe the JWST NIRSpec observations, and our data reduction is outlined in Sect. \ref{sec:reduction}. Detailed data analysis of the integrated QSO spectrum and the spatially resolved spectroscopic analysis are reported in Sects. \ref{sec:integratedspec} and \ref{sec:spatialinfo}, respectively. Section 5 also presents the new procedure developed to model and subtract the wiggle artefacts in NIRSpec IFS cubes. Finally, we present a discussion of our results in Sect. \ref{sec:results}, before concluding with a summary of our findings in Section \ref{sec:conclusions}.

Throughout, we adopt a \cite{Chabrier2003} initial mass function ($0.1-100~M_\odot$) and a flat $\Lambda$ cold dark matter cosmology with $H_0=70$~km~s$^{-1}$~Mpc$^{-1}$, $\Omega_\Lambda=0.7$, and $\Omega_m=0.3$.
In the analysis we use vacuum wavelengths, but when referring to emission lines we quote their rest-frame air wavelengths if not specified otherwise.

\section{Observations}\label{sec:observations}

\lbqs \ was observed on August 8, 2022, as part of the NIRSpec IFS GTO
programme `Galaxy Assembly with NIRSpec IFS'´´ (GA-NIFS) under programme \#1220 (PI: N. Luetzgendorf). The project is based on the use of the NIRSpec’s IFS mode, which provides spatially resolved spectroscopy over a contiguous 3.1$^{\prime\prime} \times$ 3.2$^{\prime\prime}$ sky area, with a sampling of 0.1$^{\prime\prime}$/spaxel and a spatial resolution from $\sim 0.04$\arcsec (at $\sim 1\mu$m) to $\sim 0.15$\arcsec \ (at $\sim 5\mu$m; see \citealt{Boker2022, Rigby2022}). 
The IFS observations were taken with the grating/filter pair G235H/F170LP. This results in a data cube with spectral resolution $R\sim2700$ over the wavelength range 1.7--3.1 $\mu$m.
The observations were taken with the  IRS2RAPID readout pattern with 60 groups, using a 4-point medium cycling dither pattern, resulting in a total exposure time of 3560 seconds.

\section{Data reduction}\label{sec:reduction}

The raw data were reduced with the JWST calibration pipeline version 1.8.2, using the context file {\it jwst\_1041.pmap}. All of the individual raw images were first processed for detector-level corrections using the
{\it Detector1Pipeline} module of the pipeline (Stage1 hereinafter). Then, the individual products (count-rate images) were calibrated through {\it Calwebb\_spec2} (Stage2 hereinafter), where wcs-correction, flat-fielding, and the flux-calibrations are applied to convert the data from units of count-rate to flux density. The individual Stage2 images were then resampled and co-added onto a final data cube through the {\it Calwebb\_spec3} processing (Stage3 hereinafter). A number of additional steps (and corrections in the pipeline code) were applied to improve the data reduction quality; different configurations were also used to obtain additional data products and test the pipeline robustness (e.g. of flux and spatial resolution recovery). In particular:

\begin{figure*}
\begin{center}
\includegraphics[scale=0.45,trim= 45 0 20 23,clip]{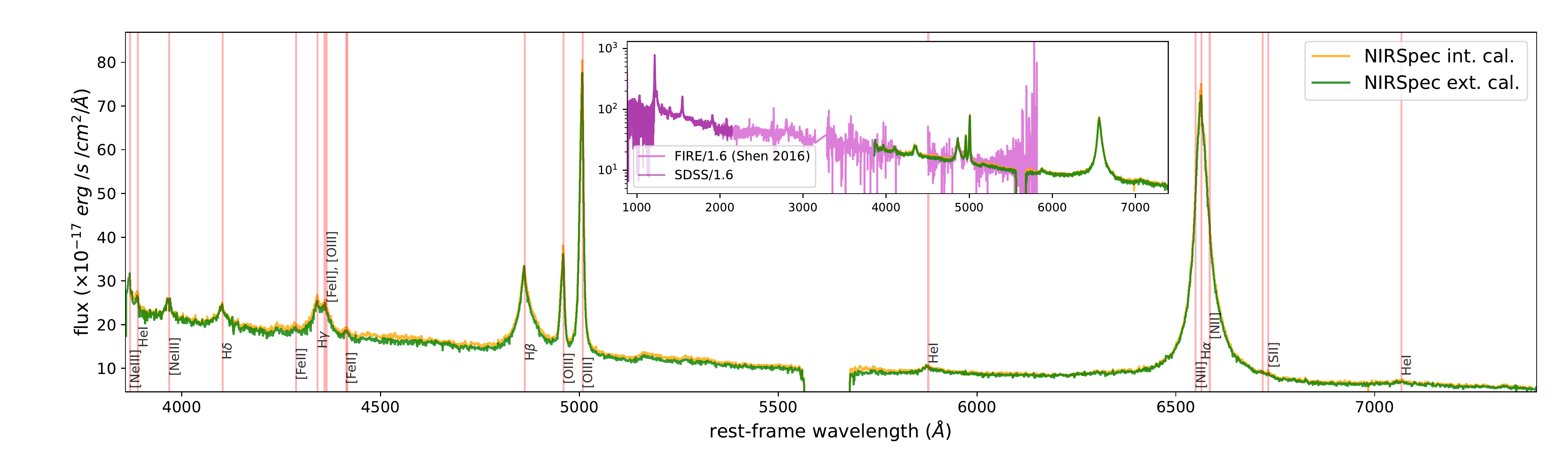}
\caption{NIRSpec spectra obtained with internal (orange curve) and external (green) flux calibration and integrated over a region of $r=1.5$\arcsec. The two NIRSpec spectra are extracted from the \drizzle \ cubes, with 0.05\arcsec spaxels. Vertical lines indicate the main emission line features detected in the NIRSpec spectrum.  The inset shows the same NIRSpec spectra compared with the Magellan/FIRE  (magenta) and the SDSS (purple) spectra, rescaled by a factor of $1.6$ to match the NIRSpec spectra in the vicinity of the \hb\ and \oiii \  lines. 
}
\label{fig:integratedspectra}
\end{center}
\end{figure*}

\begin{itemize}
    \item In order to correct for the artefacts known as a `snowballs',  caused by large cosmic ray impacts, we applied the snowball flagging for the jump during Stage 1. Sometimes this step incorrectly flags elongated streaks (due to cosmic ray impacts) as snowballs. Even though these streaks affect only a narrow region of the detector, the algorithm flags an entire circle containing the streak. This results in extended, circular regions with signal over-subtraction in the final count-rate images. To address this issue, we patched the pipeline to fit ellipses to all flagged regions consisting of five or more adjacent pixels; regions with best-fit ellipses having axis ratio smaller than 0.1 are removed from the list of snowballs.

    \item The individual count-rate frames were further processed at the end of Stage 1, to correct for different zero levels in the dithered frames: for each image, we subtracted the median value (computed considering the entire image) to get a base level consistent with zero counts per second. This step is particularly important for the very first frame obtained for \lbqs, showing (unrealistic) negative ramps in the raw (level 1b) data, and resulting in negative counts at the end of Stage 1.
    
    \item We further processed these count-rate images to subtract the $1/f$ noise (e.g. \citealt{Kashino2022}). This correlated vertical noise is modelled in each column (i.e. along the spatial axis)  with a low-order polynomial function, after removing all bright pixels (e.g. associated with the observed target) with a $\sigma$-clipping algorithm. The modelled $1/f$ noise is then subtracted before proceeding with Stage 2 of the pipeline.
    
    \item The flux calibration was performed using two different approaches: the first uses the photom step of Stage 2, and the second takes advantage of the commissioning observations of the standard star TYC 4433-1800-1 (PID 1128, o009). In the latter case, the flux calibration is performed as a post-processing correction: we reduced the star with the same pipeline version and context file, and obtained the response curve of the instrument required to convert count rates into flux densities. Hereinafter, we refer to the first approach as internal flux calibration, and to the second as external flux calibration. 

    \item The outlier\_detection step of Stage 3 is required to identify and flag all remaining cosmic rays and other artefacts left over from previous calibration steps, resulting in a significant number of spikes in the reduced data. Unfortunately, with the current version of the pipeline, this step cannot be used, because it tends to identify too many false positives and seriously compromises the data quality\footnote{At the time of this writing, the newest version of the pipeline, v1.9.4, and the latest context file, {\it jwst\_1063.pmap } are still affected by these issues.}. We therefore decided to follow two different approaches to remove the spikes: the first one uses an algorithm similar to {\sc lacosmic} \citep{vanDokkum2001} to remove outliers in individual exposures (at the end of Stage 2): because our sources are under-sampled in the spatial direction, we calculated the derivative of the count-rate maps only along the (approximate) dispersion direction. The derivative was then normalised by the local flux (or by 3$\times$ the noise, whichever was highest) and we rejected the 95\textsuperscript{th} percentile of the resulting distribution (see \citealt{DEugenio2023} for details). The second approach consists of a post-processing correction, and is done applying a $\sigma$ clipping  to exclude all spikes in the reduced data cubes (at spaxel level).

    \item Finally, we applied the {\it cube\_build} step to produce two combined data cubes: one with a spaxel size of 0.1\arcsec, obtained with the \emsm\  weighting (with higher signal-to-noise at spaxel level), and a second with a spaxel size of 0.05\arcsec, obtained with the \drizzle \ weighting; the latter has a higher spatial resolution but is more affected by point spread function (PSF) effects (see Sect. \ref{sec:wiggles}). We manually rescaled the \drizzle \ cubes by a factor of $(0.05\arcsec/0.1\arcsec)^2$ to ensure the flux conservation.
    
\end{itemize}
We patched the {\it cube\_build} script, fixing a bug affecting the \drizzle \ algorithm as implemented in the version 1.9.0\footnote{{\it cube\_build} code changes in \url{https://github.com/spacetelescope/jwst/pull/7306}}; we also patched the  {\it photom} script, applying the corrections implemented in the same version\footnote{{\it photom} code changes in \url{https://github.com/spacetelescope/jwst/pull/7319}}, which allows more reasonable flux densities to be inferred (i.e. a factor of $\sim100$ smaller with respect to those obtained with standard pipeline 1.8.2).

\subsection{Astrometric registration}

We obtained a bona fide astrometric registration matching the QSO nucleus position with that in the HST image shown in Fig. \ref{fig:LBQS_images_qsosub}, that is, applying a correction of $\Delta$RA $= -0.492$\arcsec and $\Delta$DEC $= - 0.062$\arcsec. This offset is due to an error in the reference files responsible for the coordinate transformation, then partially solved with the release of the  context file {\it jwst\_1063.pmap}\footnote{ {\it jwst\_1063.pmap} corrects for a $\sim 4$ pixels systematic offset associated with the coordinate transformation between the `OTEIP' and the world systems, but not for a smaller offset ($\sim$0.2--0.4 pixels) between the `GWA' and the `virtual slit' frame (see \citealt{Dorner2016}).}.

\begin{figure*}
\begin{center}
\includegraphics[scale=0.41,trim= 40 0 60 0,clip]{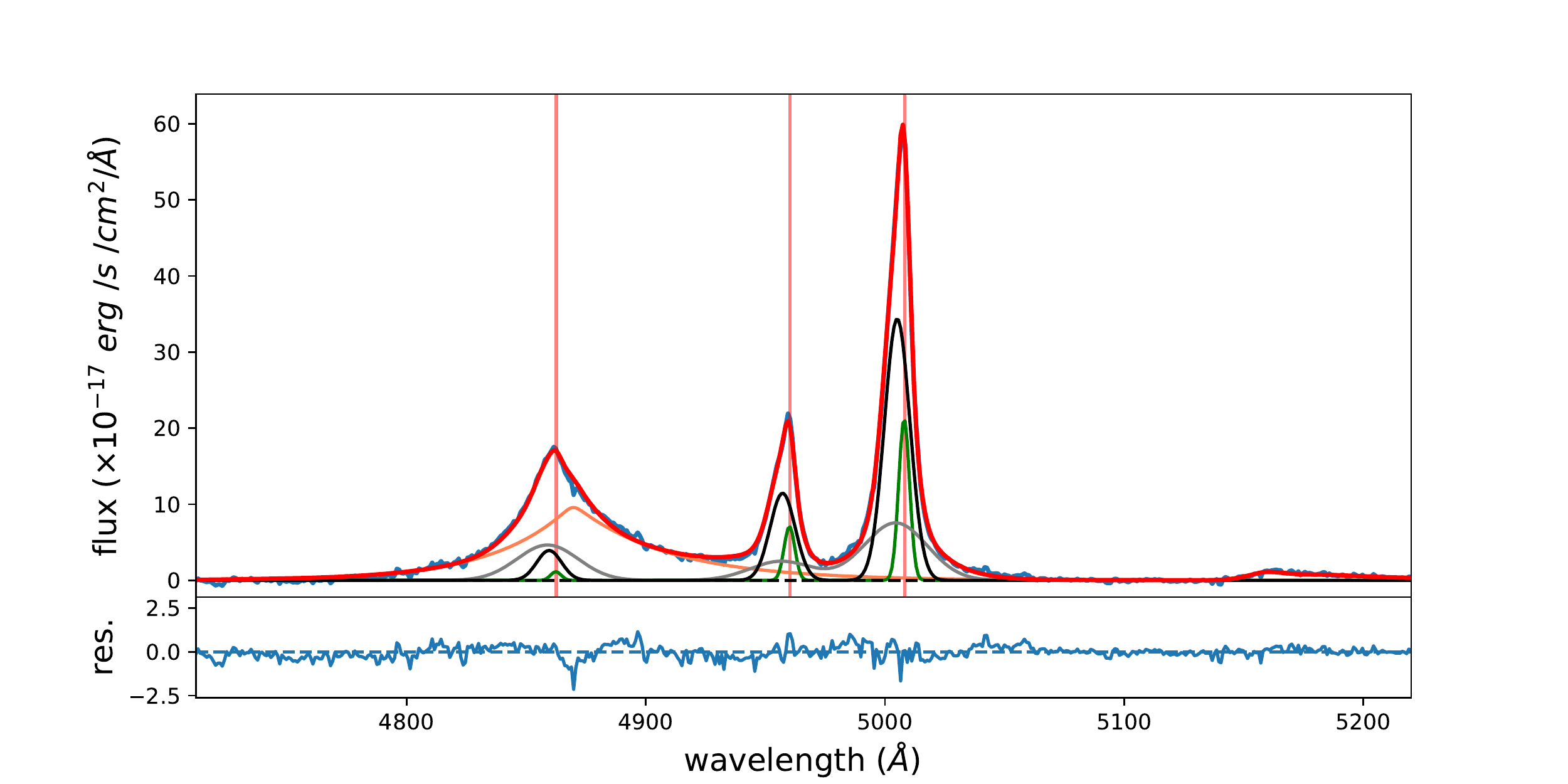}
\includegraphics[scale=0.41,trim= 40 0 60 0,clip]{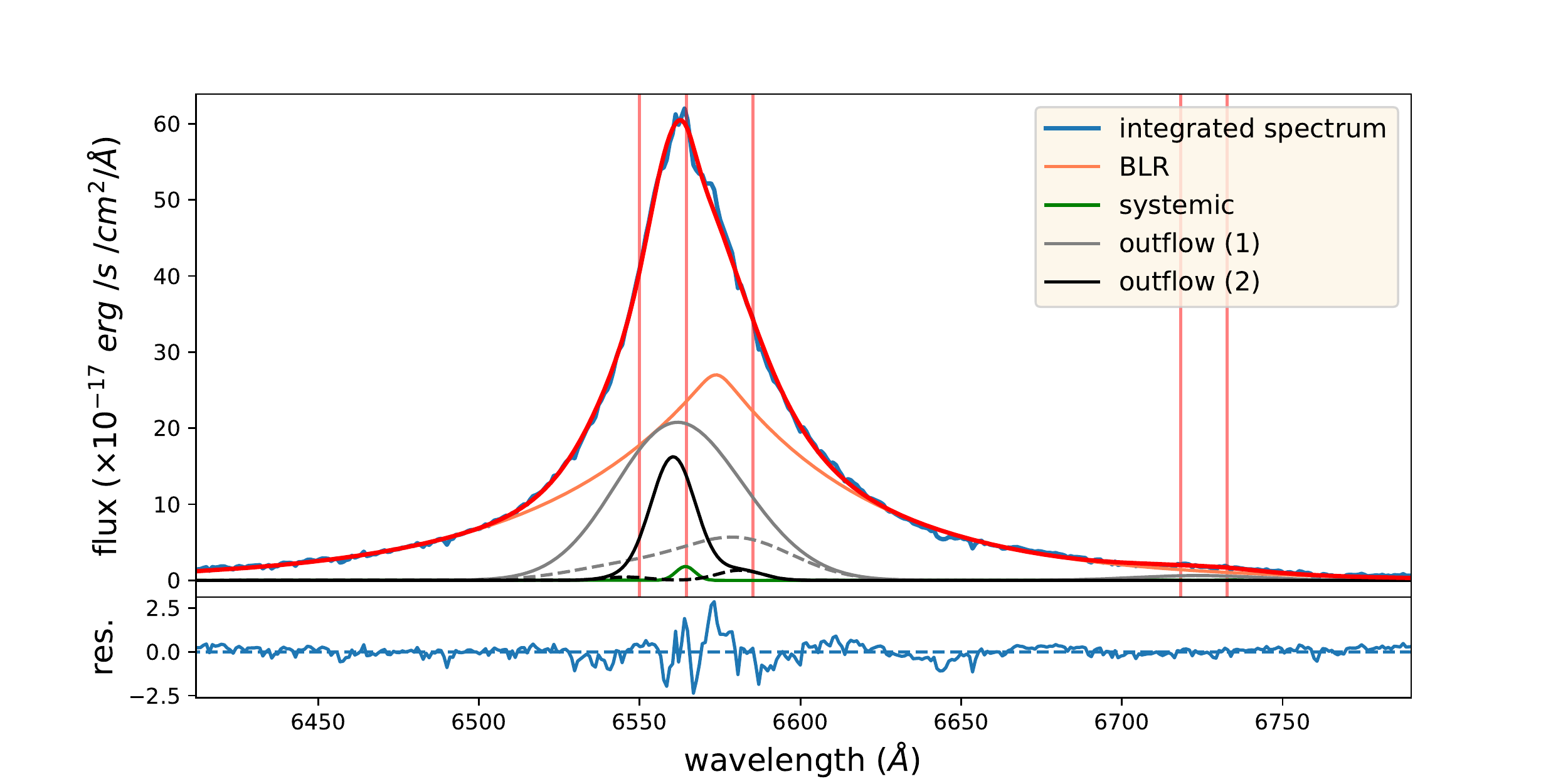}

\caption{Multi-component simultaneous best-fit results for the continuum-subtracted spectrum of \lbqs, around the \hb-\oiii \ (left) and \ha-\nii \ regions (right; integrated over a circular region with $r = 0.5\arcsec$). The blue curve represents the rest-frame NIRSpec spectrum, and the red curve indicates the best fit, with individual kinematic components shown with different colours (as labelled in the right panel). For the outflow (1) and (2) components, we also show the contribution from the only \nii \ lines with dashed lines, as the grey and black curves do not allow a clear distinction between the \ha \ and \nii \ line transitions. Vertical red lines indicate the most prominent emission lines, as in Fig. \ref{fig:integratedspectra}. The lower panels show the residual to the model fit, that is, the difference between the observed spectrum and the model.    
}
\label{fig:HaHbfit}
\end{center}
\end{figure*}

\subsection{Recovery of the QSO flux}

Figure \ref{fig:integratedspectra} shows the integrated NIRSpec spectra of \lbqs, obtained from the \drizzle \ cubes, reduced with the internal (orange curve) and external (green) flux calibration. 
The NIRSpec spectra  were extracted from a circular aperture centred at the position of the QSO nucleus, with $r= 1.5$\arcsec, hence matching the Sloan Digital Sky Survey (SDSS) fibre radius (see below). These spectra are compared in the inset with the near-infrared Magellan/FIRE spectrum (magenta, from \citealt{Shen2016}) and the SDSS spectrum (in purple); the latter is rescaled by a factor of 1.6 to match the fluxes in the vicinity of the \hb\ and \oiii \ lines. 

The agreement between NIRSpec and the spectra from other facilities is remarkable, and the small differences can be explained by taking flux calibration uncertainties into account. The small mismatch between the two integrated NIRSpec spectra (obtained with external and internal flux calibrations) is of the order of $\sim 2-3\%$, well within the nominal uncertainties of the JWST calibration pipeline (\citealt{Boker2023}). Being in the very early stages of pipeline development, we avoided investigating the discovered discrepancies further; nevertheless, we note that a larger mismatch would be present without applying all corrections reported in Sect. \ref{sec:reduction} for the internal calibration.\footnote{Systematic, wavelength-dependent discrepancies (up to $\sim 25\%$) would also appear using context files older than {\it jwst\_1023.pmap}, because of placeholder flat field corrections used during the {\it photom} step of the internal calibration.}  
 
All results described in this paper refer to the \drizzle data cubes, which we preferred over the \emsm \  cubes as the former better preserve the NIRSpec spatial resolution. Moreover, we preferred the internal to the external flux calibration, as the former also allow corrections for the flat field. Finally, we used the cubes obtained with the modified outlier detection method (see above), although there are no major differences between these and those corrected with a $\sigma$-clipping method.

\section{Spectral analysis of the integrated \lbqs spectrum}\label{sec:integratedspec}

\subsection{Spectral fit}
\label{sec:SpectralFit}
We fit the most prominent gas emission lines by using the Levenberg-Marquardt least-squares fitting code CAP-MPFIT (\citealt{Cappellari2017}). 
In particular, we modelled the \ha\ and \hb\ lines, the \oiii $\lambda\lambda$4959,5007, \nii $\lambda\lambda$6548,83, and \sii $\lambda\lambda$6716,31 doublets with a combination of Gaussian profiles, applying a simultaneous fitting procedure (e.g. \citealt{Perna2020}), so that all line features of a given  kinematic component have the same velocity centroid and full width at half maximum (FWHM). 
The modelling of the  \ha\ and \hb\  BLR emission requires the use of broken power-law components (e.g. \citealt{Nagao2006,Cresci2015a,Trefoloni2023}): they are preferred over a combination of extremely broad Gaussian profiles because the former tend to minimise the degeneracy between NLR and BLR emission. Finally, we used the theoretical model templates of \citet{Kovacevic2010} to reproduce the iron (Fe\,{\sc{ii}}) emission in the wavelength region $4000-5500\AA$.  
The final number of kinematic components used to model the spectra is derived on the basis of the Bayesian information criterion (BIC; \citealt{Schwarz1978}).

Figure \ref{fig:HaHbfit} shows the best-fit model  around the \hb-\oiii \ and \ha-\nii \ regions. The BLR emission is fitted with a broken power law; iron emission is fitted with the S and G group lines (\citealt{Kovacevic2010}). The \oiii \ doublet shows a narrow core, and prominent blue and red wings, and requires three Gaussian components. To reduce the degeneracy between BLR and NLR, we simultaneously fit four additional spectra extracted from circular regions with radius of 0.2\arcsec (4 spaxels) and centred at different positions within a few spaxels from the peak emission of the QSO: BLR profiles are tied, assuming that these emission components originate from the same unresolved region, while all other components are free to vary as originating from more extended (and likely resolved) regions. The outcomes of this simultaneous fit (reported in Fig. \ref{fig:Aintegratedspec}) are therefore used to fix the BLR parameters during the fit of the integrated spectrum shown in Fig. \ref{fig:HaHbfit}. 

We note that the integrated spectra reported in Fig. \ref{fig:Aintegratedspec} show additional peaks and/or inflection points in the \ha-\nii \ complex, due to the presence of strong \nii \ emission line components; these nitrogen features are not resolved in the integrated spectrum in Fig. \ref{fig:HaHbfit}, although they are still definable from our fit decomposition. The absence of  inflection points in Fig. \ref{fig:HaHbfit} is likely due to the more prominent BLR emission, and the stronger degeneracy between BLR and NLR kinematic components.

\subsection{Systemic redshift}
We derived the \lbqs \ redshift from the  measured wavelength of the narrow \oiii \ emission in the integrated spectrum shown in Fig. \ref{fig:HaHbfit}: $z = 3.2870\pm 0.0003$, which is in agreement with \citet{Zuo2015,Zuo2020} but at odds with other redshift measurements from the literature.  \citet{Husemann2018a} reported values in the range $3.2882-3.2887$ (for different ultraviolet lines); \citet{Coatman2019} reported $z =3.2856\pm 0.0002$ for the \oiii, and $z =3.2868\pm 0.0012$ for the \hb. All these previous measurements are within $\approx \pm 100$ \kms \ of our zero velocity (assuming $z = 3.2870$). These small discrepancies are likely due to the presence of powerful outflows, affecting all of the most prominent ultraviolet-to-optical emission line profiles (Sect. \ref{sec:integratedoutflow}).

\subsection{Velocity offset between BLR and NLR }

As shown in Fig. \ref{fig:HaHbfit}, the BLR emission line components are blueshifted  with respect to the \oiii \ core component, by $480\pm 60$ \kms. Relative redshiftings (and blueshiftings) of the peaks of the broad Balmer line emission are quite common in AGN (e.g. \citealt{Gaskell1983}). Different explanations for these offsets have been proposed: they could due to the orbital motion of a SMBH binary (e.g. \citealt{Ju2013}), to recoiling SMBHs (e.g. \citealt{Komossa2008}), or to a perturbed accretion disk around a SMBH (e.g. \citealt{Gaskell2010}). We did not investigate these scenarios further as they go beyond the goals of this paper; however, we note that each explanation is plausible given the complex environment of \lbqs.

\subsection{Black hole mass}

Assuming that the gas in the BLR is virialised, we calculated the central black hole mass from the spectral properties of the \ha \ and \hb \ BLR region following the single-epoch calibrations from \citet{Dallabonta2020},

\begin{equation}\label{eq:MBH_DB}
\begin{split}
M_{\rm{BH}}(H\beta)  &= 1.87\times 10^6  \\ &\times ~\left( \frac{ L_{\rm{H\beta}}}{10^{42}\ \rm{erg~ s}^{-1}} \right)^{0.703} \left( \frac{\rm{\sigma}_{\rm{H\beta}}}{10^{3}\ \rm{km~ s}^{-1}} \right)^{2.183} \rm{M}_\odot
\end{split}
,\end{equation}with an intrinsic scatter of $\sim 0.3$ dex, and from \citet{Greene2006}:

\begin{equation}\label{eq:MBH_GB}
\begin{split}
M_{\rm{BH}}(H\beta)  &= 3.6\times 10^6  \\ &\times ~\left( \frac{ L_{\rm{H\beta}}}{10^{42}\ \rm{erg~ s}^{-1}} \right)^{0.56} \left( \frac{\rm{FWHM}_{\rm{H\beta}}}{10^{3}\ \rm{km~ s}^{-1}} \right)^{2} \rm{M}_\odot
\end{split}
\end{equation}

\begin{equation}
\begin{split}
M_{\rm{BH}}(H\alpha) &= 2.0\times 10^6 \\
&~\times~ \left( \frac{L_{\rm{H\alpha}}}{10^{42}\ \rm{erg~ s}^{-1}} \right)^{0.55} \left( \frac{\rm{FWHM}_{\rm{H\alpha}}}{10^{3}\ \rm{km~ s}^{-1}} \right)^{2.06} \rm{M}_\odot,
\end{split}
\label{eq:MBH_HA}
\end{equation}with larger intrinsic scatters of $\sim 0.4$ dex. 
Aside from the small differences in the intrinsic scatters in the chosen relations, we stress that all single epoch relations reported in the literature have been inferred for low-$z$ and low-luminosity AGN; as a result, significant extrapolations are required for the measurement of the \lbqs \ black hole mass.

We find a H$\alpha/$H$\beta$ flux ratio of $3.88_{-0.12}^{+0.19}$ for the BLR components. 
Taking as a reference the distribution of values of BLR Balmer ratios obtained by \citet{Dong2008} for a large, homogeneous sample of $\sim$ 500 low-$z$ Seyfert 1 and QSOs with minimal dust extinction effects, H$\alpha/$H$\beta = 3.06 \pm 1.11$ (see also \citealt{Baron2016}), our Balmer decrement measurement does not suggest significant extinction in the BLR of \lbqs. Therefore, we did not perform any extinction correction for the Balmer line luminosities required to compute the M$_{\rm {BH}}$. 

The Balmer line luminosities and widths are measured from our best-fit BLR profiles shown in Fig. \ref{fig:HaHbfit} (i.e. the broken power-law components); we obtain estimates of the black hole mass of the order of $\sim 2 \times 10^9$ M$_\odot$. These values are broadly consistent with those previously reported in the literature and are based on \hb \ and \civ \ BLR measurements (with the latter being slightly larger, as commonly reported in the literature; e.g. \citealt{Coatman2017}).

We calculated the bolometric luminosity of
\lbqs \ following \citet{Dallabonta2020}, hence using the \hb \ BLR luminosity (their Eq. 25): log(L$_{\rm {bol}}$/ [erg s$^{-1}$]) $\sim 47.2$ (consistent with the value inferred from the continuum luminosity at $5100\AA$, and using the \citealt{Netzer2019} bolometric correction, $\sim 47.3$). 
This bolometric luminosity is also consistent with the one obtained starting from the intrinsic X-ray luminosity reported by \cite{Nardini2019}, and computed applying a bolometric correction $k_{bol} = 21$ (from Eq. 3 by \citealt{Duras2020}), log(L$_{\rm {bol}}$/ [erg s$^{-1}$]) $\sim 46.7$.

Using our black hole mass estimate from the \hb\ BLR (Eq. 1, which has smaller scatter than Eqs. 2 or 3), we find an Eddington ratio of $\lambda_{\rm Edd}=0.9\pm 0.1$. This value indicates that the accretion onto the central black hole is close to the Eddington limit. All measurements so far inferred, and the quantities required for their computation are reported in Table \ref{tab:integratedproperties}.

\begin{table}
\centering
\caption{Measurements of central black hole mass (with errors that include the intrinsic scatter of the single-epoch relations mentioned in the text), \oiii \ luminosity (corrected for extinction), and outflow velocity from the integrated nuclear spectrum (see Sect.~\ref{sec:integratedspec}).}
\begin{tabular}{ll}
\hline 
\hline
    Measurement & Value  \\
\hline 
    $\log(L_{\rm H\alpha}/(\rm erg/s))$ &  $45.45\pm 0.01$\\
    FWHM$_{\rm H\alpha}$ [km/s] & $3650_{-60}^{+90}$ \\
    $\sigma_{\rm H\alpha}$ [km/s] & $3260_{-120}^{+40}$ \\

    $M_{\rm BH} ({\rm H\alpha})_{GH06} [M_\odot]$ &$2.28^{+0.06}_{-0.09}\times 10^9$ \\

\hline
    $\log(L_{\rm H\beta}/(\rm erg/s))$ & $44.86\pm 0.02$ \\
    FWHM$_{\rm H\beta}$ [km/s] & $3560_{-120}^{+170} $\\
    $\sigma_{\rm H\beta}$ [km/s] & $2800_{-70}^{+120}$  \\

    $M_{\rm BH} ({\rm H\beta})_{GH06} [M_\odot]$ & $1.99^{+0.20}_{-0.16}\times 10^9 $ \\

    $M_{\rm BH} ({\rm H\beta})_{DB20,\ \sigma} [M_\odot]$ &$1.82_{-0.11}^{+0.19}\times 10^9 $ \\

\hline
    $\log(L_{\rm bol}/(\rm erg/s))_{DB20}$ & $47.22\pm 0.01$ \\
    $\lambda_{\rm Edd}$ & $0.9\pm 0.1$ \\
\hline
\hline
    $\log(L_{[OIII]}/(\rm erg/s))$ & $45.71\pm 0.03$  \\
    \oiii \ $ W80$ [km/s]& $1080\pm 30$  \\
    \oiii \ $ W90$ [km/s]& $1600\pm 55$ \\
    \oiii \ $ V10$ [km/s]& $-760\pm 50$ \\

\hline
\end{tabular} 
\footnotesize{}
\label{tab:integratedproperties}
\end{table}

\begin{figure}
\includegraphics[scale=0.56,trim= 0 0 0 0,clip]{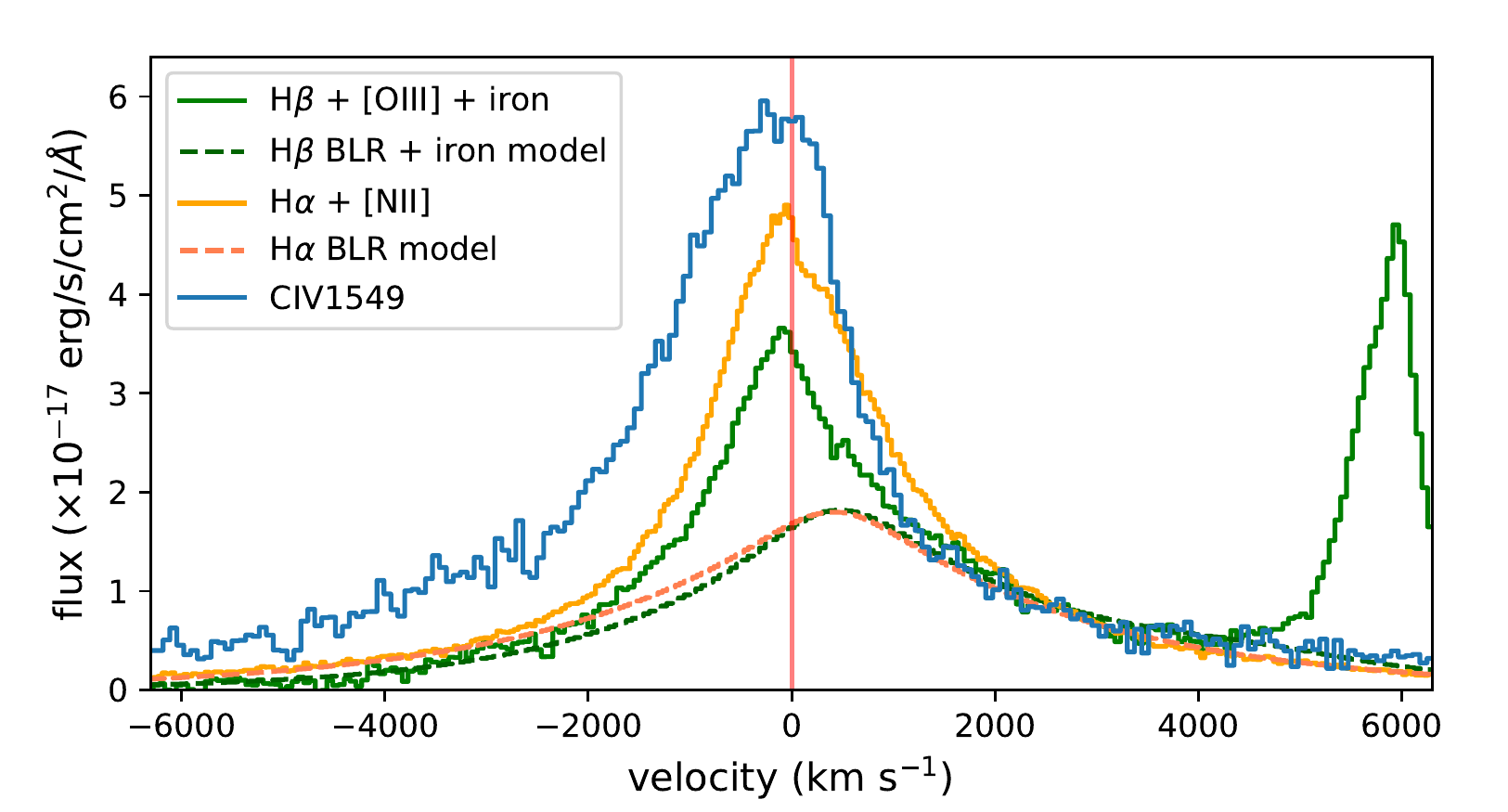}

\caption{Comparison of the BLR profiles of the \civ \ (from \citealt{Shen2016}) and \ha and \hb \ (from the integrated NIRSpec spectrum), in velocity space. For the Balmer lines, we also report the best-fit BLR profiles. All line profiles have been normalised to the flux of the BLR component in the reddest parts, which are likely less affected by BLR and NLR outflows. The \civ \  shows a significant excess in the blue part, at velocities of a few thousand \kms, which is not observed in the Balmer lines or in the \oiii \ line. This excess possibly indicates  strong BLR winds.   
}
\label{fig:BLRprofiles}
\end{figure}

\begin{figure*}
\begin{center}
\includegraphics[scale=0.63,trim= 50 0 50 50,clip]{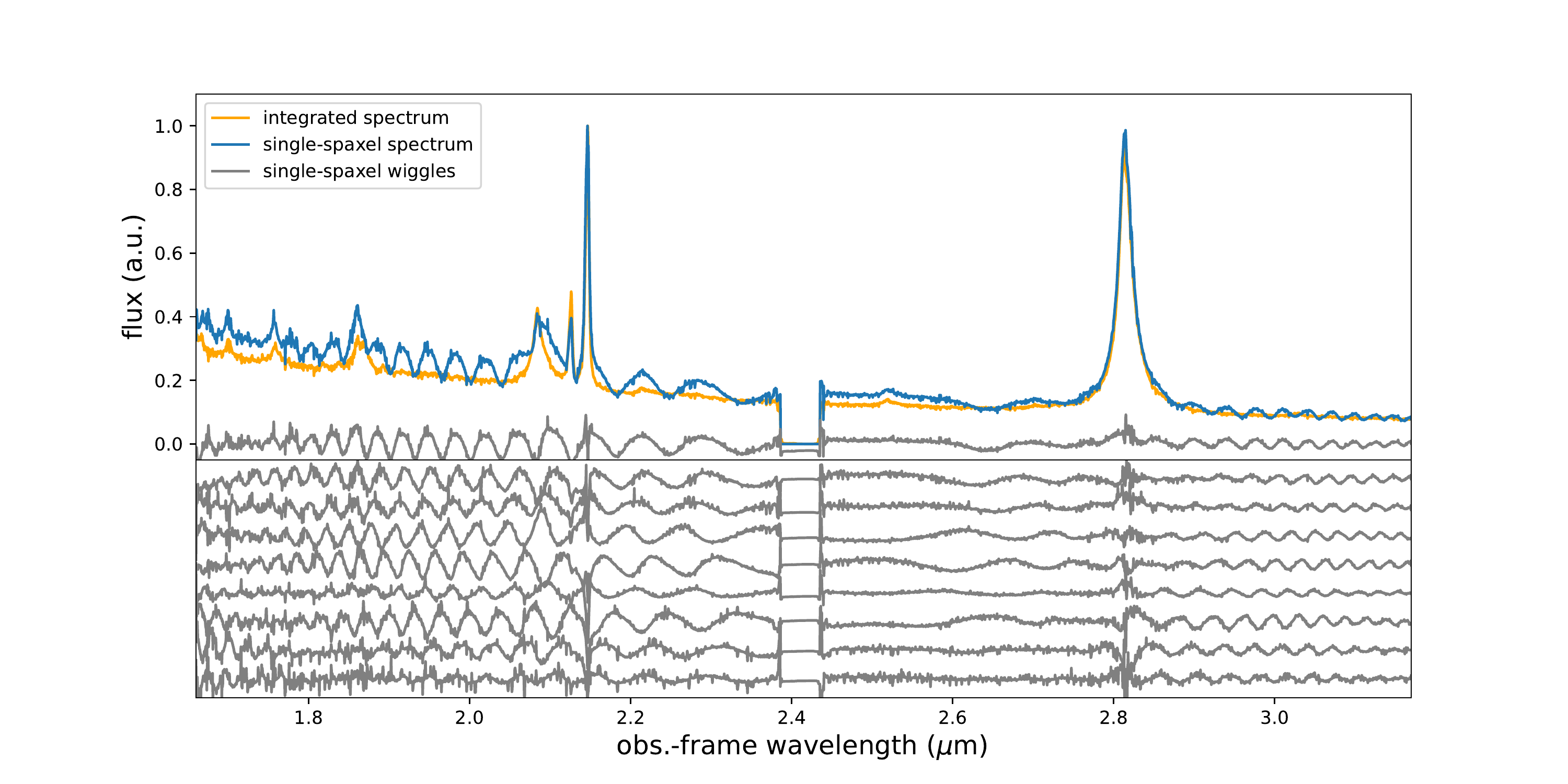}
\caption{Sinusoidal-type patterns in single-spaxel spectra extracted from the {\it drizzle} data cube with a spaxel size of 0.05\arcsec. Top panel: \lbqs \ spectrum integrated over an aperture of $r=0.5$\arcsec (orange curve), in comparison with the spectrum of the brightest spaxel (blue curve). Both spectra are normalised to their maximum values, for visualisation purposes. The wiggles affecting the single-spaxel spectrum are reported in grey and are obtained as the difference between the blue and orange curves (after subtracting a low-order polynomial function that takes the differences in the continuum level into account). Bottom panel: Wiggles obtained from the eight pixels closest to the brightest one. These wiggles strongly affect the shape of the continuum and, in particular, the \hb\ profile and the wings of the \oiii \ lines. See Fig. \ref{fig:3x3wiggles100mas} for analogous effects in the \emsm \ cube. 
}
\label{fig:3x3wiggles50mas}
\end{center}
\end{figure*}

\subsection{Outflow properties}\label{sec:integratedoutflow}

Before analysing the \oiii \ profile, we investigated the possible presence of winds in the BLR of \lbqs. In Fig. \ref{fig:BLRprofiles} we compare the Balmer line profiles with the \civ $\lambda1549$ line (from ground-based observations; \citealt{Shen2016}). All profiles are normalised to the emission in the red wing, unambiguously associated with the BLR emission for all transitions (see also Fig. 1 in \citealt{Zuo2020}). Figure \ref{fig:BLRprofiles} highlights the presence of a blue excess in the high-ionisation \civ \  line, up to a few thousand \kms, likely due to the presence of BLR winds. Such \civ \ outflows are commonly observed in luminous high-z QSOs, and often associated with large-scale \oiii \  outflows (e.g. \citealt{Coatman2019, Vietri2020}).

From the best-fit model shown in Fig. \ref{fig:HaHbfit}, we inferred the \oiii \  outflow velocity, considering different tracers commonly used in the literature: $V10 \sim -760 $ \kms, the velocity at the 10th percentile of the overall emission-line profile, $W80 \sim 1080$ \kms, defined as the line width containing 80\% of the emission line flux (obtained as the difference between the velocities at 90th and 10th percentiles), and $W90 \sim 1600$ \kms, containing 90\% of the line flux (and obtained as the difference between the velocities at 95th and 5th percentiles). All of these measurements are consistent with previous values obtained for \lbqs \ from ground-based observations (e.g. \citealt{Villar2020}).

We anticipate here that the ionised outflow is not spatially resolved in our NIRSpec observations; hence, outflow energetics, reported in Sect. \ref{sec:energetics}, have been derived on the basis of spatially integrated quantities.

\begin{figure*}
\begin{center}
\includegraphics[scale=0.34]{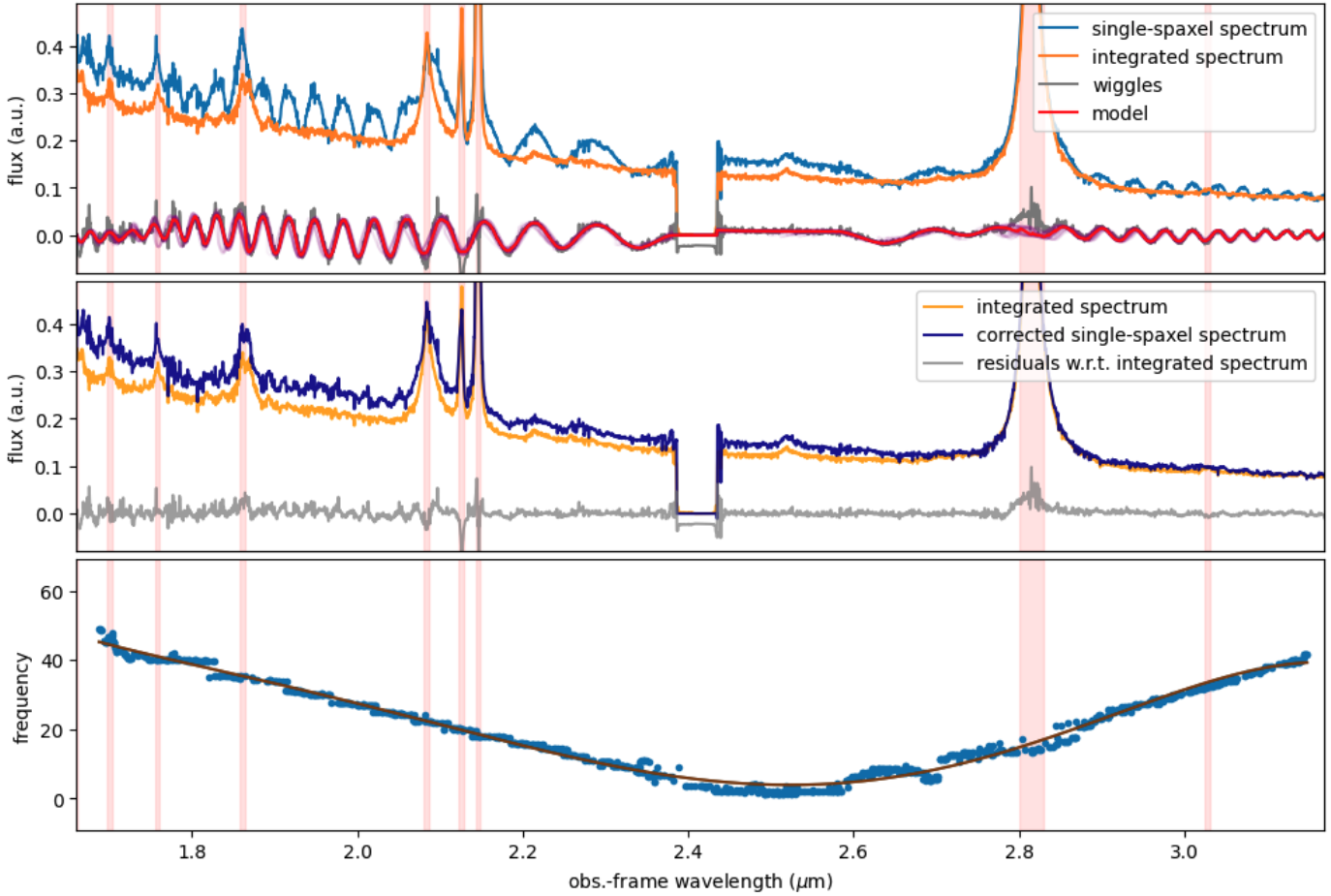}
\caption{Modelling of the wiggles in single-spaxel spectra. Top panel: integrated \lbqs spectrum (orange curve), single-spaxel spectrum (blue), and wiggles (grey) as already reported in Fig. \ref{fig:3x3wiggles50mas}. 
The red curve represents the best-fit model of the wiggles. Central panel: Single-spaxel spectrum after the correction for the wiggles (dark blue), in comparison with the integrated spectrum (orange); the grey curve represents the new residuals with respect to the integrated spectrum. Bottom panel: Best parameter for the frequency of the sinusoidal functions used to model the wiggles (blue points); a low-order polynomial function fitting these points is also reported. 
All panels display red shaded regions (associated with the QSO emission lines) that are excluded during the fit. 
}
\label{fig:LBQScentralspaxel}
\end{center}
\end{figure*}
\section{Spatially resolved spectroscopy}\label{sec:spatialinfo}

\subsection{Sinusoidal-type patterns in NIRSpec IFS}\label{sec:wiggles}

The spatial under-sampling in the NIRSpec IFS may result in apparent wiggles in the single-spaxel spectra close to the position of bright point sources, such as stars and QSOs. This effect is inherent to the cube building process, and is more pronounced in data cubes with better spatial sampling (i.e. in data cubes with spaxels of  0.05\arcsec, and constructed with the \drizzle \ weighting method). Further details about this effect, also known as `resampling noise' can be found for instance in \cite{Smith2007} and \cite{Law2023}. There is currently no correction in the pipeline for this; large spatial extraction regions are hence required to reduce the amplitude of the effect in extracted 1-D spectra. 
For isolated point sources, for which the extraction of spatially resolved information is not possible, this effect is irrelevant, as when the flux is integrated over a large aperture the wiggles disappear. However, there are situations where a point source overlaps with extended emission, thus requiring to disentangle the flux from both sources. This is the case, for instance, in studies of QSO hosts and their close environment.

Figure \ref{fig:3x3wiggles50mas} (top panel) displays the \lbqs \ spectrum integrated over an aperture of $0.5\arcsec$ (in radius), in comparison with the spectrum of the brightest spaxel extracted from the data cube constructed with the \drizzle \ weighting method (with spaxels of $0.05\arcsec$). The wiggles affecting the single-spaxel spectrum are reported in the same panel with a grey curve, and are obtained as the difference between the integrated and the single-spaxel spectra (after subtracting a low-order polynomial function taking the differences in the continuum levels into account). Similar sinusoidal-type patterns are  observed in all spaxels close to the brightest one, as shown in the bottom panel of Fig. \ref{fig:3x3wiggles50mas}: they can affect a region as large as  $r\sim 0.2-0.5$\arcsec.

The wiggles strongly limit the reconstruction and modelling of the target spectrum at single-spaxel level. In particular, they affect the determination of the continuum shape, and the modelling of permitted (e.g. Balmer) and forbidden (e.g. \oiii) emission lines. All of these components are required to remove the signal from the nuclear point source (especially its PSF wings) from the underlying extended emission (see e.g. \citealt{Husemann2013, Marasco2020}).

These limitations also affect the single-spaxel spectra extracted from \emsm \ cubes with spaxels of 0.1\arcsec (see Fig. \ref{fig:3x3wiggles100mas}), although the amplitude of their wiggles is $\approx 2-3$ times smaller than in the \drizzle \ cubes. Moreover, the use of \emsm \ implies a decrease in spatial resolution, down to $\sim 0.2\arcsec$ (\citealt{Vayner2023}). In the next section we describe our approach for modelling and subtracting these wiggles from NIRSpec data cubes; this algorithm, written in python, is available for download\footnote{\url{https://github.com/micheleperna/JWST-NIRSpec_wiggles}}.

\begin{figure*}
\centering
\begin{center}
\includegraphics[scale=0.55,trim= 0 0 0 8,clip]{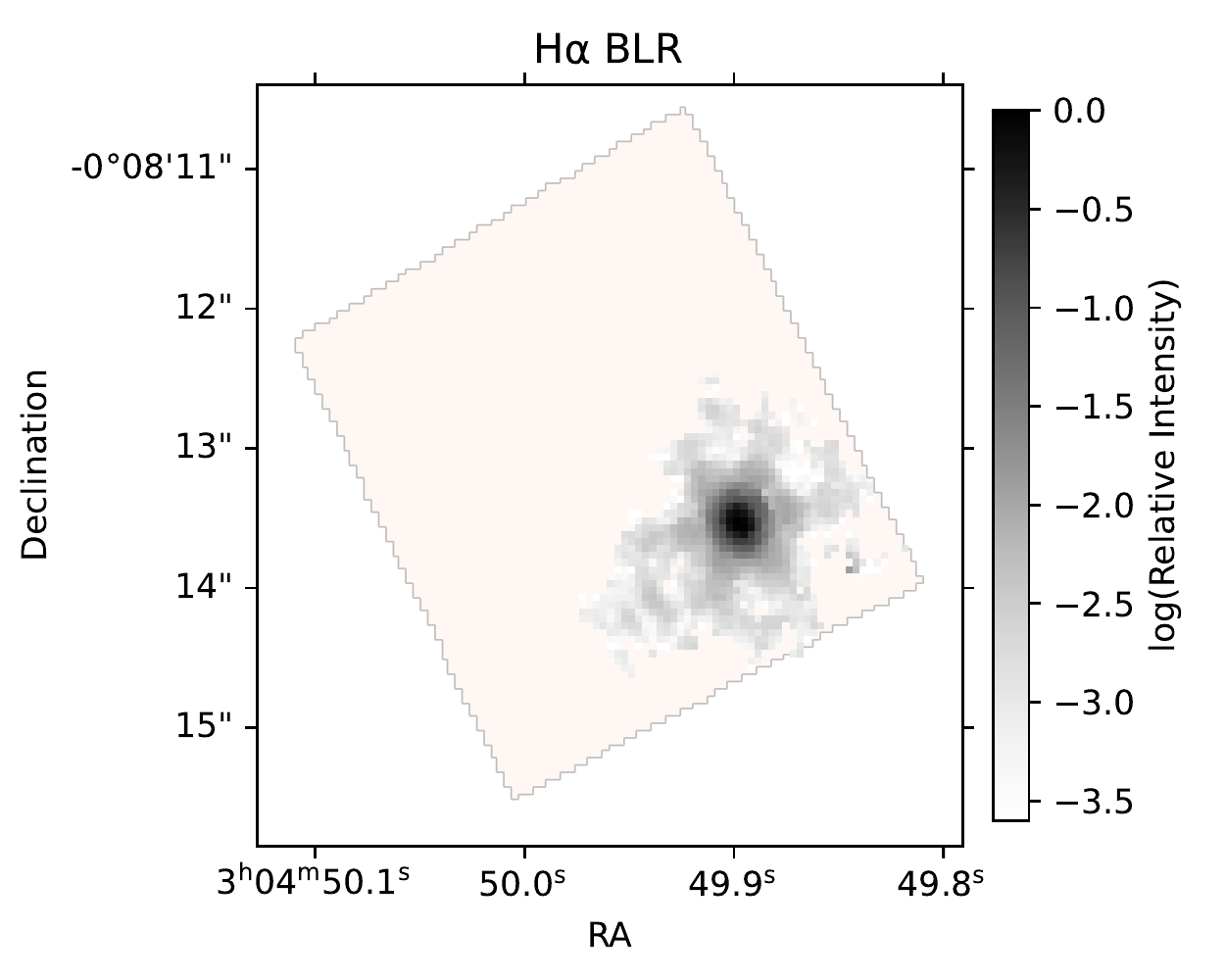}
\includegraphics[scale=0.4,trim= 5 0 0 0,clip]{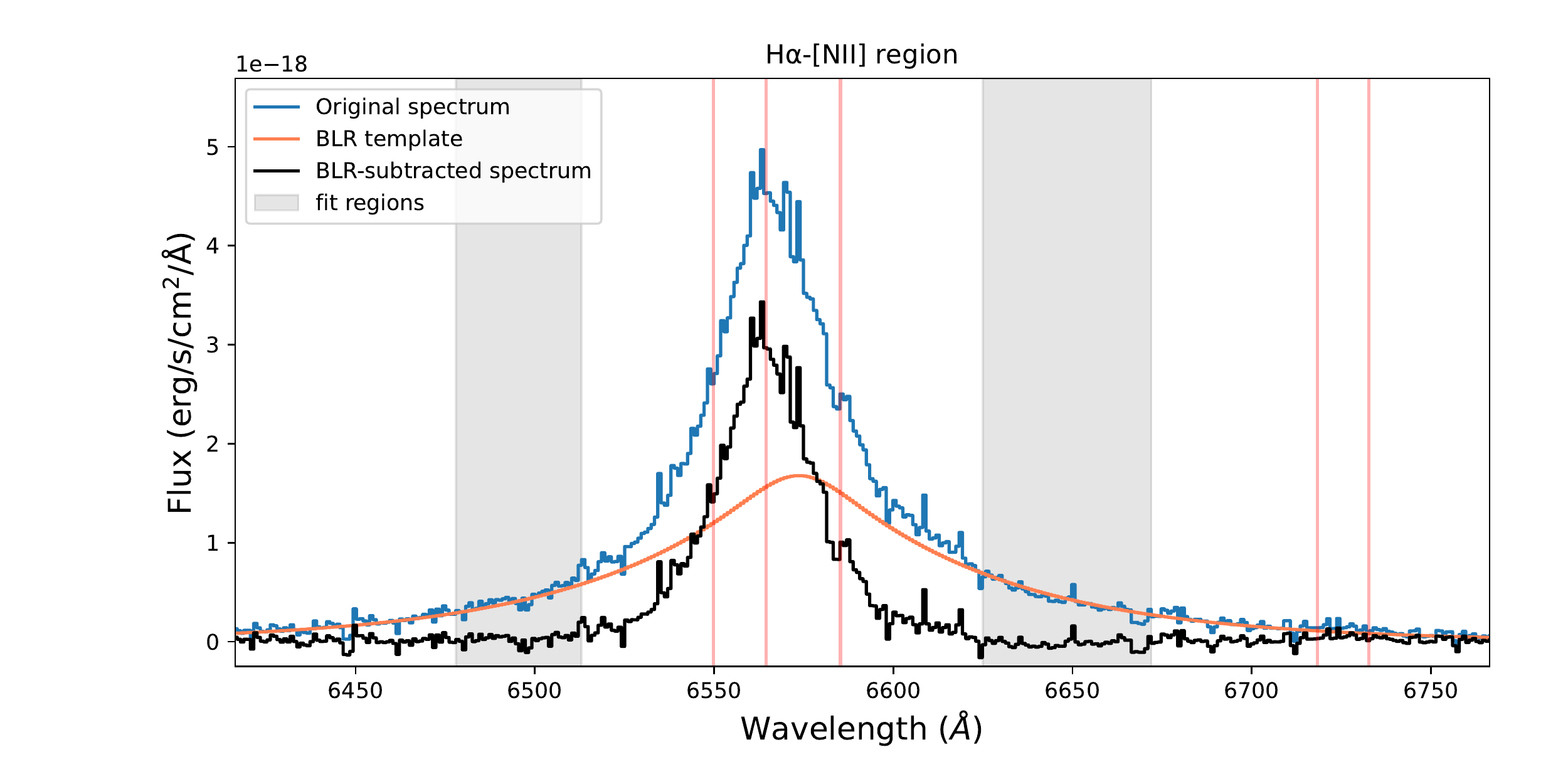}
\includegraphics[scale=0.55,trim= 0 0 0 8,clip]{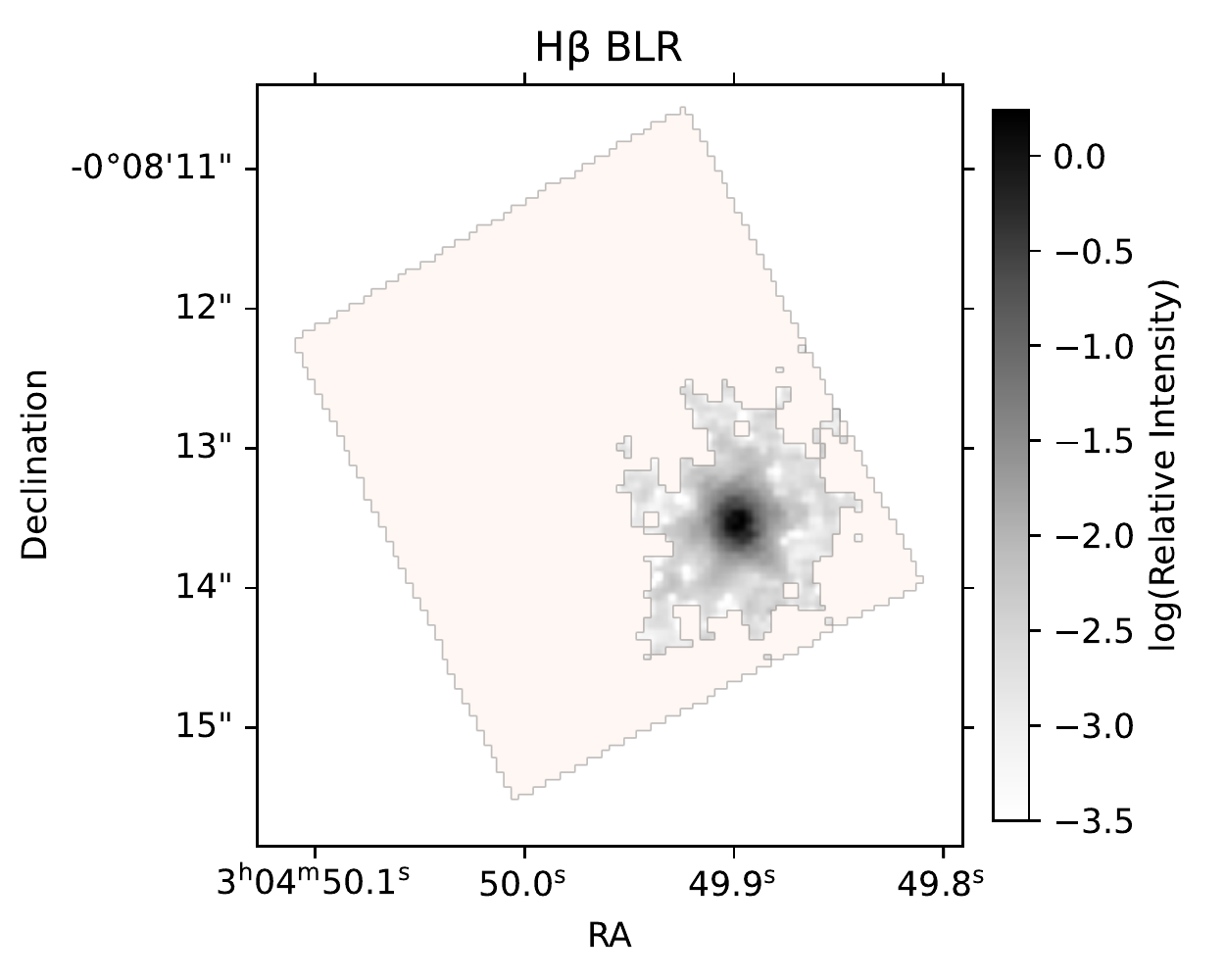}
\includegraphics[scale=0.4,trim= 5 0 0 0,clip]{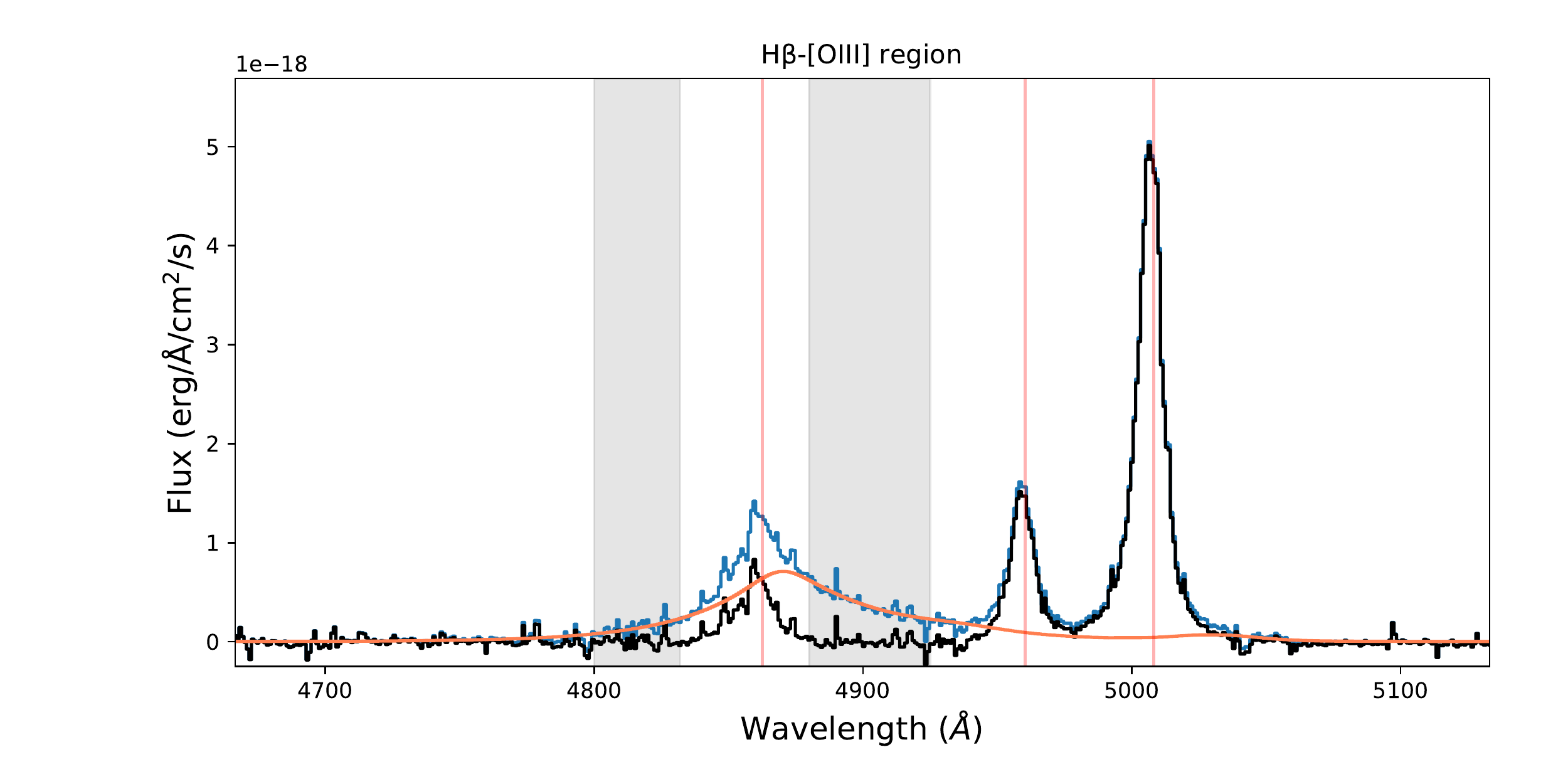}

\caption{Reconstructed PSF from the spatially unresolved BLR emission. Left: PSFs measured from the \drizzle \ cube for the \ha \ and \hb \ BLR emission, respectively, as described in Sect. \ref{sec:QSOsubtraction}. The reconstructed \hb \ PSF is less extended than \ha, being at shorter wavelengths and therefore associated with a smaller FWHM. Right: Visualisation of the BLR subtraction in an individual spaxel at 0.14\arcsec north-east of the nucleus, using the \hb\ (bottom) and \ha\ (top) BLR template. The blue spectrum is the original continuum-subtracted spectrum in the spaxel. The orange line is the BLR model. 
Using the broad spectral windows marked in grey, the BLR model is scaled to fit the original spectrum. The black curve shows the residual to that fit, which is the BLR-subtracted spectrum. }
\label{fig:PSF}
\end{center}
\end{figure*}

\subsection{Modelling of the wiggles }

Figures \ref{fig:3x3wiggles50mas} and \ref{fig:3x3wiggles100mas} show sinusoidal-type patterns with relatively constant amplitudes across the entire wavelength range, and significant variations for the phase shift and the frequency within the $3\times3$ innermost nuclear spaxels.  We note that the frequency changes smoothly along the whole wavelength range, being almost constant in relatively narrow ranges; we took advantage of this behaviour to model the wiggles.

As a first step, we fit the wiggles of the spectrum extracted from the brightest spaxel, the one with highest signal-to-noise ratio (S/N). We used a sinusoidal function to model the wiggles,
$y(w) = A \sin(2\pi f_w  w + \phi) + B$, where A is the amplitude, $f_w$ is the frequency in $1/\mu$m, $w$ is the wavelength, $\phi$ is the phase shift, and B is the continuum level; we repeated the process in small portions of the wavelength range ($\sim 0.1\mu$m) as many times as necessary to cover the entire spectrum. The combination of all best-fit sinusoidal functions is shown in the top panel of Fig. \ref{fig:LBQScentralspaxel} (red curve). The high spectral resolution and the small number of parameters to fit the wiggles allow us to get  a good representation of the wiggles across the entire wavelength range, after masking the channels associated with the most prominent emission lines and the gap between detectors.  

In the central panel of Fig. \ref{fig:LBQScentralspaxel}, we compare the integrated spectrum (orange) with the corrected one (dark blue), obtained after subtracting the best-fit model for the wiggles. The new residuals with respect to the integrated spectrum are significantly smaller than the original ones (reported in grey in the top panel).

By modelling the wiggles, we discover that the wiggle frequency, $f_w$, changes smoothly as a function of the wavelength, as shown in the bottom panel of Fig. \ref{fig:LBQScentralspaxel}: $f_w\sim 40 \ \mu$m$^{-1}$ at shortest and longest wavelengths, and $f_w\sim 5\  \mu$m$^{-1}$ in the central part of the spectrum. This $f_w$ trend is common to all single-spaxel spectra around the QSO peak, and can be used to better constrain the shape of the wiggles even for lower S/N spectra, or in masked regions (associated with strong emission lines, and the gap between the two detectors). As a final step, therefore, we fit all neighbouring spaxels using the inferred $f_w$ as a prior for the modellisation of the wiggles. Figures \ref{fig:3x3wiggles100mascorr} and \ref{fig:3x3wiggles50mascorr} show the same residuals presented in Figs. \ref{fig:3x3wiggles50mas} and \ref{fig:3x3wiggles100mas}, but after the correction described above. In Appendix \ref{sec:Awiggles} we also present some caveats of our procedure.

We stress here that the wiggles behave similarly in all the data cubes of bright point-like sources analysed so far  within the GTO programme; the procedure we described above is perfectly capable of modelling and correcting for them. As an example, we report in Fig. \ref{fig:VDEScentralspaxel} the wiggles modelling for another target from our GTO programme, VDES J0020-3656, a QSO at $z = 6.86$  observed with NIRSpec IFS with the grating--filter pair G395H--F290LP and presented in \citet{Marshall2023}. We note that data cube of this target has been obtained by combining two datasets, observed with different telescope position angles (on October 1, 2022, with $62^\circ$ and on October 16, 2022, with $160^\circ$). 
Nevertheless, the wiggles are very similar to those in \lbqs, consistent with the fact that these artefacts are inherent to the cube building process.

\subsection{QSO subtraction}\label{sec:QSOsubtraction}

Having corrected for the wiggles at single-spaxel level, we proceeded with the separation between the host and QSO emission, making use of the {\sc{QDeblend3D}  } routines  \citep{Husemann2013,Husemann2014}, which is optimised to subtract the PSF emission from NIRSpec IFS data. 

{\sc{QDeblend3D}  } considers the relative strength of the BLR lines in each spaxel to map out the spatial PSF, as the BLR is spatially unresolved. 
Due to the NIRSpec PSF dependence with wavelength, we performed the QSO subtraction twice: one for the wavelength channels around the \ha\ line, taking as a reference the \ha \ BLR emission, and one for those in the vicinity of \hb, taking as a reference the \hb \ broad wings. 

A PSF subtraction was performed following the procedure described in detail in \citet{Marshall2023}, also illustrated  in Fig. \ref{fig:PSF}. Briefly, we used the previously built model for the BLR (and iron) emission (Sec. \ref{sec:SpectralFit} and Fig. \ref{fig:HaHbfit}) as a template, rescaled in each spaxel to fit the BLR emission in broad spectral windows covering the wings of the Balmer lines (see Fig. \ref{fig:PSF}).
These broad spectral windows are free from any narrow and outflow component contributions, to avoid any bias in the measurement of the BLR strength.
Finally, we subtracted this rescaled template from each  spaxel spectrum and generated a new BLR-subtracted data cube.

\begin{figure*}
\centering
\begin{center}
\includegraphics[scale=0.55,trim= 0 10 0 8,clip]{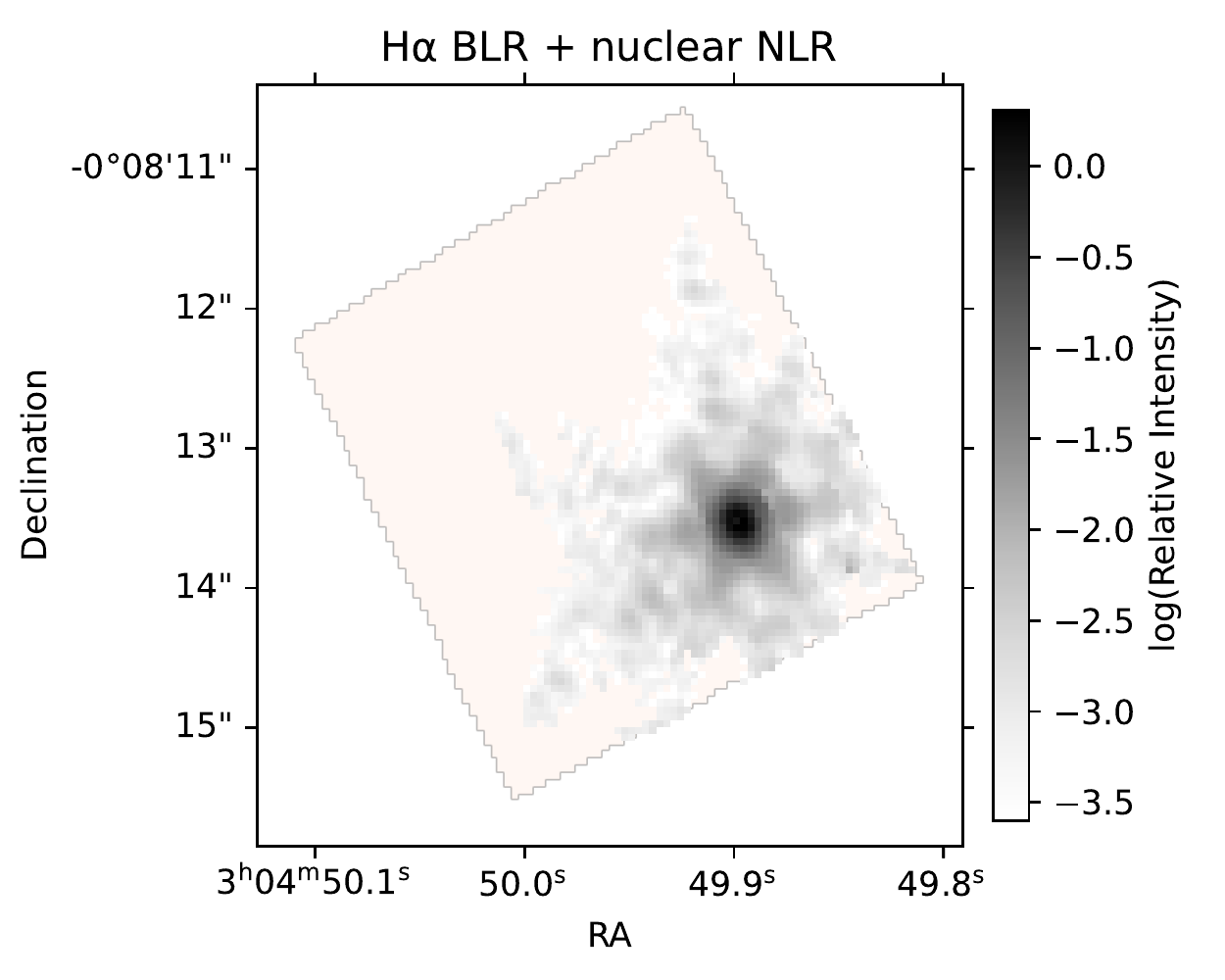}
\includegraphics[scale=0.55,trim= 0 10 0 8,clip]{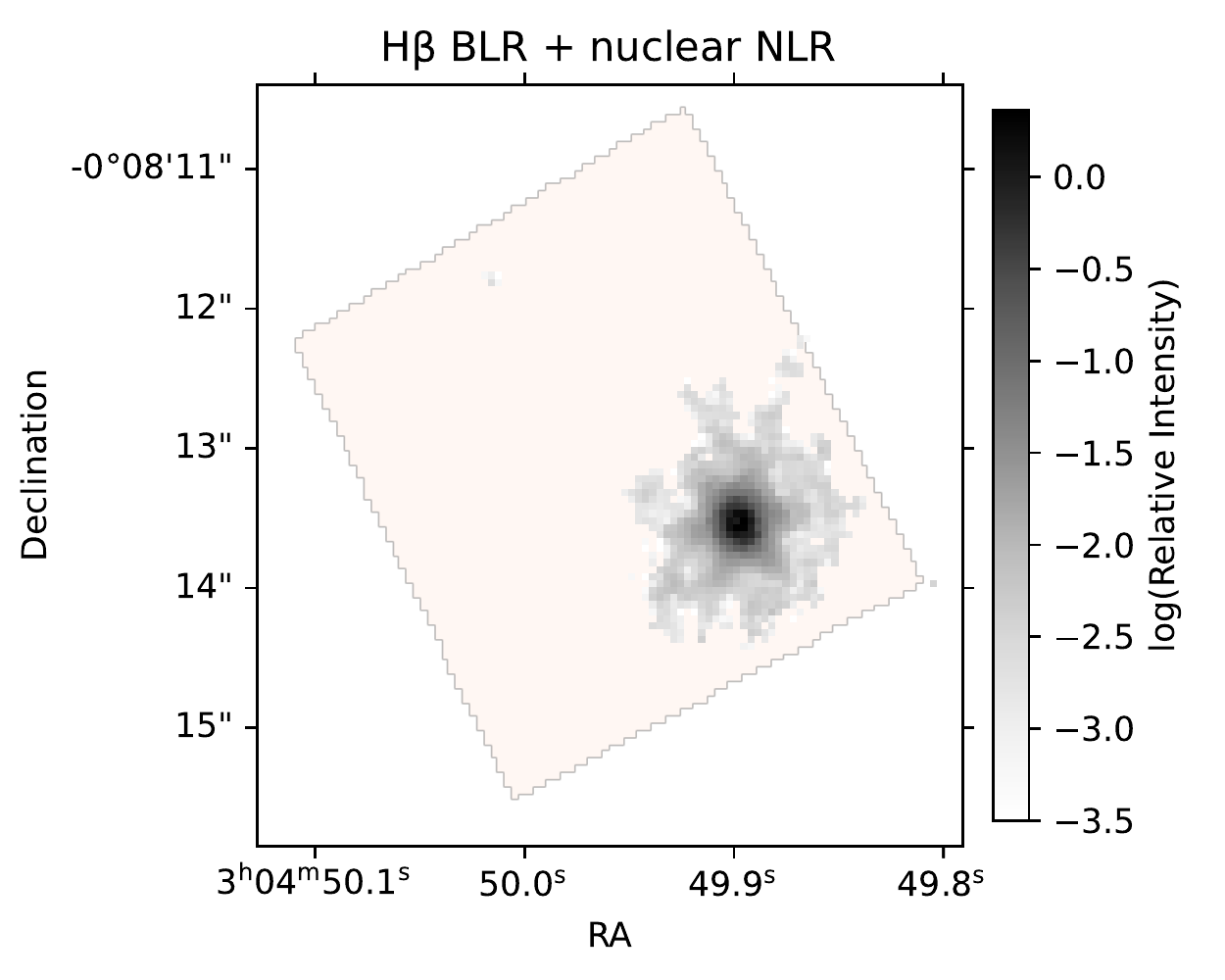}

\caption{PSFs measured from the \drizzle \ cube for the nuclear emission around \ha \ (left) and \hb \ (right) and including both the BLR and outflow components (as described in Sect. \ref{sec:QSOsubtraction}). With respect to Fig. \ref{fig:PSF}, these panels better reproduce the 2D distribution of the unresolved emission.}
\label{fig:PSF2}
\end{center}
\end{figure*}

A fractional map of the relative brightness of the spatially unresolved BLR, that is, the 2D PSF, is shown in the left part of Fig. \ref{fig:PSF}, for both \ha \ and \hb. We note that the described subtraction does not take the NLR emission into account, which is similarly spread according to the PSF shape. To take this further contribution into account, we performed a different QSO subtraction, this time using i) the integrated nuclear spectrum as a template, and ii) broader spectral windows at both sides of the Balmer lines, including the emission from high-velocity gas associated with the outflow (which is unresolved in \lbqs; see Sect. \ref{sec:results}). This new reconstructed PSF is shown in Fig. \ref{fig:PSF2}, and better reproduces the 2D distribution of unresolved emission (as the NLR outflow wings have higher S/N than the BLR wings). The cubes obtained from the subtraction of this high-velocity components (from both NLR and BLR) are not used in the analysis described in the next sections, but have been used to generate the  \oiii \ map shown in Fig. \ref{fig:LBQS_images_qsosub}.

\subsection{Line fitting}

To derive spatially resolved kinematic and physical properties of ionised gas, we fit the spectra of individual spaxels using the prescriptions already presented in Sect. \ref{sec:integratedspec}. We applied the BIC selection to determine where a multiple-Gaussian fit is required to statistically improve the best-fit model. This choice allows us to use the more degenerate multiple-component fits only where they are really needed. For the spatially resolved analysis, we used two Gaussian components at maximum, as they are perfectly capable of reproducing the line profile variations in the field of view (FOV); this limited number of components is also required to reduce the degeneracy in the fit.

\begin{figure}
\centering
\begin{center}
\includegraphics[scale=0.55,trim= 0 0 0 40,clip]{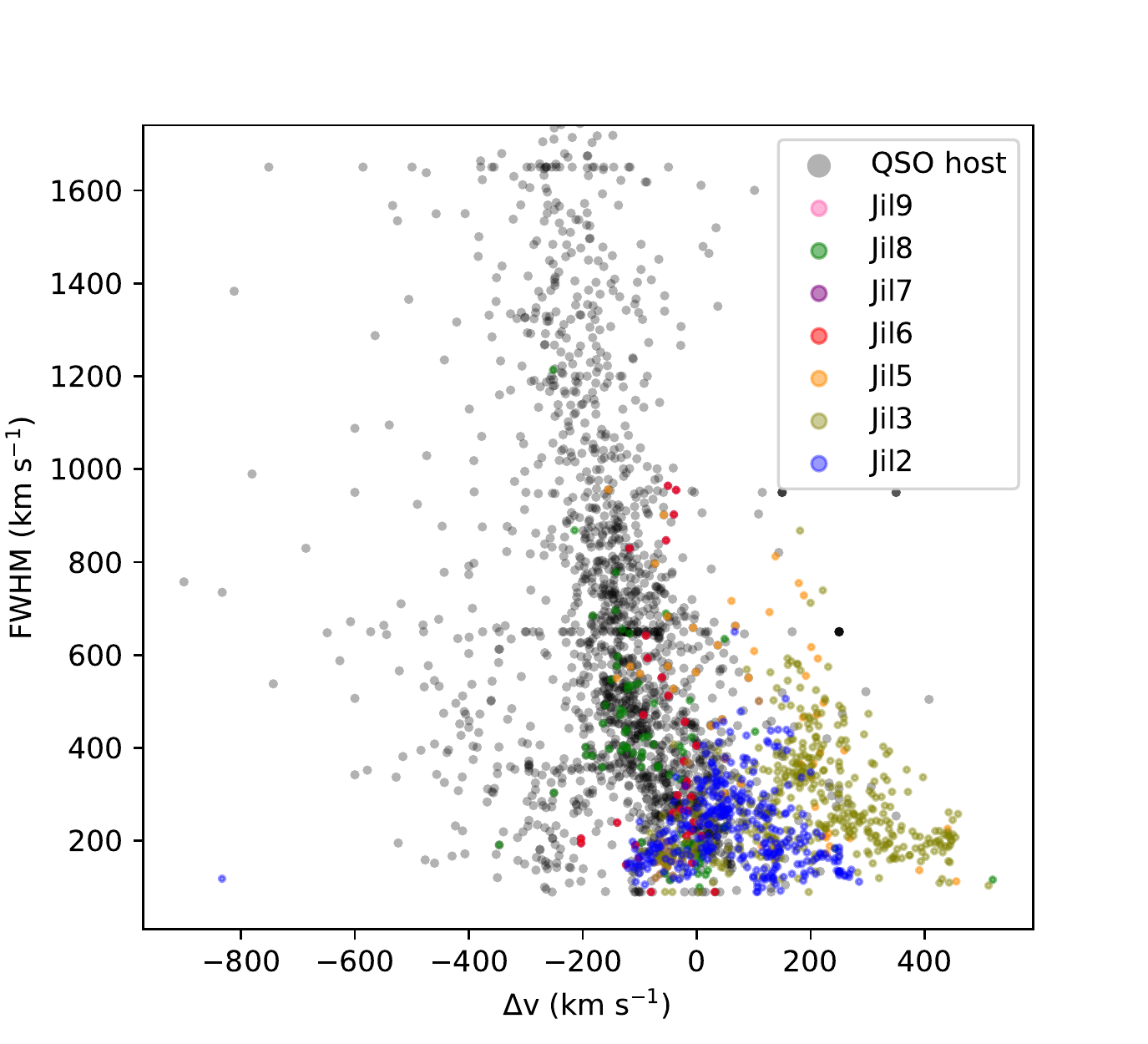}

\caption{Velocity diagram for the individual Gaussian components used to model the emission line profiles in the data cube. Different colours are used to identify different targets in the NIRSpec FOV, as labelled.}
\label{fig:components}
\end{center}
\end{figure}

Figure \ref{fig:components} shows the \lbqs\ velocity diagram, with all kinematic parameters of the Gaussian components required to fit the BLR-subtracted data cube. There is a clear trend in the figure, with the highest FWHMs ($>500$ \kms) associated with significant blueshifts ($\Delta v < -100$ \kms ), as usually observed in systems hosting AGN outflows (e.g. \citealt{Woo2016,Perna2022}).\ The Gaussian components with smaller FWHMs have relatively small offsets from the zero velocity (up to a few hundred \kms). In this figure, we use different colours to distinguish between different regions (targets) in the FOV:  while the \lbqs \ host (black points) is often associated with extreme kinematic parameters, all other companions (see the labels) show narrower profiles possibly associated with rotation. A detailed characterisation of the individual kinematic systems is reported in the next sections.

\begin{figure*}
\centering
\begin{center}
\includegraphics[scale=0.59,trim= 20 40 5 34,clip]{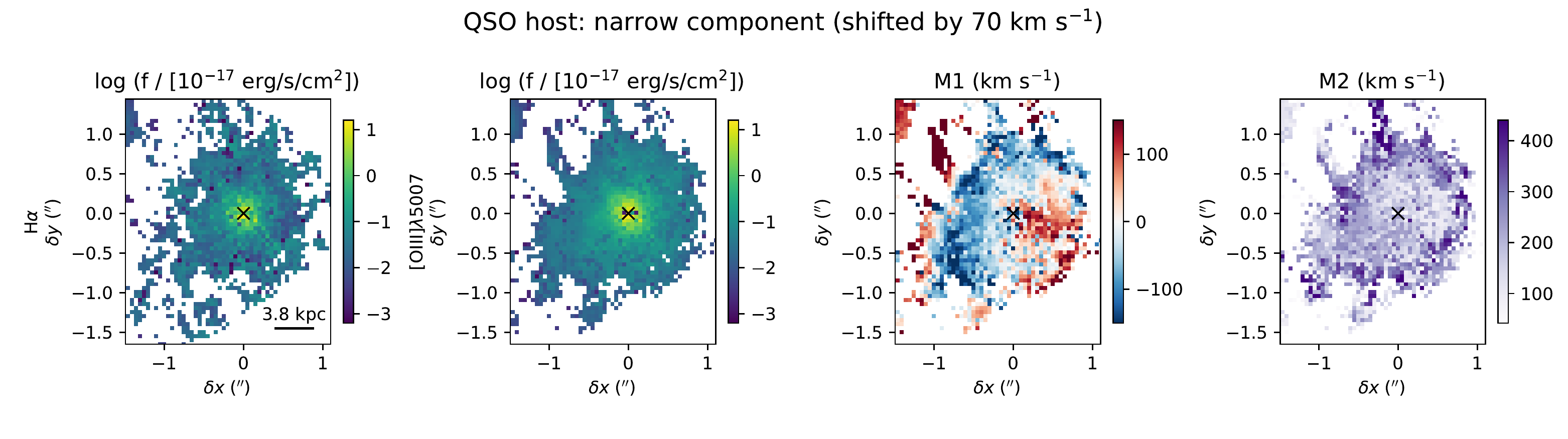}

\caption{{ From left to right: \ha\ and \oiii \ flux distributions}, and Moment 1 and Moment 2 maps of the QSO host galaxy, obtained from the narrow components of our best-fit models. Both lines show evidence of rotating gas in the QSO host.}
\label{fig:hostkinematics}
\end{center}
\end{figure*}

\section{Results}\label{sec:results}

\subsection{QSO host disk}

Figure \ref{fig:hostkinematics} shows an overview of the flux distribution and kinematics of the narrow component in the \lbqs \ host galaxy, as derived from our modelling of the \oiii \  line (top panels) and \ha \ (bottom) in the BLR-subtracted data cube. 
The flux distribution of the two lines is dominated by the nuclear emission, which spreads according to the PSF (see Fig. \ref{fig:PSF2}), although a few clumps towards the east and south-east as well as an extended plume towards the north-east (in the \oiii \ map) can be easily recognised. All of these features, reasonably associated with different sources in the QSO host environment, are discussed in the next section. 

The velocity distribution, traced by the Moment 1, shows evidence for a velocity gradient along the north-east--south-west direction, with a velocity amplitude of $\sim \pm 120$ \kms,  
possibly associated with a rotating disk. The most significant deviations from this gradient are found in the external regions, in correspondence with the clumps and the plume identified in the flux distribution panel. We also note that the \ha \ velocity field is noisier than the \oiii \ one, because of the BLR subtraction step, and the degeneracy between \ha\ and \nii\ lines. 

The \oiii \ and \ha \ line widths, traced by the Moment 2 map, do not show significant variations across the host. However, elevated dispersions in the central region of the galaxy in both the \ha\ and \oiii \ maps might be present.

As the \ha \ maps are probably more affected by PSF artefacts and BLR-subtraction, we decided to use the \oiii \ line to model the gas kinematics with {\sc{3D-Barolo}} (\citealt{DiTeodoro2015}), following the procedure described in \citet{Perna2022}, to test whether the QSO host kinematics are compatible with a rotation-supported system and to infer the host dynamical mass. 
The main assumption of the {\sc{3D-Barolo}}  model is that all the emitting material of the galaxy is confined to a geometrically thin disk, and its kinematics are dominated by pure rotational motion. The possible presence of residual components associated with the outflow, as well as the presence of additional kinematic components associated with close companions might affect the modelling. Nevertheless, this model enables us to assess the presence of such disks and to infer a simple kinematic classification through the standard $v_{\rm{rot}}/\sigma_0$ ratio, where $v_{\rm{rot}}$ is the intrinsic maximum rotation velocity (corrected for inclination, $v_{\rm{rot}} = v_{LOS}/sin(i)$) and $\sigma_0$ is the intrinsic velocity dispersion of the rotating disk, related to its thickness. In this work, we define $\sigma_0$ as the measured line width in the outer parts of the galaxy, corrected for the instrumental spectral resolution (e.g. \citealt{Schreiber2018}). The {\sc{3D-Barolo}} best-fit plots are shown in Fig. \ref{fig:baroloHost}. From them, we infer an inclination $i = 10\pm 7^\circ$, a $v_{\rm{rot}} = 360\pm 80$ \kms, and a $\sigma_0 = 190\pm 45$ \kms \ (at $\sim 0.3\arcsec$, i.e. $\sim 2.4$ kpc from the nucleus, as the more external regions are more affected by noise). 
The rotation-to-random motion ratio $v_{\rm{rot}}/\sigma_0\approx 2$ indicates that this galaxy is associated with a dynamically warm disk, consistent with  $z\sim 2$ galaxies presented in \citet{Schreiber2018}, with $v_{\rm{rot}}/\sigma_0$ spanning the range from 0.97 to 13 (with a median of 3.2), as inferred from \ha \ gas kinematics (see also e.g. \citealt{Wisnioski2019}). 

The {\sc{3D-Barolo}} best-fit velocity maps also show significant residuals in the receding part, at $\sim 0.15\arcsec$ south-west of the nucleus, with velocities $\approx 100$ \kms; they might be associated with a plume, or a further companion on the LOS. This kinematic component might also be  present in the integrated spectrum in Fig. \ref{fig:HaHbfit}: the significant residuals in the red part of the \ha \ line, if due tp \ha \ line, would correspond to L(\ha) $\approx 10^{43}$ erg/s, consistent with the luminosity of other Jil companions (see Table \ref{tab:Jilproperties}).

From the {\sc{3D-Barolo}} best fit, we also inferred a tentative estimate for the dynamical mass, assuming that the source of the gravitational
potential is spherically distributed (following e.g. \citealt{Perna2022}):  M$_{\rm {dyn}}$ = $(14\pm 6)\times 10^{10}$ M$_\odot$, within a radius of $2.4 \pm 0.6$ kpc (corrected for the PSF, and containing 85\% of the \oiii\ total flux, as inferred from the QSO-subtracted cube). Combining this measurement with the M$_{\rm {BH}}$ derived in Sect. \ref{sec:integratedspec}, we obtained a M$_{\rm {BH}}/$M$_{\rm {dyn}} \approx 0.014$. 
This places the \lbqs \ host galaxy slightly above the local black hole-–host mass relation (\citealt{Kormendy2013}), consistent with other high-$z$
QSOs reported in the literature (see e.g. \citealt{Marshall2023} and references therein).

We finally investigated the dominant ionisation source for the emitting gas across the \lbqs\ host, using the classical `Baldwin, Phillips \& Terlevich' (BPT)  diagram (\citealt{Baldwin1981}). The distributions of the flux ratio diagnostics are almost constant across the host galaxy extension, with log([NII]/\ha) \ $= -0.3\pm 0.2$ and log(\oiii /\hb) \ $= 0.7\pm 0.2$. These values place the \lbqs \ host in the AGN-dominated region of the BPT diagram (\citealt{Baldwin1981, Kewley2013}; Fig. \ref{fig:BPT}).

\begin{figure*}
\centering
\begin{center}
\includegraphics[scale=0.59,trim= 20 40 5 34,clip]{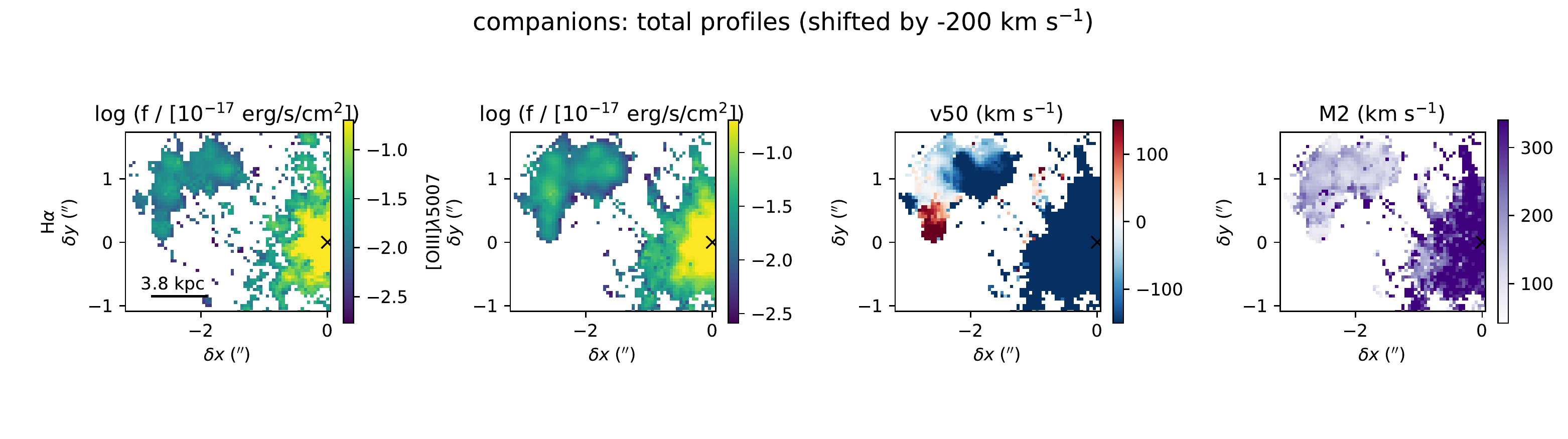}

\caption{From left to right: \ha\ and \oiii\  flux distributions, and Moment 1 and Moment 2 maps of the Jil companions, obtained from the total profiles of our best-fit models. Both lines show evidence of rotating gas in the north-east companions.}
\label{fig:companions_totalprofiles}
\end{center}
\end{figure*}

\subsection{QSO outflow energetics}\label{sec:energetics}

The outflow component used to model the QSO host is not spatially resolved, and is therefore not reported in the figures. In this section we measure the mass of the ionised outflow as inferred from the
blueshifted outflow component of \hb. We used the equation

\begin{equation}
    \dot M_{\rm{out}}(H\beta) = 8.6 \times \frac{L_{41}(H\beta) \ v_{\rm{out}}}{n_e\ R_{\rm{out}}} M_{\odot} yr^{-1}
\end{equation}

\noindent from \citet{Cresci2015a}, where L$_{41}$(H$\beta$) is the \hb \ luminosity associated with the outflow
component in units of $10^{41}$ erg s$^{-1}$,
$n_e$ is the electron density, $v_{\rm{out}}$ is the outflow velocity, and $R_{\rm{out}}$ is
the radius of the outflowing region in units of kiloparsecs.

In general, $n_e$ can be estimated from the \sii \  doublet ratio (e.g. \citealt{Osterbrock2006}), using the high-velocity components of the \sii \  lines. Unfortunately, these components are only barely detected in our integrated spectra, and cannot be used to infer the outflow electron density. We therefore conservatively considered an electron density of 1000 cm$^{-3}$, inferred from the study of large samples of AGN both at low redshift ($z < 0.8$, \citealt{Perna2017b}) and at $0.6 < z < 2.7$ (\citealt{Schreiber2019}). A factor of $\sim 3$ higher mass rate would be obtained for instance using the electron density measured in the outflowing gas of the QSO XID2028 at $z\sim 1.5$ (i.e. $360\pm 180$ cm$^{-3}$), as measured from recent JWST/NIRSpec IFS observations (\citealt{Cresci2023}). 

Here we consider the L$_{41}$(H$\beta$) to be the luminosity of the \hb \ outflow component
as measured from our full integrated spectral fit described in Sect. \ref{sec:integratedspec} and shown in Fig. \ref{fig:HaHbfit}, as the outflow is not resolved in our data cube: 
log(L(H$\beta$)/[erg/s]) $= 45.17_{-0.13}^{+0.09}$. The luminosity has been corrected for the extinction considering the colour excess for the same outflow component, inferred from the Balmer decrement and assuming a Milky Way extinction law (\citealt{Cardelli1989}): $E(B-V) = 0.58_{-0.15}^{+0.07}$.

The identification of the BLR component in \lbqs \ suggests that the outflow could be primarily orientated towards us; this could also explain why the ejected gas is not spatially resolved, regardless the exquisite NIRSpec resolution ($\sim 800$ pc). Under this assumption, the observed velocity offset of the outflow components with respect to the BLR systemic is close to the true outflow velocity (e.g. \citealt{Harrison2012}); as the outflow component in the integrated spectrum requires the use of two Gaussian components, we decided to use as velocity offset the v50 inferred from the total outflow profile. We therefore derive a $v_{\rm{out}} = 930_{-110}^{+60}$ \kms.

The last ingredient required for the computation of the mass rate is the outflow extension; as this component is not spatially resolved in our NIRSpec cube, we assumed that the outflow is propagating at constant velocity (e.g. \citealt{Brusa2015, Fiore2017}), 
and that its dynamical time ($t_d$) is equal to the AGN phase inferred by \citet{Worseck2021}, $t_d > 11$ Myr. This is very close to the $t_d$ usually inferred from observations of ionised outflows (e.g. \citealt{Greene2012,Perna2015a}). We therefore estimate $R_{\rm{out}} = t_d \times v_{\rm{out}} \gtrsim 9$ kpc.
This estimate is compatible with the extension of ionised outflows observed in other QSOs at high $z$, in the range $\approx 2-15$ kpc (\citealt{Carniani2015, Kakkad2020, Cresci2023}).   
Because of that, we considered the inferred lower limit as an order of magnitude estimate for the outflow extension. 

We therefore obtain an outflow mass rate 
$\dot M_{\rm{out}}(H\beta) \sim 10^4$ M$_\odot$ yr$^{-1}$. This value, although significantly larger than other mass rates reported in the literature, is still consistent with the general expectations inferred from the scaling relations presented, for instance, in \citet{Fiore2017} and \citet{Fluetsch2019}. The inferred value is $\sim 3$ times larger than the one obtained from the \oiii\ gas, following \cite{Carniani2015}; similar discrepancies are often reported in the literature (e.g. \citealt{Carniani2015,Perna2015a,Perna2019,Marshall2023}) and are probably due to the ionisation structure of the \oiii\ and \hb \ clouds in the NLR of an AGN. The kinetic and momentum powers are $\dot E_{\rm{out}} = 1/2 \dot M_{\rm{out}}v_{\rm{out}}^2  \sim 4\times 10^{45}$ erg s$^{-1}$ and $\dot P_{\rm{out}} = \dot M_{\rm{out}}v_{\rm{out}} \sim  8\times 10^{37}$ dyne, respectively. Hence, the kinetic power is $\sim  2$\% of the radiative luminosity of the AGN, while the momentum rate is in excess of $\sim  15$ times the radiative momentum flux (L$_{\rm {bol}}$/c), consistent with the energetics of other QSOs in the literature (see e.g. \citealt{Perna2015b, Bischetti2017, Tozzi2021}).

\subsection{Further considerations of the outflow extension}

We report here two further arguments to better justify the assumed outflow extension ($\approx 9$ kpc). On the one hand, greater extensions would be at odds with the fact that the outflow is unresolved in our data cube: high collimation (with a half opening angle $\alpha_{out}$ of a few degrees) would be required to explain the presence of a spatially unresolved ($\lesssim 0.8$ kpc, i.e. below the spatial resolution of our data) and highly extended outflow ($> 9$ kpc) along our LOS, at odds with the reconstructed geometry of other outflows at lower redshifts (with $\alpha_{out}\approx 10-60^\circ$; e.g. \citealt{Muller2011,Meena2021,Cresci2023}). On the other hand, by assuming that the outflow has an extension $< 0.8$ kpc, we would obtain outflow energetics that are ten times higher (e.g. $\dot M_{\rm{out}} \sim 10^5$ M$_\odot$ yr$^{-1}$). Both the scenarios are quite unlikely. We therefore conclude that the measurements reported in the previous section can represent rough estimates of the outflow energetics for \lbqs.

\subsection{QSO environment: The Jil objects}

Figure \ref{fig:companions_totalprofiles}  shows an overview of the flux distribution and kinematics of the ionised gas in \lbqs \ companion sources, as derived from our modelling of the \oiii \ line (top panels) and \ha \ (bottom). The flux distribution shows multiple clumps in the north-east regions, as well as plumes and irregular structures within $\approx 1\arcsec$ of the \lbqs \ nucleus. All of these sources have relative velocity shifts up to a few hundred \kms \ with respect to the QSO systemic; this implies that they are not artefacts induced by the nuclear PSF. The velocity distribution, traced by the Moment 1 of the total fitted profiles, shows evidence for gradients with velocity amplitudes of $\sim \pm 200$ \kms. The velocity width (Moment 2) in these companions is significantly smaller than the ones in the QSO host. 

\begin{figure*}
\centering
\includegraphics[scale=0.33,trim= 0 70 60 0,clip]{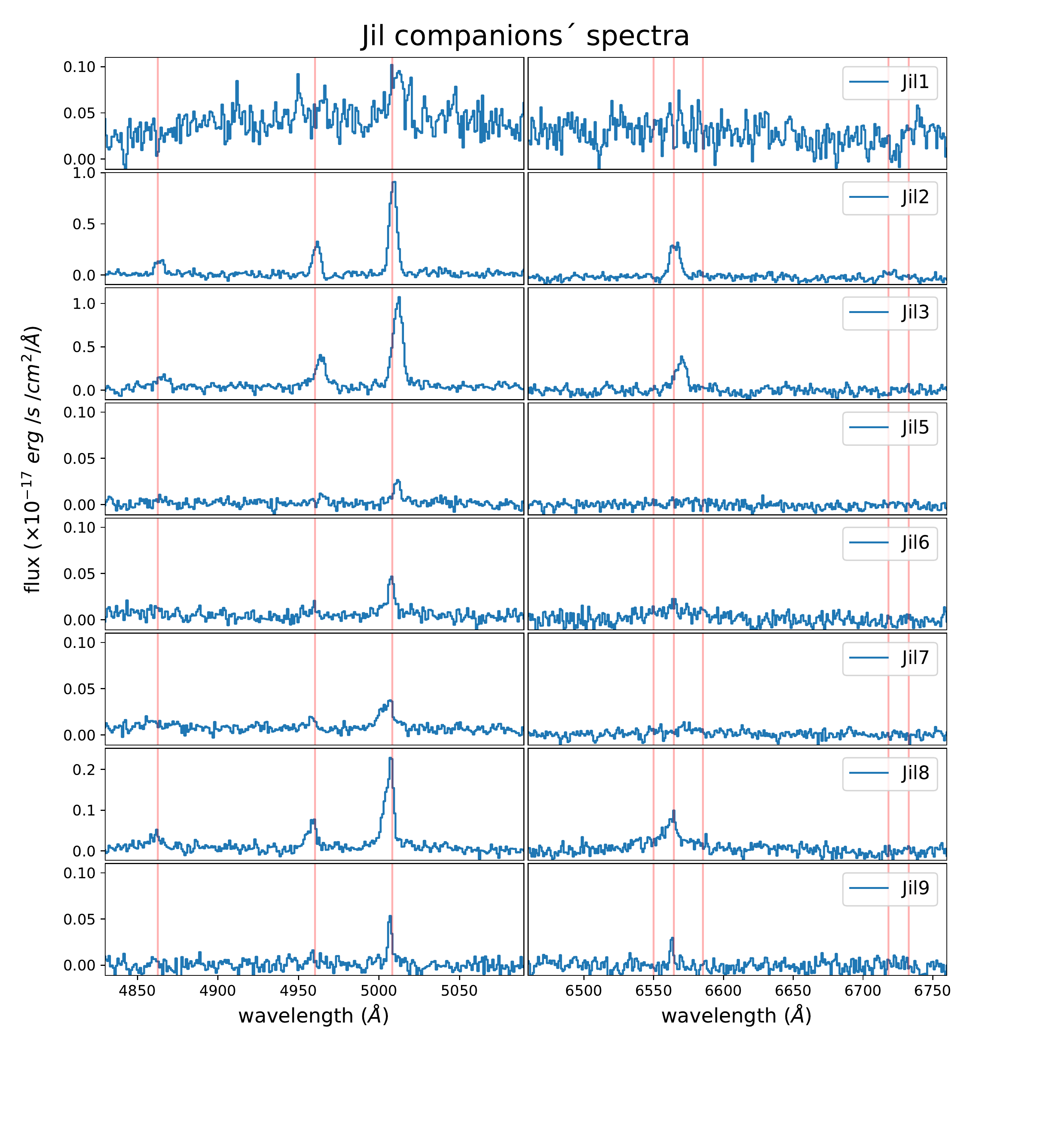}
\includegraphics[scale=0.52,trim= 33 30 40 35,clip]{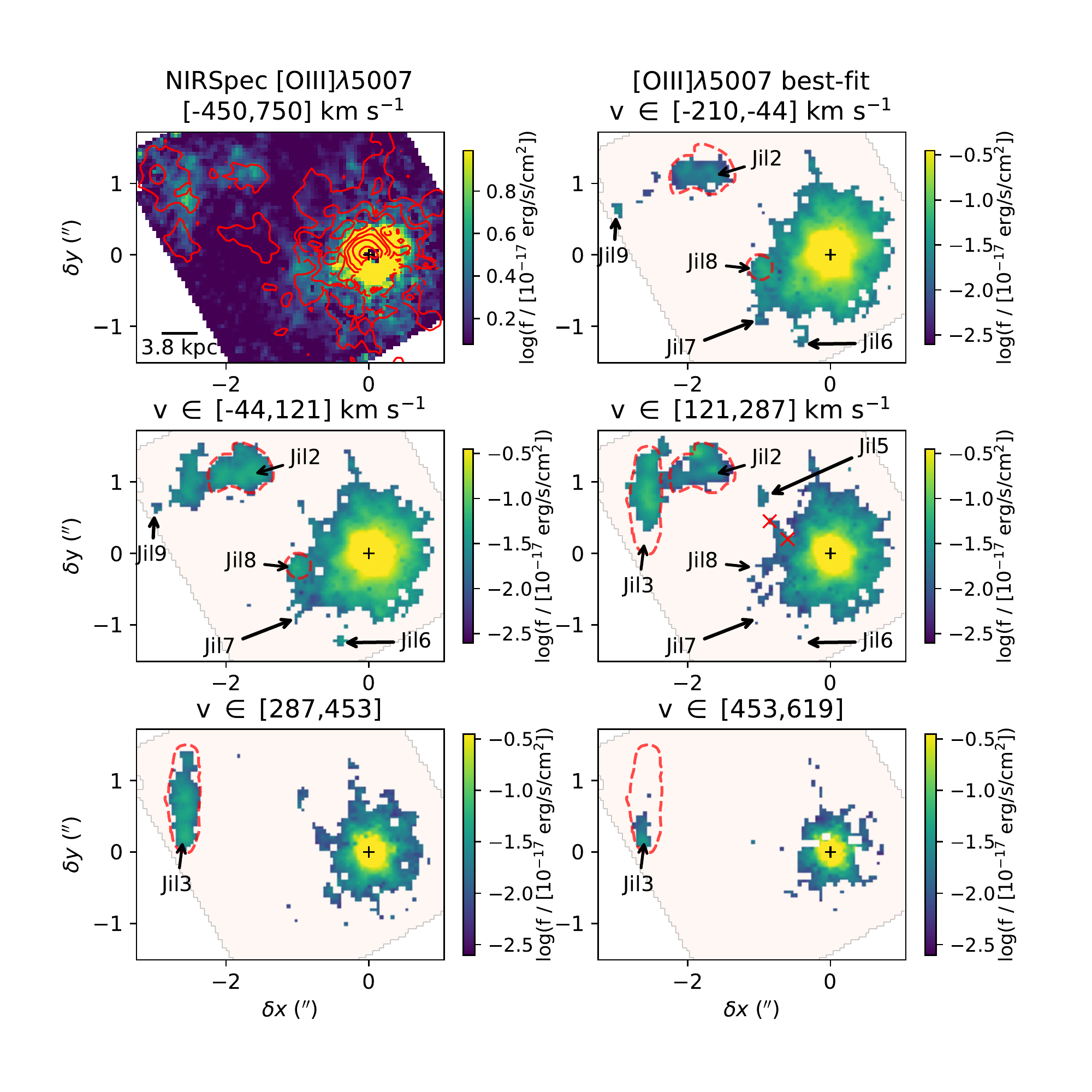}

\caption{Jil spectra and spatial distribution. Left panels: Jil companions' spectra. Jil2, Jil3, and Jil8 spectra have been obtained by integrating over the entire extension of the targets (with \oiii \ detected above $8\sigma$). For the remaining sources, we considered $3\times 3$ spaxel integration (Jil5, Jil6, Jil7, and Jil9), or $5\times 5$ spaxels (Jil1). All profiles are relatively narrow and redshifted with respect to the QSO systemic; this ensures that we are not affected by PSF contamination. The continuum emission is never detected; the continuum observed in the Jil1 spectrum is likely due to data reduction artefacts. Right panels: \oiii \ flux distributions, obtained by integrating over different velocity channels. The top-left panel has been obtained by integrating over a large velocity range, using the QSO NLR$+$BLR-subtracted cube (Sect. \ref{sec:QSOsubtraction}); all other velocity channel maps have been extracted from the best-fit \oiii \ profiles. The top-left panel shows the HST contours presented in \citet{Husemann2021}; all remaining panels show the Jil companions, as detected in NIRSpec. The dashed red lines identify the regions from which the integrated Jil2, Jil3, and Jil8 spectra, reported on the left part of the figure, have been extracted.}
\label{fig:Jilspectra}
\end{figure*}

In order to better identify all possible companions around \lbqs, in Fig. \ref{fig:Jilspectra} we show a few narrow-band images for the best-fit \oiii \ emission line, with overlaid contours from the HST image (already reported in Fig. \ref{fig:LBQS_images_qsosub}, left): these narrow-band images clearly show several clumps at different velocities. Some of them are associated with the Jil companions already identified by \citet{Husemann2021}: Jil1, Jil2, and Jil3. However, Jil1 is barely detected in our data as it resides on the very edge of the NIRSpec FOV, where the noise is higher and the data reduction generates unreliable spectral features. We also note that NIRSpec \oiii \ emission slightly differs from the flux distribution in the near-infrared HST, the former being more extended and clumpier; this also makes it difficult to separate the Jil sources.
Additional \oiii \ clumps not detected in the HST image are here dubbed Jil5, Jil6, Jil7, Jil8, and Jil9, following Husemann et al. Their integrated spectra are shown in the left part of Fig. \ref{fig:Jilspectra}.

The emission line properties of each companion, inferred from spectroscopic analysis, are reported in Table \ref{tab:Jilproperties}. Here we give a brief description of the specific properties inferred from each companion.

\begin{table*}
\centering
\begin{minipage}[!h]{1\linewidth}
\centering
\caption{Properties of the companion sources in the \lbqs \ environment. }
\begin{tabular}{lccccccl}
\hline 
\hline
    target & $z$ ($\Delta v$ [\kms]) & log($\frac{\rm{L([OIII])}}{\rm{erg\ s}^{-1}}$) & log($\frac{\rm{L(H\alpha)}}{\rm{erg\ s}^{-1}}$) & E(B$-$V) & log(\oiii \ / H$\beta$) & log(\nii \ / H$\alpha$) & $W80$ (\kms) \\
\hline 

    Jil1 &  $3.2898\pm 0.0003$ (195) & $41.5 \pm 0.1$ & $<40.9$ & $-$ & $>0.72$ & $-$ & $390\pm 30$\\
    Jil2 &  $3.2877\pm 0.0002$ (50) & $42.9 \pm 0.1$ & $42.6\pm 0.1$ & $0.2\pm 0.1$ & $0.81\pm 0.05$ & $<-1.04$ & $360\pm 30$ \\
    Jil3 &  $3.2904\pm 0.0002$ (240)& $43.6\pm 0.2$ & $43.0\pm 0.2$ & $0.5\pm 0.2$ & $1.00\pm 0.04$ & $< -0.82$ & $470\pm 30$\\
    Jil5 &  $3.2886\pm 0.0002$ (110) & $40.9\pm 0.1$ & $<40.5 $ & $-$ & $>0.7$ & $-$& $220_{-30}^{+140}$\\
    Jil6 &  $3.2866\pm 0.0004$ (-30)& $41.2\pm 0.1$ & $40.5\pm 0.2$ & $-$ & $0.67_{-0.11}^{+0.37}$ & $0.11_{-0.37}^{0.25}$ & $165_{-55}^{+30}$ \\
    Jil7  & $3.2836\pm 0.0004$ (-240)& $41.4\pm 0.3$ & $<40.6$ & $-$ & $0.52_{-0.04}^{+0.36}$ & $-$ &$580\pm 80$\\
    Jil8 & $3.2848\pm 0.0004$ (-150)& $42.1\pm 0.2$ & $41.3\pm 0.2$ & $-$ & $0.69_{-0.07}^{+0.14}$ & $<-0.31$ & $440_{-55}^{+30}$\\
    Jil9 & $3.2858\pm 0.0004$ (-80)& $41.0\pm 0.1$ & $40.4\pm 0.2$ & $-$ & $> 0.7$ & $<-0.10$ & $170_{-5}^{+30}$\\
\hline 

\hline
\end{tabular} 
\label{tab:Jilproperties}
\end{minipage}

\small{\justifying 
Notes: For each target, in the second column we report the redshift and the velocity offset with respect to the \lbqs \ host galaxy. Integrated \oiii \ and \ha \ luminosities have been corrected for extinction, when E(B$-$V) could be estimated, assuming a Milky Way extinction law (\citealt{Cardelli1989}). For targets with no \hb \ detection, we measured the log(\oiii/H$\beta$) lower limit assuming that the \hb \ upper limit is three times smaller than \ha. The non-parametric velocity $W80$ refers to the \oiii \ line profile. \par}

\end{table*}

Jil1 is detected in  \oiii \ at $\sim 3\sigma$. This is the only companion for which we could detect continuum emission; however, it has to be considered a spurious measure, because of its position at the edges of the FOV and the known issues with the data reduction. 

Jil2 is detected in \oiii, \ha, and \hb, but not in \nii \ or \sii \ lines. It shows a clumpy morphology, with an extension over $\sim 4$ kpc in projection. A velocity gradient with amplitude of $\sim \pm 100$ \kms \ is observed. 
To test whether Jil2 is compatible with a rotationally supported system, we modelled the \oiii \ line with {\sc{3D-Barolo}}, as done for the QSO host. The Jil2 best-fit models are shown in Fig. \ref{fig:baroloJil2}. The significant residuals in the maps are likely due to the clumpy morphology of this system, as well as the superposition with Jil3. These arguments likely explain the measured rotation-to-random motion ratios, $v_{\rm{rot}}/\sigma_0\approx 0.7$, and question the presence of a rotating disk. Nevertheless, we infer a tentative dynamical mass for this system, log(M$_{\rm {dyn}}$) = $(8_{-6}^{+16})\times 10^{9}$ M$_\odot$, considering a circular velocity of 75 \kms (corrected for an inclination $i = 80^\circ \pm 15^\circ$), as measured with {\sc{3D-Barolo}}, and a radius of 1 kpc (as order of magnitude size, given the clumpy morphology of this source). This dynamical mass is $\approx 3\times$ the stellar mass estimate inferred by \citet{Husemann2021} (from spectral energy distribution analysis), but still consistent within the errors.

\begin{figure*}
\centering
\includegraphics[scale=0.4,trim= 0 10 0 0,clip]{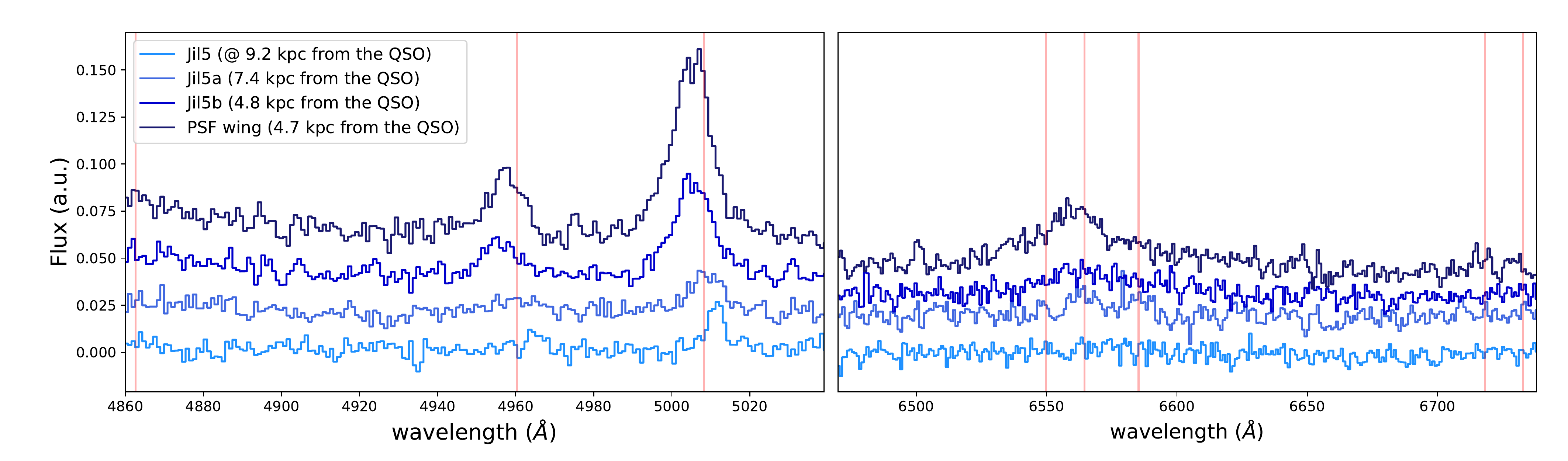}

\caption{Jil5 companion spectrum (light blue), together with two additional spectra extracted from the region between Jil5 and the QSO host galaxy (labelled as Jil5a and Jil5b, at 7.4 and 4.8 kpc from the QSO nucleus, respectively, also indicated in Fig. \ref{fig:Jilspectra} with red crosses). The dark blue spectrum has been extracted from a region at a distance of 4.8 kpc from the nucleus (as for Jil5b) but covering the PSF wing extending towards the north. In order to ease the visualisation, we added vertical offsets to the spectra.  This figure highlights a velocity gradient of a few hundred \kms\ across a few kiloparsecs (see also Fig. \ref{fig:companions_totalprofiles}), possibly indicating feeding processes or a tidal tail due to the interaction between Jil5 and the QSO host galaxy. 
}
\label{fig:Jil5spectra}
\end{figure*}

The presence of an obscured AGN in Jil2, initially proposed by \citet{Husemann2018a} on the basis of the presence of bright ultraviolet lines in the MUSE cube, will be discussed in the next section. Here we briefly mention that the measured log(\oiii /\hb) $\sim 0.8$, slightly higher than the value obtained for the NLR gas associated with \lbqs, is compatible with the presence of an AGN in this companion.

Jil3 is detected in \oiii, \ha, and \hb, but not in \nii \ or \sii\ lines. It shows an elongated morphology, with a clear and regular gradient with amplitude $\pm 200$ \kms \ over $\sim 8$ kpc. This extension, and the clumpy morphology, could suggest the presence of multiple systems; nevertheless, we also provide a {\sc{3D-Barolo}} model for the \oiii \ emission for this source. The best-fit results are reported in Fig. \ref{fig:baroloJil3}. We infer a tentative dynamical mass log(M$_{\rm {dyn}}$) = $(1.3\pm 0.5)\times 10^{10}$ M$_\odot$, assuming a circular velocity of 70 \kms (corrected for an inclination $i = 80\pm 10$, as measured with {\sc{3D-Barolo}}) and a radius of 1 kpc (as for Jil2). 
From the integrated spectrum of Jil3, we measure a high log(\oiii/\hb) $\sim 1$, consistent with possible presence of an AGN in its vicinity (i.e. in Jil2).

No further characterisation can be obtained for Jil4, which falls outside the NIRSpec FOV. 

Jil5 is located at $\sim 10$ kpc north-east of the \lbqs \ nucleus, and is connected with the QSO host  by a filamentary structure showing a clear velocity gradient (see Fig. \ref{fig:companions_totalprofiles}). To highlight the presence of such a gradient, in Fig. \ref{fig:Jil5spectra} we show the Jil5 spectrum in comparison with those extracted from intermediate positions along this elongated structure (identified by red crosses in the velocity channels at $\sim 200$ \kms in Fig. \ref{fig:Jilspectra}). We can therefore speculate that this companion is contributing to the feeding of the QSO host. For this companion we measure a log(\oiii /\hb) $> 0.7$, consistent with flux ratios measured in the QSO host, and hence likely ionised by the QSO radiation. 

Jil6 is located at $\sim 10$ kpc south-east of the QSO, and is detected in   \oiii \ and \ha \ (and in \hb \ and \nii \ at S/N $\sim 2-3$). Both log(\oiii /\hb) $\sim 0.7$ and log(\nii /\ha) $\sim  0.1$ suggest a QSO ionisation. 
 
Jil7 is located at $\sim 10$ kpc south-east of the QSO, and is detected in \oiii \ and \hb. It shows a prominent blue wing in the \oiii \ ($V10 \sim - 550 \pm 50$ \kms ), likely due to the superposition of different kinematic components along the LOS, and relatively high line ratios (log(\oiii /\hb)  $\sim 0.52$).

Jil8  is located at $\sim 8$ kpc east of the QSO nucleus, with an extension of $\sim 2$ kpc. It is detected in \oiii, \ha, and \hb. The broad components in the emission lines are due to PSF artefacts. For this companion, log(\oiii /\hb) $\sim 0.7$ suggests a QSO ionisation. 

Jil9 is located at $\sim 23$ kpc north-east of the blue QSO, and is detected in \oiii \ and \ha. It shows a velocity offset of $\sim 300$ \kms \ from Jil3, and narrow line profiles ($W80\sim 170$ \kms). In this case, we detect a lower limit for log(\oiii /\hb) $> 0.7$, consistent with the presence of high ionisation.

\subsubsection{Dual QSO with 20 kpc separation}

\begin{figure}
\centering
\includegraphics[scale=0.55,trim= 0 0 10 30,clip]{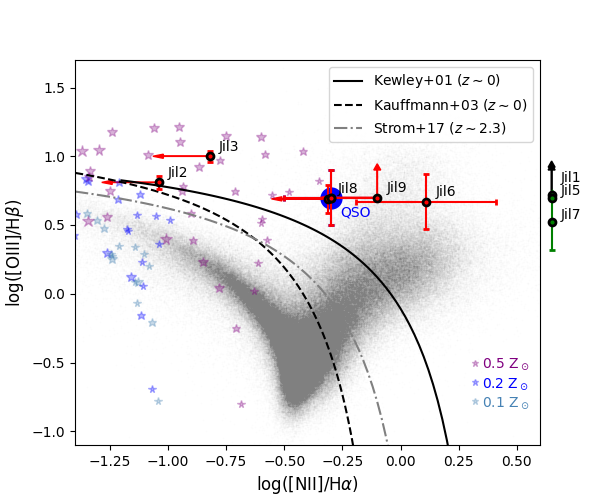}

\caption{ BPT diagnostic diagram. Red points represent  flux ratios inferred from the integrated Jil spectra, while the blue point indicates the QSO host ratios. For Jil1, Jil5, and Jil7, \oiii/\hb \ ratios are reported outside of the BPT, as \ha \ and \nii \ are undetected for these companions.  Local galaxies from SDSS DR7 (\citealt{Abazajian2009}) are indicated in grey, while small stars represent model predictions for low-metallicity AGN from \citet[][see this paper for a plethora of physical parameters related to gas and AGN properties, such as ionisation and accretion disk temperature]{Nakajima2022}, as labelled. The dashed line indicates the demarcation by
\citet{Kauffmann2003} between star-forming galaxies (left) and AGN (right) at low $z$; the solid line from \citet{Kewley2001} includes more
extreme starbursts and composite objects among the star-forming galaxies at low $z$; the dot-dashed grey line from
 \citet{Strom2017} shows the locus of star-forming galaxies at $z\sim 2$. 
}
\label{fig:BPT}
\end{figure}

All flux ratios so far inferred for the Jil targets (and for the QSO host galaxy) are reported in Fig. \ref{fig:BPT}. These constraints locate almost all Jil sources in the AGN regions of the BPT diagram; for the remaining sources not included in the diagram,  Jil1, Jil5, and Jil7, for which we cannot detect \ha\ or \nii, we can likely assume physical conditions similar to those in the other Jil companions, because of the similarly high \oiii /\hb \ ratios.  

The line ratio diagram also shows that Jil2 and Jil3 galaxies are associated with very stringent upper limits for the log(\nii /\ha), of the order of $\lesssim -1$. This may indicate that they are metal-poor AGN or galaxies, consistent with model predictions ($Z \lesssim 0.5 Z_\odot$; e.g. \citealt{Groves2004,Baron2019b}; see e.g. the predicted ratios from \citealt{Nakajima2022} reported in the figure), and the ultraviolet diagnostics (\citealt{Husemann2018a}). On the other hand, the \lbqs \ host might be associated with a higher metallicity ($Z \approx  Z_\odot$; according to the same grid models), because of the higher \nii /\ha.

The relative proximity of Jil5, Jil6, Jil7, and Jil8 to the QSO is a likely explanation for the high \oiii/\hb \ in such targets. On the other hand, the high \oiii /\hb \ in Jil1, Jil2, Jil3, and Jil9 can be explained by the presence of an AGN in Jil2, inferred by \citet{Husemann2018a} on the basis of ultraviolet diagnostics. In support of this scenario, we used the \heii $\lambda4686$ diagnostics (\citealt{Shirazi2012, Nakajima2022, Ubler2023, Tozzi2023}). Since the \heii $\lambda4686$ is undetected in NIRSpec, we used the ratio \heii $\lambda1640$/\heii $\lambda4686 = 7.2$, expected for recombination (\citealt{Seaton1978}), to infer the \heii $\lambda4686$ flux in Jil2 (correcting for extinction). This gives for Jil2 a log(\heii $\lambda4686 /$\hb) $= -0.22$, consistent with AGN ionisation (see Fig. 7 in \citealt{Ubler2023}). We stress that the detection of \heii$\lambda1640$ emission line in the surroundings of QSOs (i.e. at scales $> 10$ kpc) is not common: for instance, this line has been tentatively detected (at $\sim 2\sigma$) by stacking MUSE data cubes of 27 bright QSOs at $z = 3-4.5$   by \citet[][to be compared with the $>10\sigma$ detection in Jil2]{Fossati2021}. 

For Jil2, we also report a $\sim 4\sigma$ detection of the \sii \ doublet, and hence a log(\sii/\ha) $= -0.52\pm 0.07$. This value places Jil2 in the Seyfert-like region of the line ratio diagnostic diagram  \oiii/\hb\ versus \sii/\ha\ (\citealt{Veilleux1987}).

We infer for Jil2 an AGN bolometric luminosity log(L$_{\rm {bol}}$/ [erg/s]) $\sim 45.8$, from the narrow \hb \ luminosity (corrected for extinction; see Table \ref{tab:Jilproperties}), following \citealt{Netzer2019}. 
This result is consistent with the predictions reported in \citet{Husemann2018a, Husemann2021}, to explain the presence of \heii $\lambda1640$ in the Jil2 spectrum. 
All the arguments raised so far  therefore further support the scenario of a dual QSO in this complex system at $z\sim 3.3$.

\subsubsection{Mergers as drivers for rapid SMBH growth?}

Although the detailed physical connections among the eight companions - and with the QSO host - is difficult to establish with the present data, it is remarkable that \lbqs \ has this set of Jil galaxies within a (projected) distance of $\sim 20$ kpc, all within  a velocity range of $\sim \pm 250$ \kms \ from the QSO host systemic velocity. 
A blank field at $z \sim 3$ is expected to have a space density of 
 $\sim 0.01$ \oiii \ emitters (with L(\oiii) $>10^{41}$ erg/s) per Mpc$^{-3}$ (\citealt{Khostovan2015,Hirschmann2022}); this corresponds to $5\times 10^{-4}$ expected galaxies within a $\sim 3\arcsec\times3\arcsec$ region (the NIRSpec FOV), and within the narrow redshift range associated with the Jil companions ($z = 3.286-3.290$). We conclude therefore that \lbqs \ is sitting in a ultra-dense environment, being its space density many orders of magnitude higher than the general field.

Interestingly, both ground- and space-based observations of $z>3$ QSOs have shown that the presence of companions is common:  for instance, sub-millimetre galaxies and \lya \ emitters in the vicinity of high-$z$ QSOs have been identified with ALMA (e.g. \citealt{Trakhtenbrot2017,Venemans2020, Bischetti2021,Garcia2022}) and MUSE (e.g. \citealt{Fossati2021}), respectively. Indeed, almost all luminous high-$z$ QSOs so far observed with  JWST/NIRSpec IFS (\lbqs; SDSS J1652+1728 in \citealt{Wylezalek2022}; DELS J0411$-$0907 and VDES J0020$-$3653 in \citealt{Marshall2023}; GS\_3073 in \citealt{Ubler2023}) and JWST/NIRCam WFSS (SDSS J0100+2802 in \citealt{Kashino2022}) are surrounded by newly discovered companions. 

These results clearly support the idea that mergers can be important drivers for rapid early SMBH growth (e.g. \citealt{Hopkins2008,Zana2022}). Indeed, NIRSpec IFS, thanks to its high sensitivity and angular resolution ($\sim 0.8$ kpc  in a FOV of $25\times 25$ kpc$^2$ at $z\sim 3$), is revealing tidal bridges and tails at kiloparsec scales connecting such companions, hence allowing the study of galaxy interactions at such high redshifts.  

\section{Conclusions}\label{sec:conclusions}

We have presented JWST/NIRSpec integral field spectroscopy of the blue QSO \lbqs \ at $z=3.2870$. These observations cover a contiguous sky area of $\sim 3\arcsec \times 3\arcsec$ (23 $\times$ 23 kpc$^2$), which allowed us to map the extension of the QSO host as well as characterise its environment with a spatial sampling of $\sim 0.4$ kpc.
The main results of our analysis focussed on the QSO host are summarised below.

\begin{itemize}
    \item By analysing the integrated QSO spectrum, we measured the black hole mass from the \hb\ and \ha\ broad lines: $M_{\rm {BH}}\approx 2\times 10^9$ M$_\odot$. With a bolometric luminosity of log(L$_{\rm {bol}}$/ [erg s$^{-1}$]) $\sim 47.2$, this QSO is accreting material close to the Eddington limit ($\lambda_{\rm Edd}=0.9\pm 0.1$). 
    \item We have presented and make available for download a new procedure to model and subtract the apparent wiggles in single-spaxel spectra due to the spatial under-sampling of the PSF in NIRSpec IFS observations (see Figs. \ref{fig:3x3wiggles50mas} and \ref{fig:LBQScentralspaxel}). This correction is essential for performing spatial analyses of extended emission sitting below a point source, such as for studies of QSO hosts and close environments.
    \item We performed a QSO–host decomposition using models of the QSO broad lines, and used multi-component kinematic decomposition of the optical emission lines to infer the physical properties of the emitting gas in the \lbqs \ host, as well as in its environment. 
    \item We revealed a broadly regular velocity field in the QSO host, which is possibly tracing a warm rotating disk with $v_{\rm{rot}}/\sigma_0 \approx 2$, as inferred from {\sc{3D-Barolo}} modelling. We also derived a tentative  dynamical mass for the host, M$_{\rm{dyn}} = (14\pm 6)\times 10^{10}$ M$_\odot$; this places our galaxy slightly above the local black hole--host mass relation \citep{Kormendy2013}, consistent with other high-$z$ QSOs.  
    
    \item We identified a powerful outflow, with a velocity $v_{\rm{out}} \sim 1000$ \kms and a mass rate $\dot M_{\rm{out}} \sim
 10^4$ M$_\odot$ yr$^{-1}$. Its kinetic and momentum powers are compatible with the general predictions of AGN feedback models (e.g. \citealt{Harrison2018}).
    
    \item Standard BPT line ratios indicate that the central QSO dominates the ionisation state of the gas, with no obvious sign of a contribution from young stars in the host galaxy. 

\end{itemize}

We also studied the complex, ultra-dense environment of \lbqs\  thanks to the large FOV of our IFS observations, covering three out of the four companions already discovered by \citet{Husemann2021}.\ Our main results are as follows.

\begin{itemize}
    \item We detected eight Jil companion objects close to \lbqs, three of which were already discovered with MUSE and HST observations (\citealt{Husemann2018a,Husemann2021}), for a total of nine companions within 30 kpc of the QSO. All of these companions are within $\pm 250$ \kms \ of the QSO systemic velocity. 
    \item Regular velocity gradients, possibly tracing rotating gas, were detected in Jil2 and Jil3. For these targets, we derived tentative dynamical masses of the order of $10^{10}$ M$_\odot$. However, we caution that the observed velocity gradients may also be due to merger processes between different companions.  
    \item  Though difficult to determine, some  morpho-kinematic structures suggest that the Jil companions may be connected with the QSO \lbqs, so we can speculate  that they contribute to its feeding.  In particular, Jil5 shows evidence of gravitational interaction with the QSO host. 
    \item  All BPT line ratios measured for Jil companions are compatible with AGN ionisation. 
    \item We provide further evidence for the presence of an obscured QSO at $\sim 20$ kpc from \lbqs\  on the basis of \oiii /\hb, \sii /\ha, and \heii /\hb \ line ratios. This QSO is likely responsible for the gas ionisation in the surroundings of Jil2.   
\end{itemize}

This work has explicitly demonstrated the exceptional capabilities of the JWST/NIRSpec IFS to study the QSO environments in the early Universe. With a total exposure time of $\sim 1 $ hour, we unveiled in unprecedented detail the interstellar properties of the \lbqs \ host galaxy and those of its multiple companions in its immediate vicinity. 

The study of the \lbqs \ host galaxy was limited by PSF artefacts; before we could subtract them, we had to address the wiggles. We have shown that wiggles can be modelled and subtracted, taking advantage of the fact that their frequency, $f_w$, changes smoothly as a function of the wavelength and, most importantly, that $f_w$ does not show spaxel-to-spaxel variations. However, this step adds further difficulties in the analysis of the NIRSpec data cubes. We note that the amplitude of these artefacts decreases as the number of exposures increases. This information should be taken into consideration by observers when planning NIRSpec IFS observations. 

\vspace{1cm}

\begin{acknowledgements}

This work is based on observations made with the NASA/ESA/CSA James Webb Space Telescope. The data were obtained from the Mikulski Archive for Space Telescopes at the Space Telescope Science Institute, which is operated by the Association of Universities for Research in Astronomy, Inc., under NASA contract NAS 5-03127 for JWST. These observations are associated with program \#1220, as part of the NIRSpec Galaxy Assembly IFS GTO program.

We are grateful to the anonymous referee for a constructive report that helped to improve the quality of this manuscript. We thank Kimihiko Nakajima for providing the theoretical model grids published by \cite{Nakajima2022}, and David Law and Bartolomeo Trefoloni for helpful comments on an earlier version of this manuscript.
MP, SA, and BRP  acknowledge support from the research project PID2021-127718NB-I00 of the Spanish Ministry of Science and Innovation/State Agency of Research (MICIN/AEI). MP also acknowledges support from the Programa Atracci\'on de Talento de la Comunidad de Madrid via grant 2018-T2/TIC-11715.
MAM acknowledges the support of a National Research Council of Canada Plaskett Fellowship, and the Australian Research Council Centre of Excellence for All Sky Astrophysics in 3 Dimensions (ASTRO 3D), through project number CE170100013.

RM, JS and FDE acknowledge support by the Science and Technology Facilities Council (STFC), from the ERC Advanced Grant 695671 ``QUENCH''.
RM and JS also acknowledge funding from a research professorship from the Royal Society. 
GC acknowledges the support of the INAF Large Grant 2022 ``The metal circle: a new sharp view of the baryon
cycle up to Cosmic Dawn with the latest generation IFU facilities''.
H{\"U} gratefully acknowledges support by the Isaac Newton Trust and by the Kavli Foundation through a Newton-Kavli Junior Fellowship.
AJB, GCJ and AJC acknowledge funding from the ``FirstGalaxies'' Advanced Grant from the European Research Council (ERC) under the European Union’s Horizon 2020 research and innovation program   (Grant agreement No. 789056). 
SC acknowledges support from the European Union (ERC, WINGS,101040227).
PGP-G acknowledges support  from  Spanish  Ministerio  de  Ciencia e Innovaci\'on MCIN/AEI/10.13039/501100011033 through grant PGC2018-093499-B-I00.
IL acknowledges support from PID2022-140483NB-C22 funded by AEI 10.13039/501100011033 and BDC 20221289 funded by MCIN by the Recovery, Transformation and Resilience Plan from the Spanish State, and by NextGenerationEU from the European Union through the Recovery and Resilience Facility.

\\  

This research has made use of NASA's Astrophysics Data System, QFitsView, and SAOImageDS9, developed by Smithsonian Astrophysical Observatory. It also had made use of Python packages and software
AstroPy \citep{Astropy2013},
Matplotlib \citep{Matplotlib2007},
NumPy \citep{Numpy2011},
OpenCV \citep{bradski2000}, opencv-python,
Photutils \citep{photutils},
Regions \citep{Bradley2022},
QDeblend3D \citep{Husemann2013,Husemann2014}.

\end{acknowledgements}

%
%
\bibliographystyle{aa}
\bibliography{46649corr.bib}

\begin{appendix}

\section{Nuclear spectra}

\begin{figure*}
\includegraphics[scale=0.41,trim= 20 0 60 40,clip]{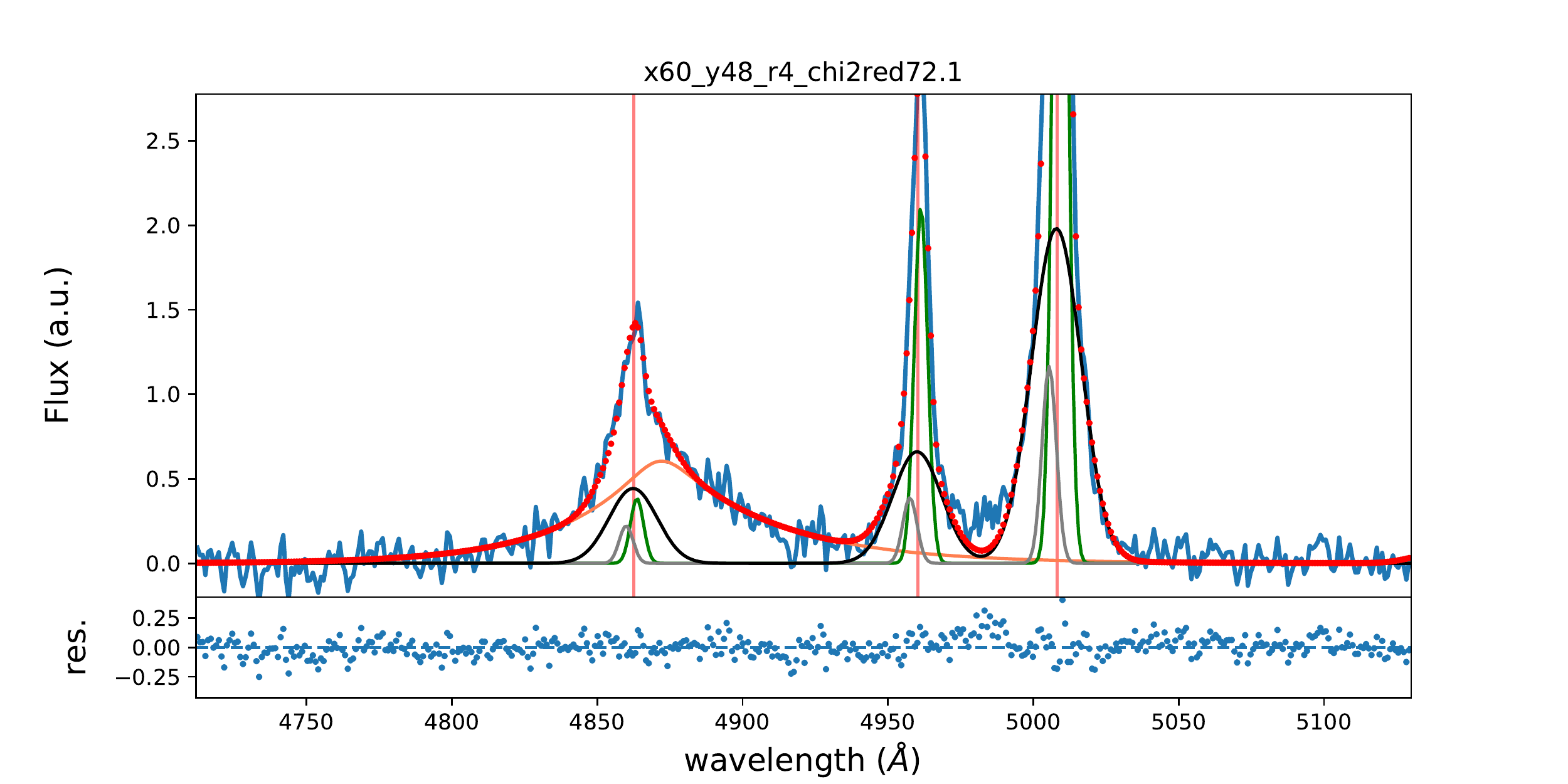}
\includegraphics[scale=0.41,trim= 42 0 60 40,clip]{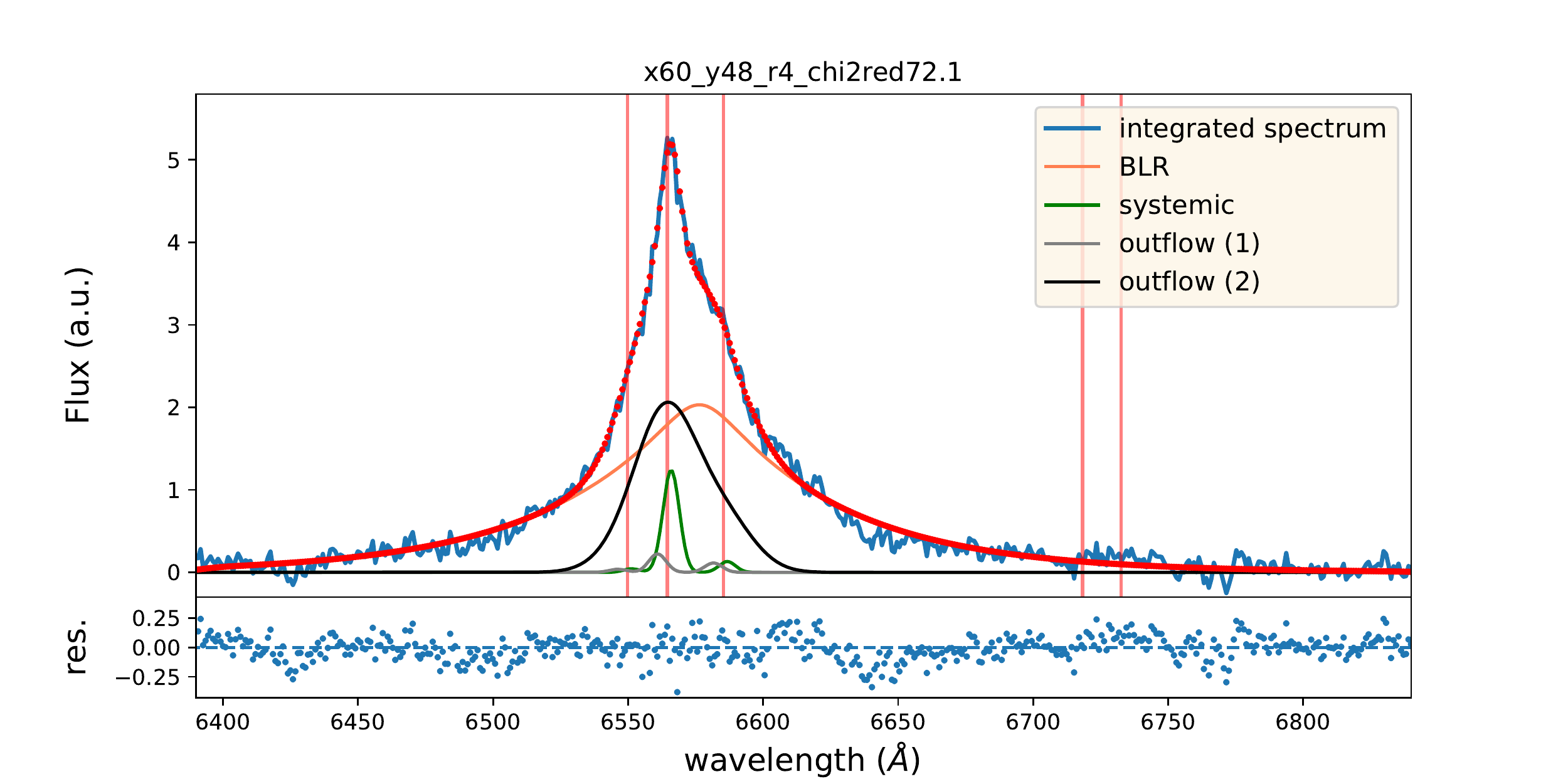}

\includegraphics[scale=0.41,trim= 20 0 60 40,clip]{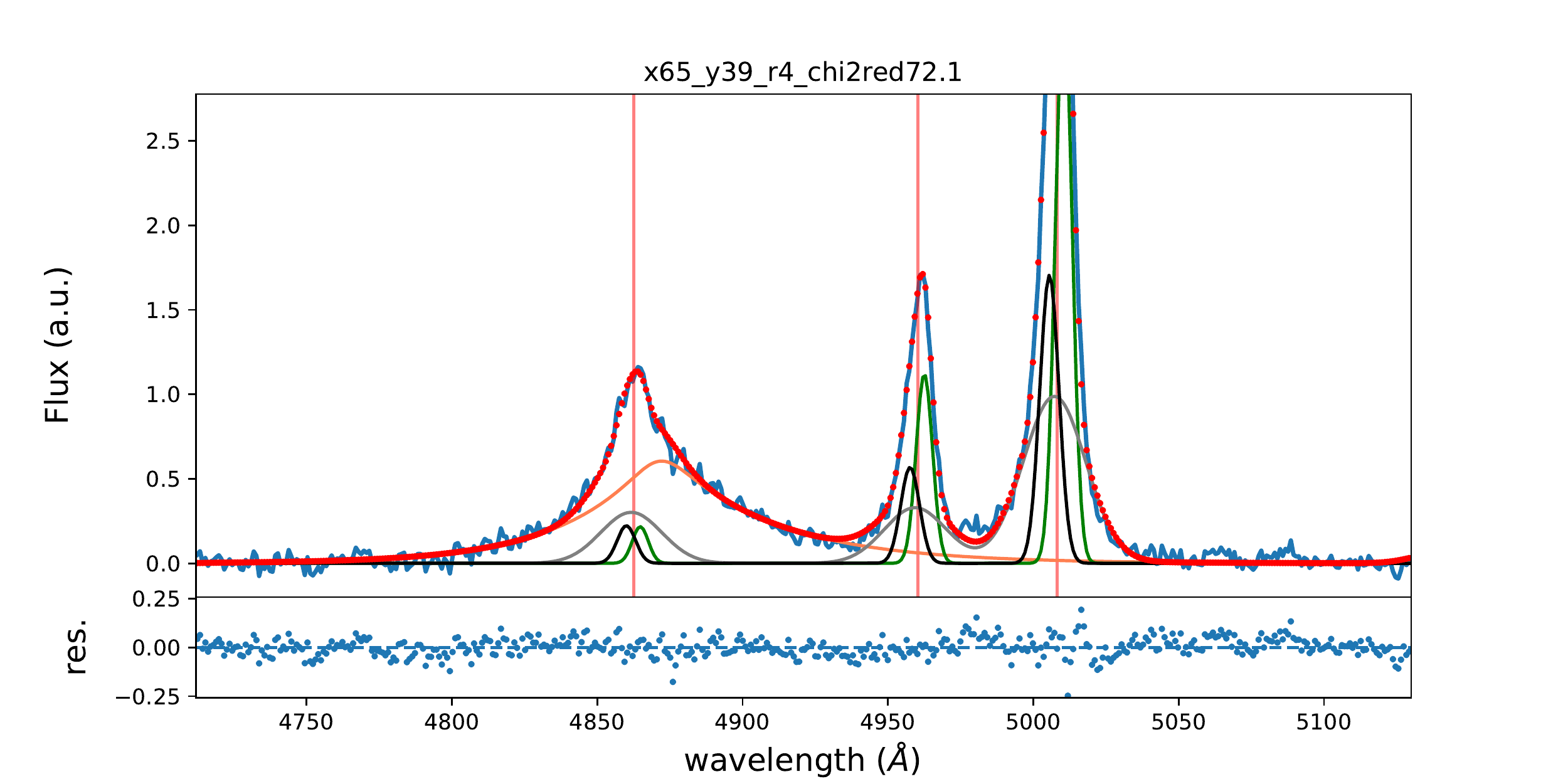}
\includegraphics[scale=0.41,trim= 42 0 60 40,clip]{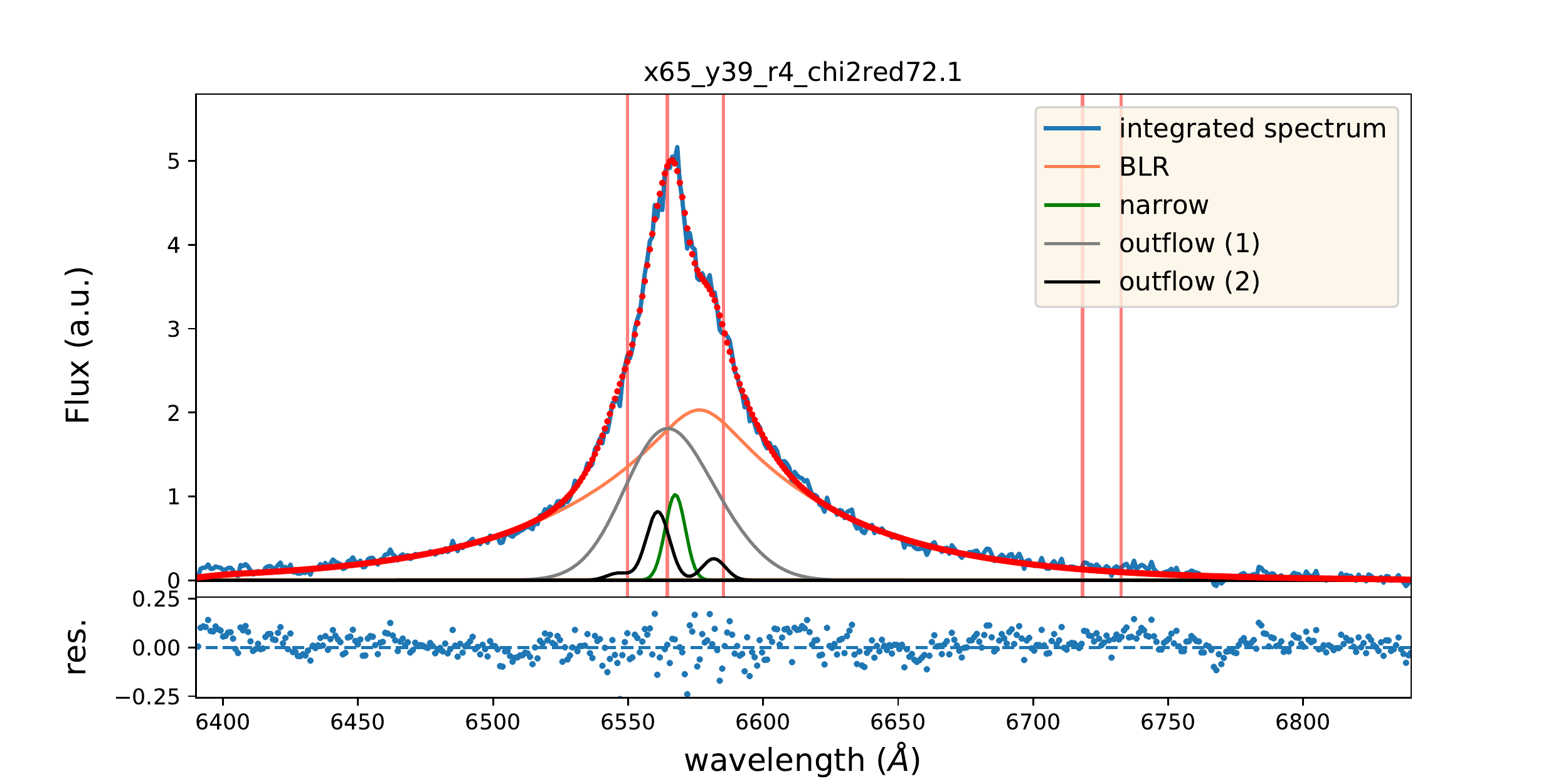}

\includegraphics[scale=0.41,trim= 20 0 60 40,clip]{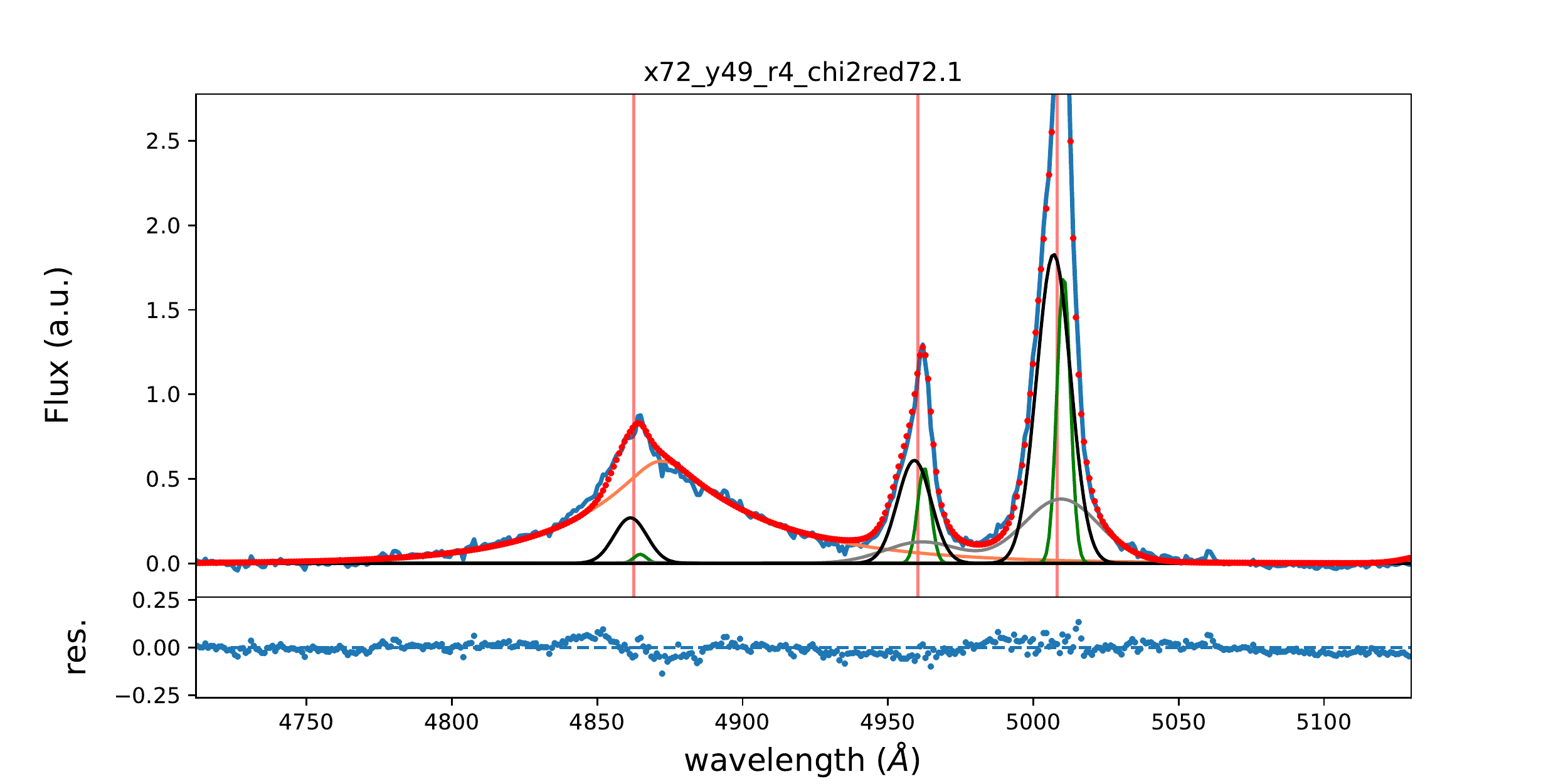}
\includegraphics[scale=0.41,trim= 42 0 60 40,clip]{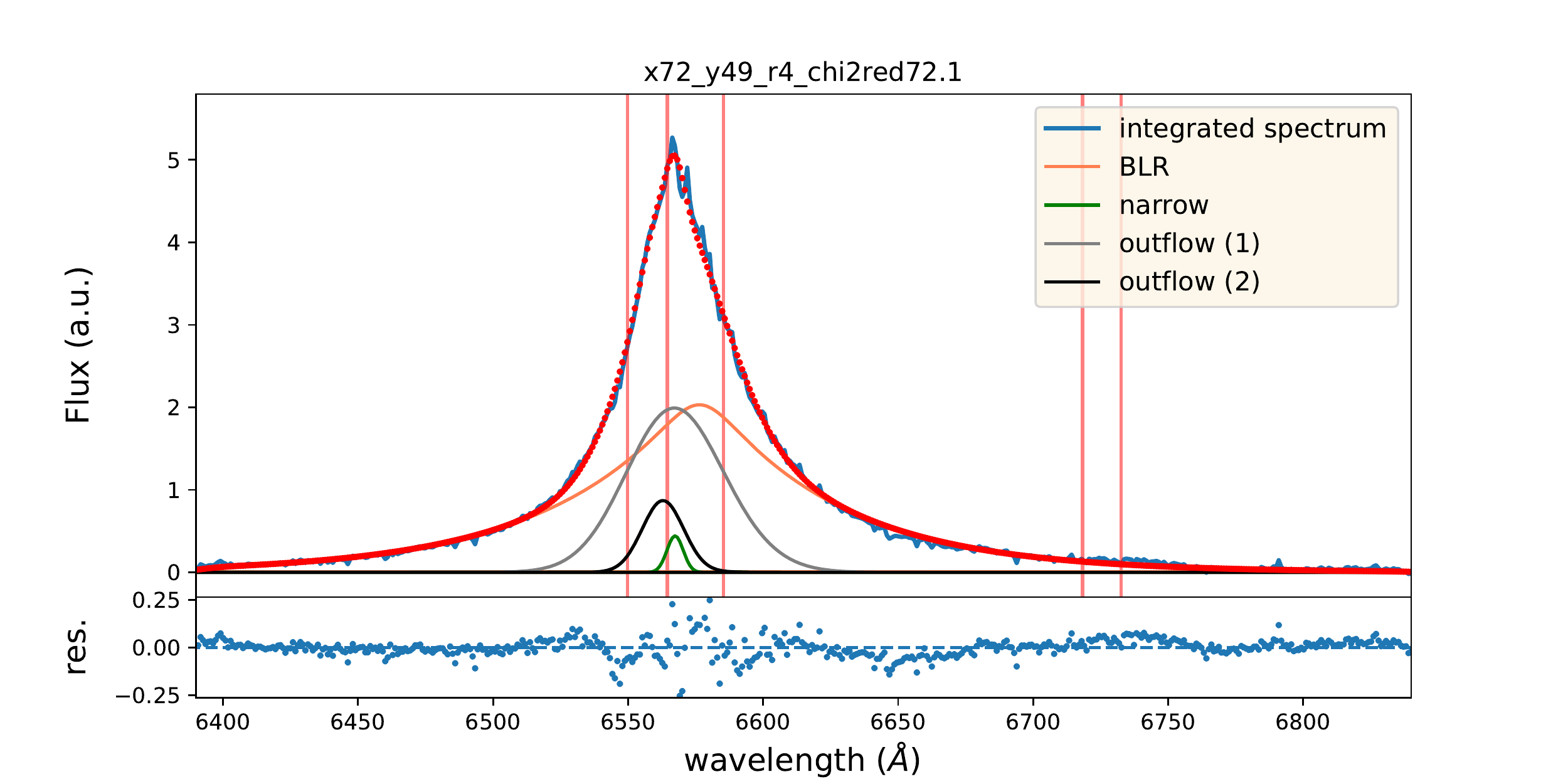}

\includegraphics[scale=0.41,trim= 20 0 60 40,clip]{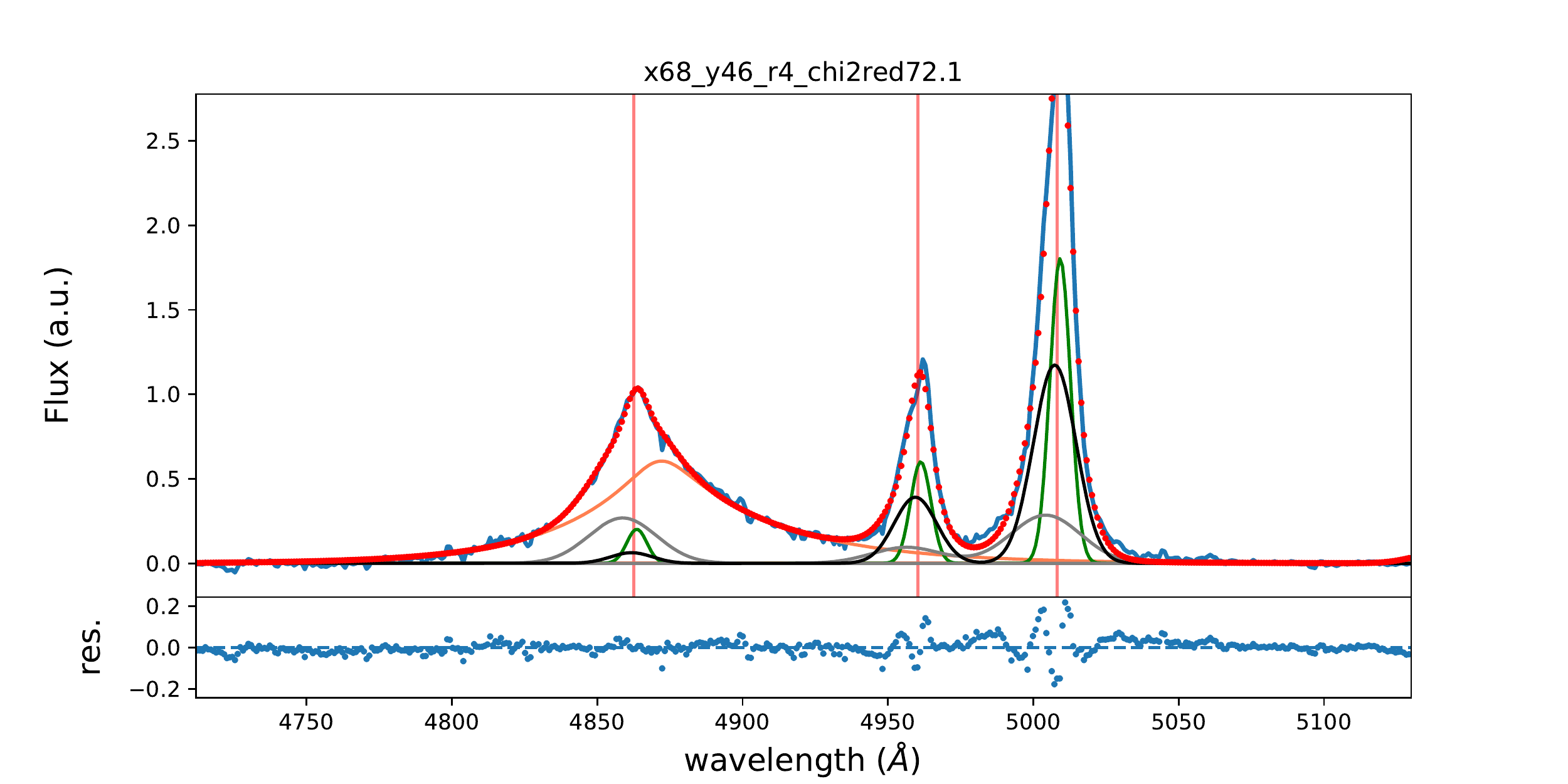}
\includegraphics[scale=0.41,trim= 42 0 60 40,clip]{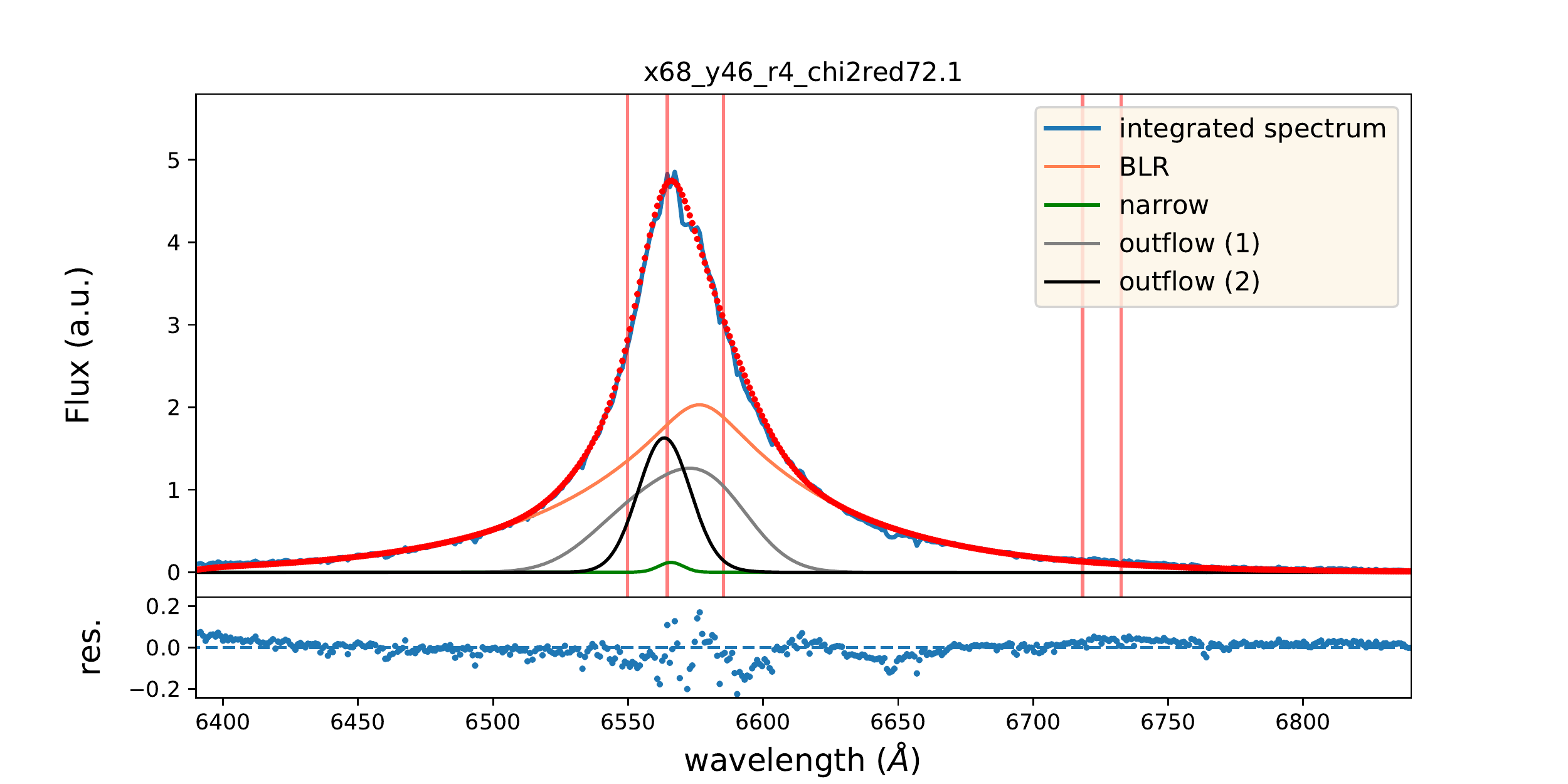}

\caption{ Integrated spectra extracted from circular regions with a radius of 0.2\arcsec and centred at different positions within a few spaxels of the peak emission of the QSO. The best-fit models shown here were obtained by fitting the four spectra with the same BLR profiles, as explained in Sect. \ref{sec:integratedspec}. 
}
\label{fig:Aintegratedspec}
\end{figure*}

Figure \ref{fig:Aintegratedspec} shows the simultaneously fit of four spectra extracted from circular regions with radius of 0.2\arcsec (4 spaxels) and centred at different positions within a few spaxels from the peak emission of the QSO, used to reduce the degeneracy between BLR and NLR. All spectra are normalised so that the BLR wings of the \ha \ and \hb \ have the same fluxes, and can be fitted with the same broken power-law functions. During the fit, BLR profiles are therefore tied, assuming that these emission components originate from the same unresolved region. All other components are free to vary as they originate from more extended and likely resolved regions. The small aperture radius is required to observe significant variations in the \ha-\nii \ complex (e.g. with respect to the integrated spectrum in Fig. \ref{fig:HaHbfit}). 

\section{Sinusoidal-type patterns}\label{sec:Awiggles}

In Sect. \ref{sec:wiggles}, we proved that our approach is capable of modelling and correcting for the spurious wiggles in the single-spaxel spectra. However, it has important limitations. For instance, some residual wiggles are present in a few spaxels (see e.g. Fig. \ref{fig:3x3wiggles50mascorr}). 
Moreover, in the very innermost nuclear regions, the \ha \ BLR emission covers a significant number of wavelength channels, preventing a proper modellisation of the underlying wiggles in the vicinity of the \ha \ line (see Fig. \ref{fig:LBQScentralspaxel}). This can affect the reconstruction of the \ha \ kinematics. 

Another aspect is related to the emission line fluxes: an improper correction of the wiggles implies an incorrect reconstruction of the emission line profile and, as a consequence, an incorrect measurement of its integrated flux. For \lbqs, we check that the \oiii$\lambda\lambda4959,5007$ line ratio is preserved at 1:3, which is consistent with theory (\citealt{Osterbrock2006}). Figure \ref{fig:Aoiiiratios} shows the nuclear spectra extracted from different areas (integrating over circular regions with radius from 1 to 5 spaxels), from the original cube (top panel) and the one corrected for wiggles (bottom). All spectra are continuum-subtracted and normalised to the \oiii \ peak; the inset in the bottom panel shows that the \oiii$\lambda4959$ peaks at $\sim 0.33$, consistent with the expectations. We note however that significant deviations (up to $\sim 50\%$) are observed in individual spaxels, both in the original and in the corrected spectra, although the corrected ones have line ratios closer to the theoretical 1:3 ratio. 
We also checked that our corrections preserve the shape of the spectrum and integrated fluxes, as shown in Fig. \ref{fig:wigglesfluxcons}.

Therefore, we caution that the presence of wiggles might affect both the kinematics and flux ratio measurements; a proper modellisation and subtraction of the wiggles is required to mitigate their effects. In fact, as shown in Fig. \ref{fig:Aoiiiratios}, off-centred integrated spectra are always affected by these wiggles.

\begin{figure*}[t]
\begin{center}
\includegraphics[scale=0.63,trim= 50 0 50 50,clip]{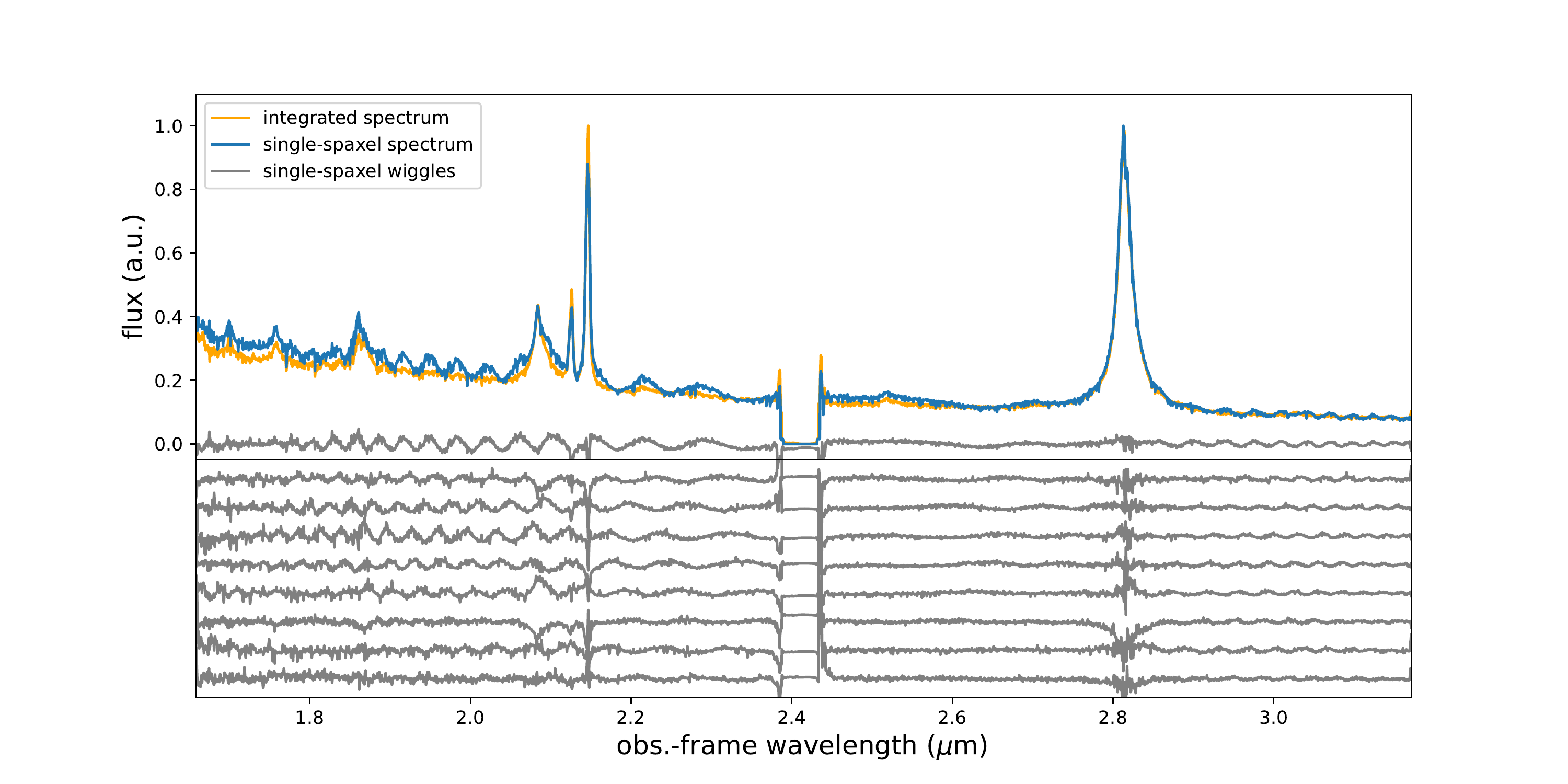}
\caption{Sinusoidal-type patterns in single-spaxel spectra extracted from the {\it emsm} data cube with a spaxel size of 0.1\arcsec. Top panel: \lbqs \ spectrum integrated over an aperture of $r=0.5$\arcsec (orange curve), in comparison with the spectrum of the brightest spaxel (blue curve). Both spectra are normalised to their maximum values for visualisation purposes. The wiggles affecting the single-spaxel spectrum are reported in grey and are obtained as the difference between the blue and orange curves (see Fig. \ref{fig:3x3wiggles50mas} for details). Bottom panel: Wiggles obtained from the eight pixels closest to the brightest one.  
}
\label{fig:3x3wiggles100mas}
\end{center}
\end{figure*}

\begin{figure*}
\begin{center}
\includegraphics[scale=0.55]{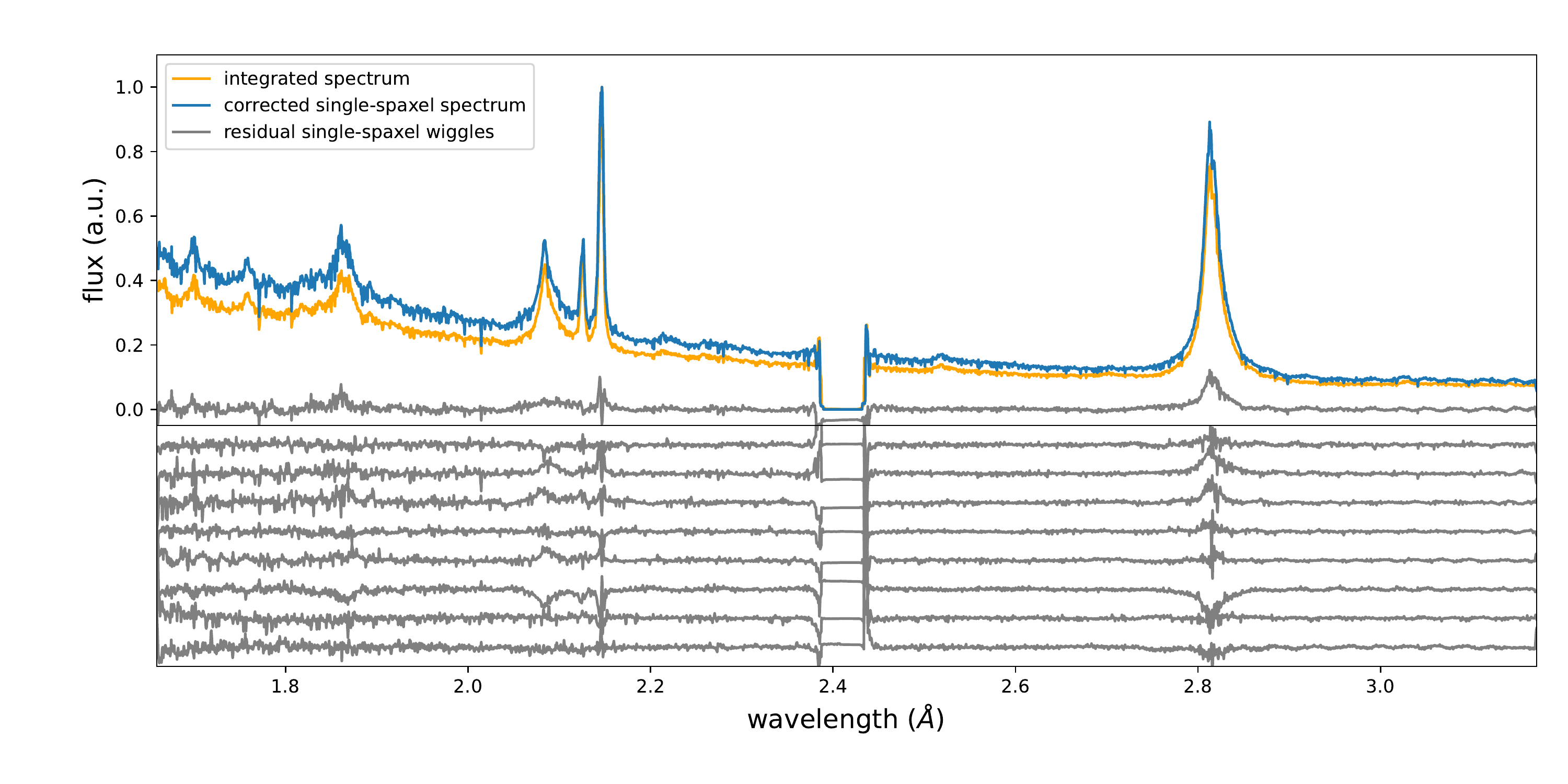}
\caption{Wiggle-corrected spectra extracted from the {\it emsm} data cube with a spaxel size of 0.1\arcsec. Top panel: \lbqs \ spectrum integrated over an aperture of $r=0.5$\arcsec (orange curve), in comparison with the spectrum of the brightest spaxel, after the wiggle subtraction (blue curve). Both spectra are normalised to 1 for visualisation purposes. The residuals are reported in grey and are obtained as the difference between the blue and orange curves (see Fig. \ref{fig:3x3wiggles50mas} for details). Bottom panel: Residuals obtained from the eight spaxels closest to the brightest one. The most significant residuals are found at the position of the brightest emission lines: they are not due to the wiggles, but to the line profile variations.  
}
\label{fig:3x3wiggles100mascorr}
\end{center}
\end{figure*}

\begin{figure*}
\begin{center}
\includegraphics[scale=0.55]{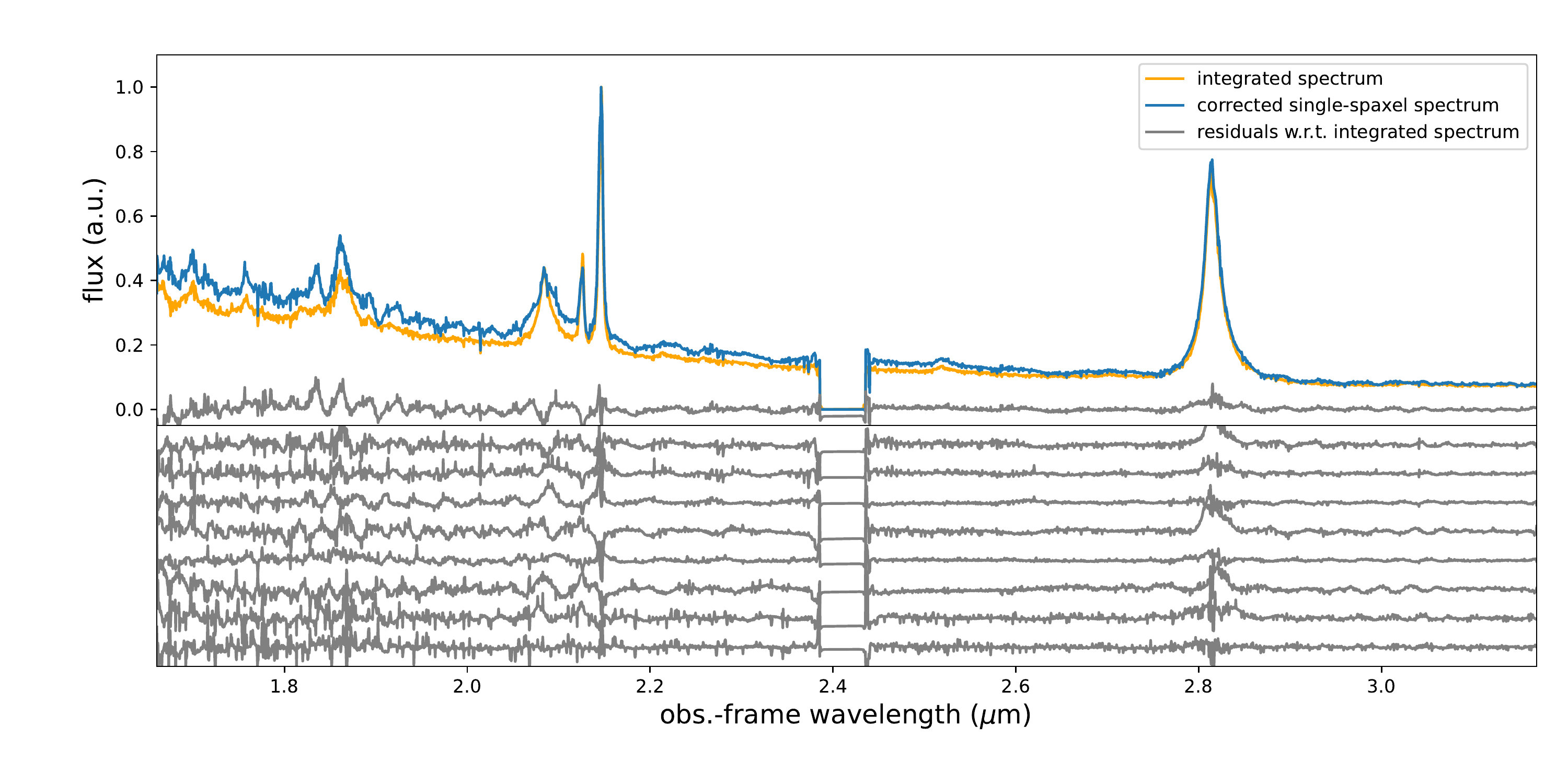}
\caption{Same as Fig. \ref{fig:3x3wiggles100mascorr}, but for the \drizzle \ data cube, with spaxels of 0.05\arcsec.  
}
\label{fig:3x3wiggles50mascorr}
\end{center}
\end{figure*}

\begin{figure*}
\begin{center}
\includegraphics[scale=0.35]{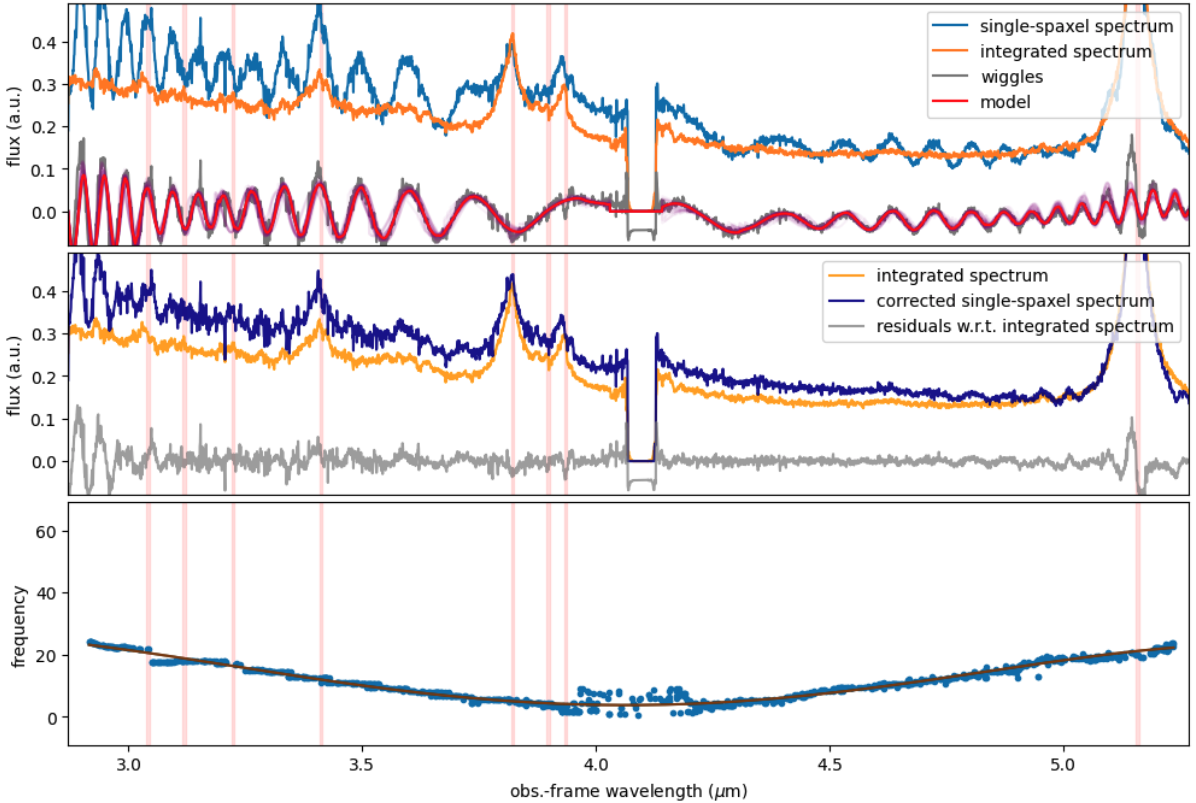}
\caption{Same as Fig. \ref{fig:LBQScentralspaxel}, but for the \drizzle \ data cube of the QSO VDES J0020-3656 (\citealt{Marshall2023}), with spaxels of 0.05\arcsec. 
}
\label{fig:VDEScentralspaxel}
\end{center}
\end{figure*}

\begin{figure*}
\begin{center}
\includegraphics[scale=0.45,trim= 45 0 20 0,clip]{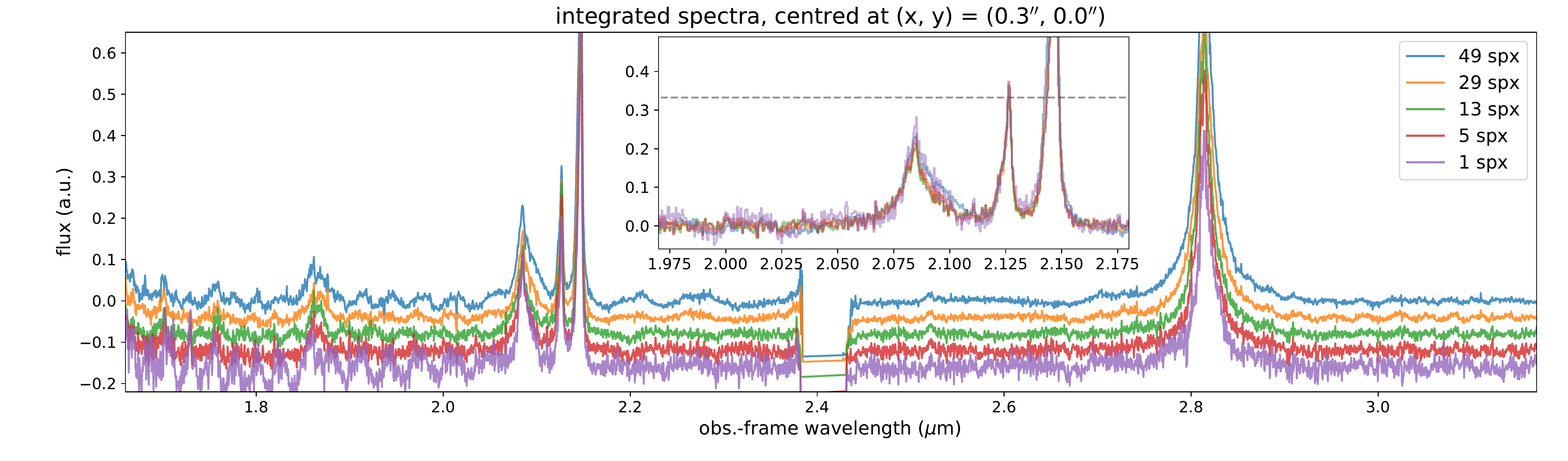}
\includegraphics[scale=0.45,trim= 45 0 20 0,clip]{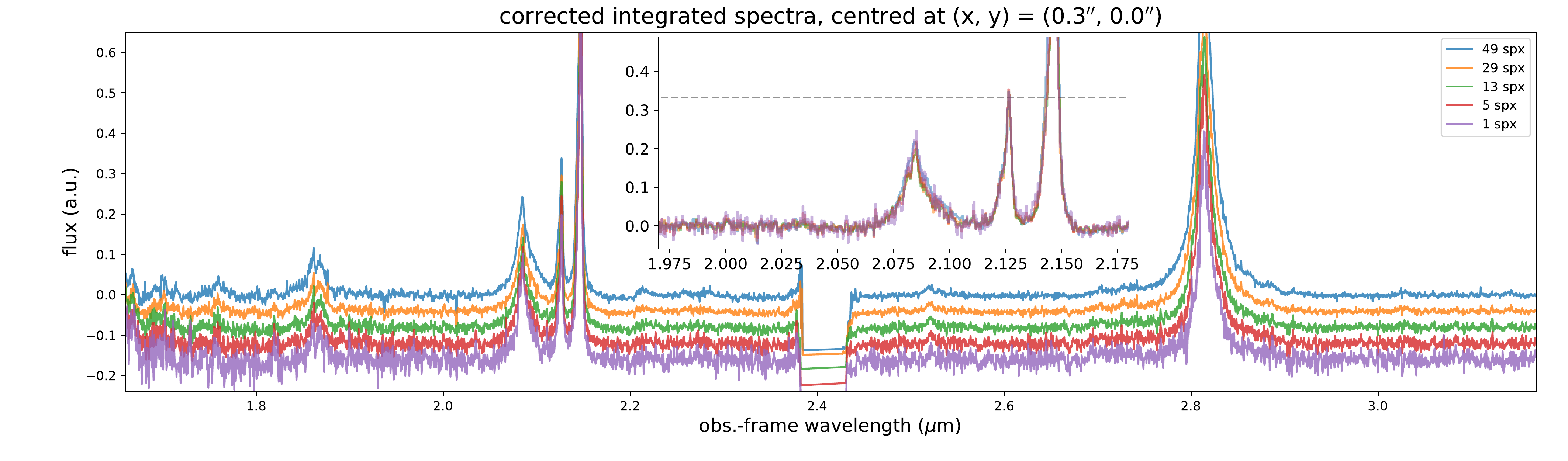}
\caption{Integrated spectra extracted from circular regions containing 1 to 49 spaxels (corresponding to radii of 1 to 5 spaxels), centred at 0.3\arcsec  east of the \lbqs \ nucleus (from the \drizzle \ cubes with spaxels of 0.05\arcsec). The top panel shows the original spectra, while the bottom panel shows the same spectra after the correction for the wiggles at the spaxel level (Sect. \ref{sec:wiggles}). All spectra are continuum-subtracted and are normalised to the peak of \oiii; for those extracted from regions with radii $< 5$ spaxels, we added vertical offsets to ease the visualisation. The insets show a zoomed-in view of the vicinity of the \oiii \ and \hb \ lines, without any vertical offset; these spectra show that the \oiii$\lambda4959$ peaks at $\sim 0.33$ (indicated by the horizontal dashed line), consistent with theoretical expectations.  
}
\label{fig:Aoiiiratios}
\end{center}
\end{figure*}

\begin{figure*}
\begin{center}
\includegraphics[scale=0.65,trim= 45 0 40 0,clip]{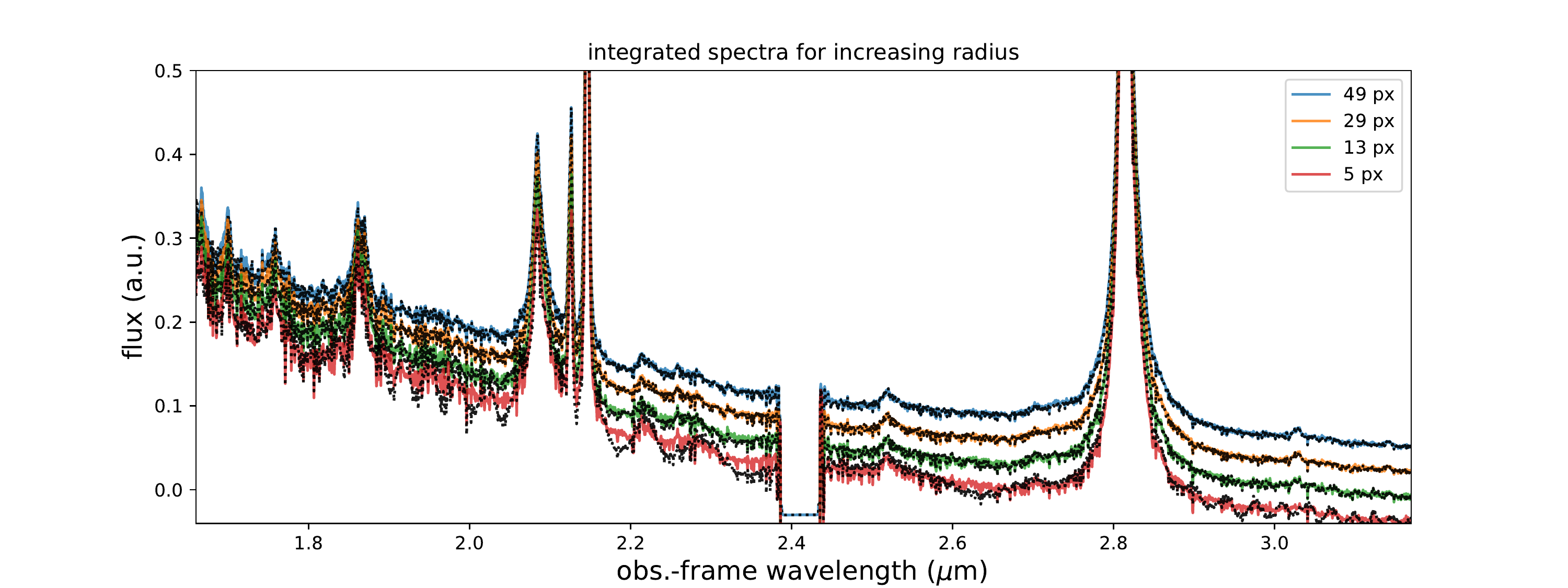}
\caption{ Integrated spectra extracted from circular regions containing 5 to 49 spaxels (corresponding to radii of 2 to 5 spaxels), centred on the \lbqs \ nucleus (from the \drizzle \ cubes with spaxels of 0.05\arcsec). The solid lines show the spectra after the wiggle subtraction, while the dotted lines show the original spectra.  All spectra are normalised to the peak of \oiii; for those extracted from regions with radii $< 5$ spaxels, we added vertical offsets to ease the visualisation.  The figure proves that our correction preserves the integrated fluxes and the shape of the spectrum. 
}
\label{fig:wigglesfluxcons}
\end{center}
\end{figure*}

\section{3D-Barolo fit}

\begin{figure}
\begin{center}
\includegraphics[scale=0.25]{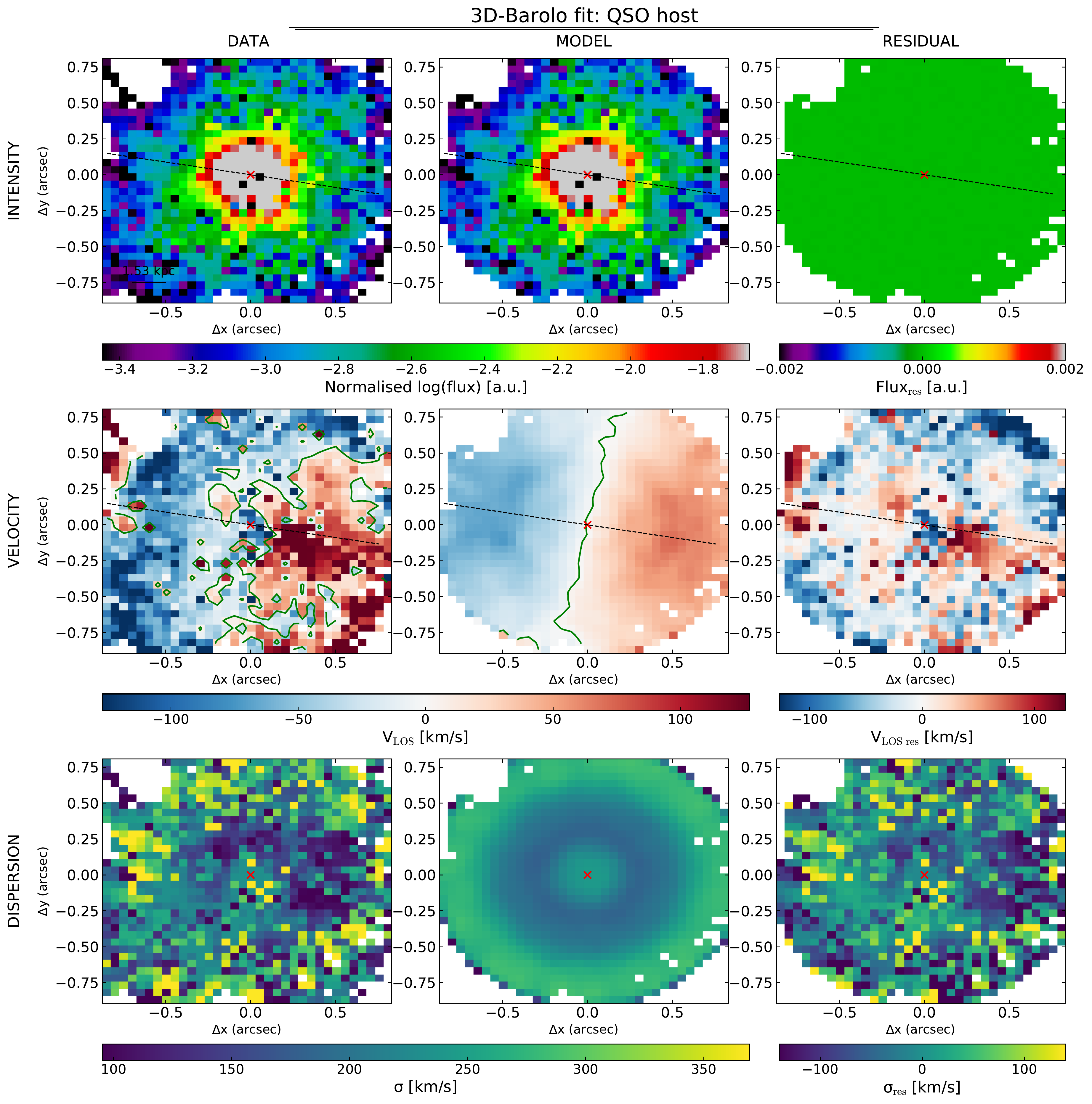}
\caption{\lbqs \  host galaxy disk kinematic best fit of Moment 0, 1, and 2 (first to third rows). These best fits are inferred from the analysis of the narrow \oiii \ component obtained from our multi-component Gaussian fit (i.e. all components with FWHM $< 600$ \kms in Fig. \ref{fig:components}). The black and green lines identify the major axis and the zero-velocity curve, respectively; the red cross identifies the QSO position.  
}
\label{fig:baroloHost}
\end{center}
\end{figure}

\begin{figure}
\begin{center}
\includegraphics[scale=0.25]{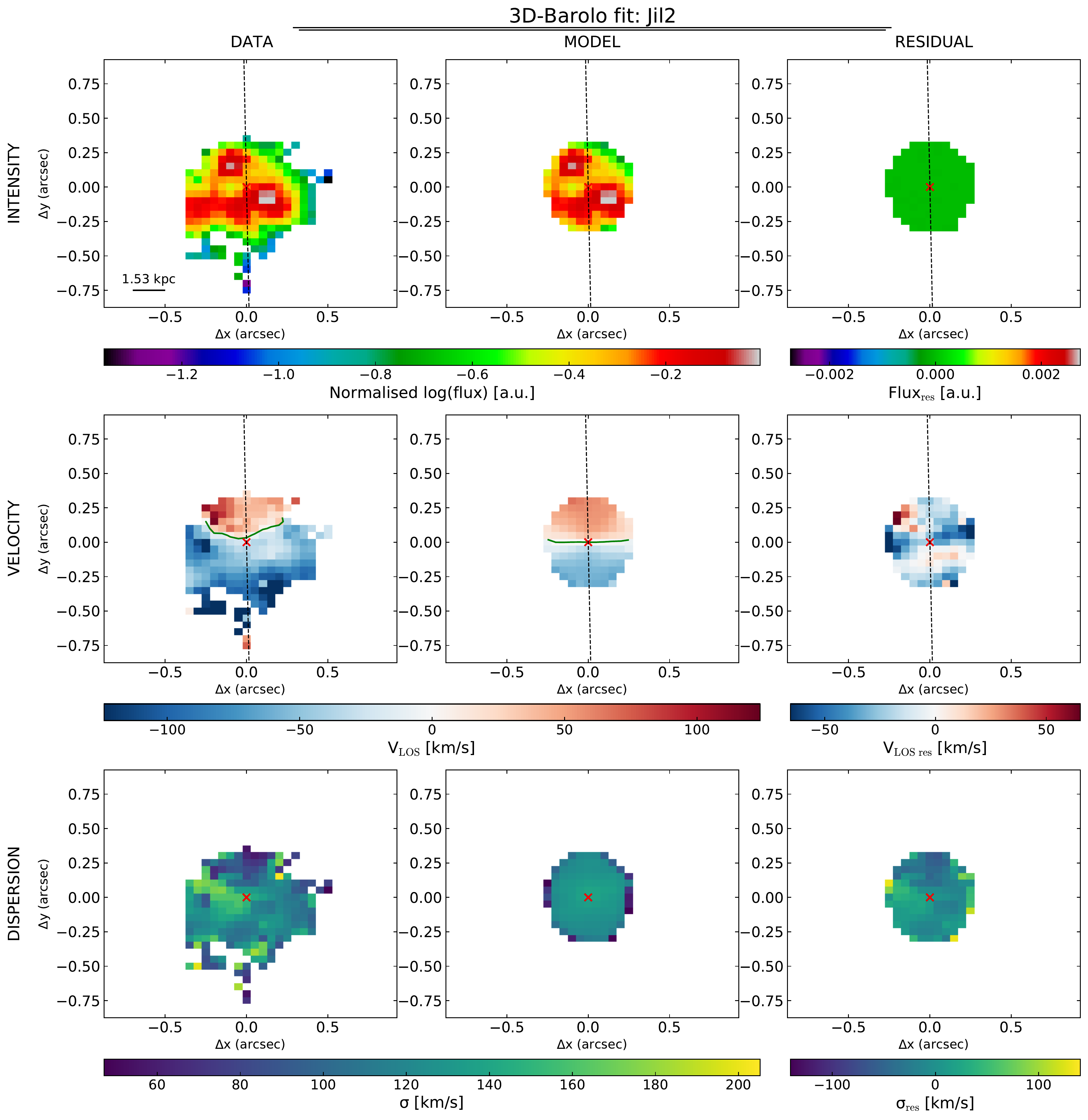}
\caption{Jil2 Moment 0, 1, and 2 and the {\sc{3D-Barolo}}  disk kinematic best fit.  See Fig. \ref{fig:baroloHost} for details.
}
\label{fig:baroloJil2}
\end{center}
\end{figure}

\begin{figure}
\begin{center}
\includegraphics[scale=0.25]{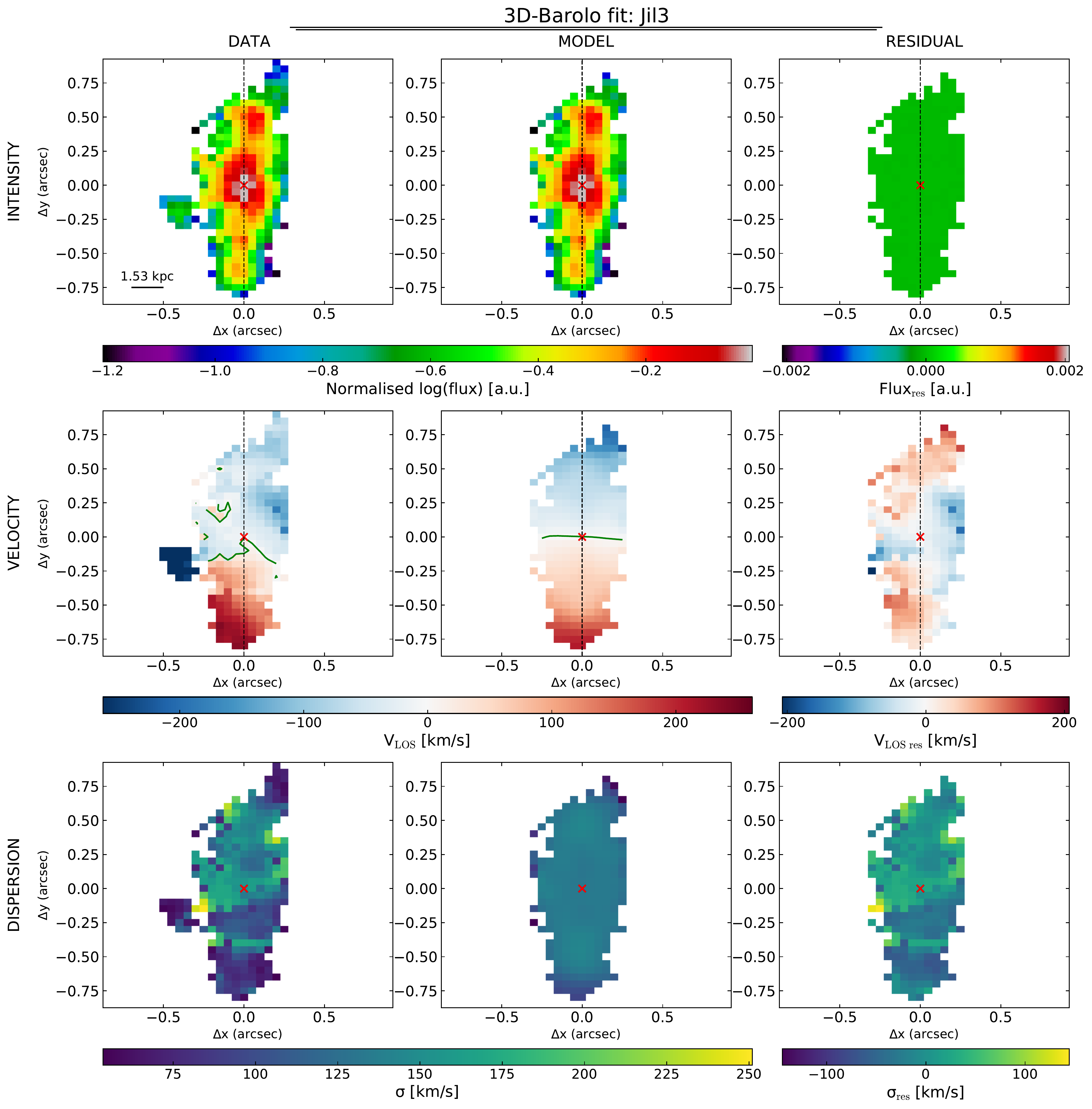}
\caption{Jil3 Moment 0, 1, and 2 and the {\sc{3D-Barolo}} disk kinematic best fit.  See Fig. \ref{fig:baroloHost} for details.
}
\label{fig:baroloJil3}
\end{center}
\end{figure}

Figures \ref{fig:baroloHost}, \ref{fig:baroloJil2}, and \ref{fig:baroloJil3} show the {\sc{3D-Barolo}} best-fit modellisation for the three targets that display broadly regular velocity gradients. We caution that the significant residuals, likely due to the superposition of different kinematic components associated with distinct clumps (or targets) on the same LOS, call into question the reliability of the inferred best-fit parameters. 

\end{appendix}

\end{document}